\documentclass[twocolumn,numberedappendix,appendixfloats]{openjournal}

\usepackage{xcolor}
\usepackage{textgreek}
\usepackage[utf8]{inputenc}
\usepackage[english]{babel}

\usepackage{hyperref}
\hypersetup{
    unicode, 
    colorlinks=true,
    linkcolor=linkcolor,
    citecolor=linkcolor,
    filecolor=linkcolor,
    urlcolor=linkcolor,
}
\usepackage{color,colortbl}
\definecolor{linkcolor}{rgb}{0.0,0.3,0.5}
\usepackage{tensind}
\tensordelimiter{?}
\DeclareGraphicsExtensions{.bmp,.png,.jpg,.pdf}
\usepackage{verbatim}
\usepackage[normalem]{ulem}
\usepackage{orcidlink}
\usepackage{natbib}
\usepackage{soul}
\usepackage{amsfonts}
\usepackage{amsmath}
\usepackage{amssymb}
\usepackage{graphicx}
\usepackage{setspace}

\graphicspath{ {./figs/} }

\begin{document}
\makeatletter
\long\def\frontmatter@title@above{%
  \vspace*{-\headsep}\vspace*{\headheight}
  \footnotesize
  {\footnotesize\textsc{Submitted to the Open Journal of Astrophysics}}\par
  \vspace*{-\baselineskip}\vspace*{0.625in}
}
\makeatother
\makeatletter
\submitted{}
\makeatother
\title{Constraining the Inner Dark-Matter Slope of Sculptor: A Comparative Analysis of Dynamical Methods}

\author{Rishi Sanjeev\orcidlink{0009-0003-7595-639X}}
\email{rishsanjeev@gmail.com}
\affiliation{Independent researcher, Wilmington, Delaware, USA}

\begin{abstract}

    The core-cusp problem is a discrepancy between simulations and observations of the slopes of inner dark matter density profiles. Dwarf spheroidal galaxies are the cleanest test due to dark matter dominance at all radii, as well as having low baryonic mass. The two-population estimator, separating dwarf spheroidals into metallicity-split subcomponents, is a standard method to break the mass-anisotropy degeneracy. Four dynamical methods are applied to the 1339-star \citet{Tolstoy_I} VLT/FLAMES Sculptor catalog: spherical Jeans, GravSphere, the \citet{2011ApJ...742...20W} two-population estimator, and a continuous $f(\boldsymbol{J},[\text{Fe}/\text{H}])$ action-based distribution function. This work finds $\Gamma=2.49^{+0.23}_{-0.20}$, with 99.6\% of the posterior lying above $\Gamma=2$ (84.4\% after $\lambda$ calibration), robust to membership cuts and metal-poor contamination; GravSphere gives $\gamma=0.46^{+0.40}_{-0.28}$. Controlled mocks that hold tracer geometry fixed while breaking the metallicity-kinematics link attribute $\sim77$--$83\%$ of the recovered bias to geometry rather than decomposition itself; thus, mock-based claims about population splitting must control for geometry. The continuous distribution function encodes a metallicity--action coupling $k_J$ inaccessible to any split-based method; a direct action-space computation on 575 stars with Gaia proper motions finds $[\text{Fe}/\text{H}]$ declining by $\sim0.3\,\text{dex}$ across three decades in radial action. On the central-density diagnostic of \citet{2019MNRAS.484.1401R}, the same posteriors give $\rho_\text{DM}(150\,\mathrm{pc})=1.77\times10^8\,M_\odot\,\mathrm{kpc^{-3}}$, agreeing with their independent Sculptor measurement at $0.74\sigma$ and placing Sculptor in their cusp-like class. The shallow $\gamma$ of this work is an asymptotic quantity; the local slope at $150\,\mathrm{pc}$ is consistent with the Navarro--Frenk--White profile under the spherical models used here, though a recent axisymmetric analysis using resolved 3D velocities finds a shallow slope over that range and identifies an inclination degeneracy to which spherical models are blind.

\end{abstract}


\maketitle

\section{Introduction}
\label{sec:intro}

In the standard model of cosmology, the Lambda-Cold Dark Matter ($\Lambda$CDM) model \citep{1998AJ....116.1009R,Perlmutter_1999}, $N$-body simulations predict dark matter (DM) to form sharply increasing density ``cusps'' in the centers of galaxies \citep{1991ApJ...378..496D,1996ApJ...462..563N,2008MNRAS.391.1685S}. The Navarro--Frenk--White (NFW) profile quantitatively represents the cusp with local density $\rho$ being a function of radius $r$ \citep{1997ApJ...490..493N}: \begin{equation}\label{eq:nfw}
\rho(r) = \frac{\rho_s}{x\,(1+x)^2}, \quad x \equiv r/r_s\end{equation}which in the limit $r \ll r_s$ reduces to $\rho(r) \propto x^{-1}$. $\rho_s$ represents the characteristic scale density of the galaxy, which normalizes the overall density, and $r_s$ the scale radius, which is defined as the radius from the center of the halo where $\gamma=2$ (see Equation~\ref{eq:inner-slope}). The inner logarithmic slope of a density profile is defined as: \begin{equation}\label{eq:inner-slope}\gamma \equiv -\frac{d\ln\rho}{d\ln r}\end{equation} where $\gamma=1$ corresponds to a cusp and $\gamma=0$ a core. However, observations of dwarf galaxies have instead shown that DM tends to form a core, where density plateaus near the center of the object \citep{Oh_2015}.

A proposed baryonic feedback process to explain the inconsistency between observed and predicted inner slope values is supernovae, which cause gas to be expelled from the interstellar medium, changing the gravitational potential in the area in which the supernova occurred, for example, the inner halo \citep{doi:10.1126/science.1148666,Pontzen_2014}. The lack of accounting for baryonic feedback processes in simulations may be a factor in the inaccurate representation of inner DM halos; modern hydrodynamic simulations incorporate baryons and feedback to showcase how cores can be reproduced under specific conditions \citep{10.1093/mnras/stw713,pascale:hal-04212524}. Another cause could be dynamical friction, which, when occurring in fragmented gas clouds in the pre-feedback phase, may create variable gravitational potential wells, causing a disparity in observed versus predicted inner halo DM density profiles \citep{El-Zant_2001,2015MNRAS.446.1820N}. One alternative model modifies the collisionless cold DM model via the theory of self-interacting DM (SIDM) \citep{1992ApJ...398...43C,Spergel_2000}, in which DM particles frequently collide and transfer energy, eventually leading to isothermality. Another $\Lambda$CDM modification proposal is fuzzy DM \citep{PhysRevLett.85.1158}, where DM particles are extremely light scalar bosons that use Heisenberg uncertainty to avoid cusps due to their kiloparsec-scale de Broglie wavelengths.

Dwarf spheroidal galaxies (dSphs) are low-baryonic-mass galaxies (V-band luminosities $L_V\sim10^{3}\text{–}10^{7}L_{\odot}$, \citealp{1998ARA&A..36..435M}) that are ideal for the study of the core-cusp problem due to being DM-dominated at all radii, with mass-to-light ratios in the V band reaching $M/L_V \gtrsim 100\,M_\odot/L_\odot$ \citep{2007ApJ...663..948G}. The inferred DM profile has the least error when the baryonic term is negligible, as in dSphs, because it is calculated by subtracting the baryonic mass profile from the dynamical mass profile. Using dSphs also partially sidesteps the need to account for baryonic processes altering halos, due to low stellar mass and thus less total feedback energy available for halo reshaping. There are many nearby ``classical'' dSphs with resolved stellar kinematics through measurements from \citet{Walker_2007,2009ApJ...704.1274W} and cross-matched catalogs combining Gaia Data Release 3 (DR3) and VLT/FLAMES data \citep{Tolstoy_I}, making dSphs well-suited candidates for analysis.

The Collisionless Boltzmann Equation (CBE) is a relation between $f(\vec{r},\vec{v})$, the distribution function (DF) in six-dimensional (6D) phase space, of a tracer to its gravitational potential \citep{2008gady.book.....B}: \begin{equation}\label{eq:cbe}\frac{\partial f}{\partial t} + \vec{v}\cdot\nabla f - \nabla\Phi\cdot\frac{\partial f}{\partial\vec{v}} = 0\end{equation} As stars exist in 6D phase space, with a three-dimensional (3D) position component and a 3D velocity component, the CBE theoretically could be used to model dSphs; however, individual proper motions (PMs) at typical dSph distances are too imprecise for dynamical modeling, so only line-of-sight (LOS) measurements remain: 2D projected position and radial velocity. As transforming the CBE to 3D has proven to be computationally intensive \citep{2002MNRAS.330..778W}, Jeans equations, which integrate the CBE over velocity space, are favored for dSph modeling \citep{2004ApJ...611L..21W,2009ApJ...704.1274W,2011MNRAS.411.1013B}: \begin{equation}\label{eq:sph-jeans-unprojected}\frac{d(\nu\sigma_r^2)}{dr} + \frac{2\beta(r)}{r}\nu\sigma_r^2 = -\nu\frac{d\Phi}{dr}\end{equation} The spherical Jeans equation is preferred due to the dispersion-dominated kinematics and weak or absent systemic rotation in dSphs \citep{1998ARA&A..36..435M,2024A&A...692A.195A}, as well as computational efficiency and tractability in comparison to axisymmetric or triaxial models.

Because only LOS velocities are available, the Jeans analysis constrains a degenerate combination of the mass profile and the anisotropy; thus, a cuspy halo with tangential orbits and a cored halo with radial orbits produce near-identical $\sigma_{\text{los}}(R)$ \citep{2008gady.book.....B,1982MNRAS.200..361B}. \citet{2011ApJ...742...20W} break the degeneracy by using two chemodynamically distinct populations, each with its own half-light radius $r_h$ and mass estimate ($M(r_h)=5r_h\sigma^2/(2G)$), yielding a model-independent slope. Each population is a separate tracer of the same potential, and because their half-light radii differ, the enclosed mass is derived at two distinct points, providing a direct slope without the need for fitting a profile.

Sculptor is generally described as two-population on independent grounds: photometry (\citealp{1999ApJ...520L..33M} find two distinct red giant branch (RGB) bumps in BV data to $V = 22$ mag, as well as a horizontal-branch luminosity discontinuity), spatial and kinematic distributions \citep{Tolstoy_2004}, and joint chemodynamical modeling \citep{2024A&A...692A.195A}. However, the decomposition of the stellar component into two populations is a modeling choice with an untested effect on $\gamma$, and, as this work shows, whether decomposing the sample improves or degrades the recovered slope depends on whether the underlying distribution is discrete or continuous, which is not known a priori (see Section~\ref{sec:decomp-bias}). The two-population approach has been applied to Sculptor with several methodologies with mixed conclusions of coreness and cuspiness \citep{Battaglia_2008,2012MNRAS.419..184A,Strigari_2017}. Even if a dSph possesses a bimodal metallicity distribution, mass estimators that assume spherical symmetry, such as that of \citet{2009ApJ...704.1274W}, are themselves contested; \citet{2018MNRAS.474.1398G} show that analyses in the style of \citet{2011ApJ...742...20W} return spurious core detections ($\Gamma\geq2$) in $17.8\%$ of cases for simulated cuspy dSphs, a failure they primarily attribute to misalignment between the two elongated metallicity populations. A chemodynamically motivated split cannot measure a gradient in metallicity across orbits; doing so requires a DF with metallicity as a continuous variable.

$f(\boldsymbol{J})$ action-integral-based DFs \citep{2010MNRAS.401.2318B,2011MNRAS.413.1889B} are automatically steady-state solutions of the CBE and create the foundation for one approach to the metallicity gradient measurement problem. \citet{2015MNRAS.449.3479S} build the first continuous chemodynamical model, though it is a disc model with an interstellar-medium metallicity gradient, star-formation history, radial migration, and a thin/thick decomposition, none of which are present in a pressure-supported dwarf. They note that \citet{2011ApJ...742...20W} establish that incorporating chemical information into dynamical models is valuable for dwarf spheroidals but instead apply their framework to Galactic dynamics.

Four dynamical methods were implemented on the \citet{Tolstoy_I} spectroscopic catalog of Sculptor: a generalized NFW (gNFW) spherical Jeans analysis, GravSphere \citep{2017MNRAS.471.4541R}, a \citet{2011ApJ...742...20W} style two-population estimator, and a continuous chemo-dynamical action-DF model. $\gamma$ is constrained by GravSphere and the two-population estimator, with the spherical Jeans fit providing a third measurement that sits higher (see Section~\ref{sec:core-like-robustness}) and the continuous DF being the dSph generalization that \citet{2015MNRAS.449.3479S} pointed toward, targeting $k_J$, a gradient measurement inaccessible to any split-based method. This method is corroborated independently in action space. Mocks quantify how the choice of decomposition and the tracer structure each bias the slope. Fornax serves as a cross-galaxy validation of the two-population methodology implemented from \citet{2011ApJ...742...20W}.

\section{Methods}
\label{sec:methods}

\subsection{Data}
\label{sec:data}

\subsubsection{Sculptor}
\label{sec:sculptor-data}

The \citet{Tolstoy_I} VLT/FLAMES LR8 survey of Sculptor is used for all subsequent analyses, with 1604 stars with $v_{\text{los}}$ measurements, 1339 of which also have reliable $[\text{Fe}/\text{H}]$ measurements at the Ca II triplet wavelengths, selected from the 1701 rows of the published catalog by membership flag and $[\text{Fe}/\text{H}]$ quality flag (see Appendix~\ref{ap:data-overview-figs} for an overview of the sample). It is built from 2257 new as well as 2389 archival spectra, comprising VLT/FLAMES LR8 spectra for $\sim55\%$ of Gaia DR3 RGB stars at $G<20$, rising to $>70\%$ for the brightest stars at $G<18.75$ (see Table~\ref{table:tolstoy-props} for global properties). The Plummer-fit values are adopted from \citet{2018ApJ...860...66M} for consistency with the Plummer tracer models used throughout (see Table~\ref{table:munoz-params} for structural parameters). \citet{Tolstoy_I} independently report a spatial metallicity gradient ($-0.7\,\mathrm{dex\,deg^{-1}}$, consistent with the $-0.69\pm0.06$ fitted by \citealp{2024A&A...692A.195A}) and a kinematically distinct very-metal-poor group ($[\text{Fe}/\text{H}]<-2.5$).

All methods use the projected elliptical radius (semi-major axis) of each star, \begin{equation}\label{eq:elliptical-radius}R=\sqrt{x_{\rm maj}^{2}+\left(\frac{y_{\rm min}}{1-e}\right)^{2}},\end{equation} where $x_{\rm maj}$ and $y_{\rm min}$ are the sky offsets from the galaxy center rotated into the major- and minor-axis frame by the position angle, and $e$ is the projected ellipticity. This follows \citet{2024A&A...692A.195A}; the structural parameters $e$ and PA are taken from \citet{2018ApJ...860...66M} for Sculptor (see Table~\ref{table:munoz-params}) and \citet{1995MNRAS.277.1354I} for Fornax (see Table~\ref{table:irwin-params}), with the adopted distance setting the physical scale.

\begingroup
    \setlength{\tabcolsep}{10pt}
    \renewcommand{\arraystretch}{1.5}
    \setlength\extrarowheight{2pt}
    \begin{table}
        \centering
        \begin{tabular}{ c c c c }  
            $N_{[\text{Fe}/\text{H}]}$ & $\langle[\text{Fe}/\text{H}]\rangle$ & $N_{v_{\text{los}}}$ & $\langle v_{\text{los}}\rangle$ (km/s) \\
            \hline \hline     
            $1339$ & $-1.82\pm0.45$ & $1604$ & $111.2\pm0.25$ \\
            \hline \hline
        \end{tabular}  
        \caption{Global properties from the \citet{Tolstoy_I} VLT/FLAMES LR8 Sculptor catalog.}
        \label{table:tolstoy-props}
    \end{table}
\endgroup

\begingroup
    \setlength{\tabcolsep}{10pt}
    \renewcommand{\arraystretch}{1.5}
    \setlength\extrarowheight{2pt}
    \begin{table}
        \centering
        \begin{tabular}{ c c c c }  
            $e$ & $D$ (kpc) & $r_h$ (arcmin) & $\text{PA}$ (deg) \\
            \hline \hline     
            $0.33$ & $83.9$ & $11.17$ & $92.0$ \\
            \hline \hline
        \end{tabular}
        \caption{Structural parameters from the \citet{2018ApJ...860...66M} MegaCam survey.}
        \label{table:munoz-params}
    \end{table}
\endgroup

\subsubsection{Fornax}
\label{sec:fornax-data}

\begin{figure*}[!htbp]
    \centering
    \includegraphics[width=\textwidth]{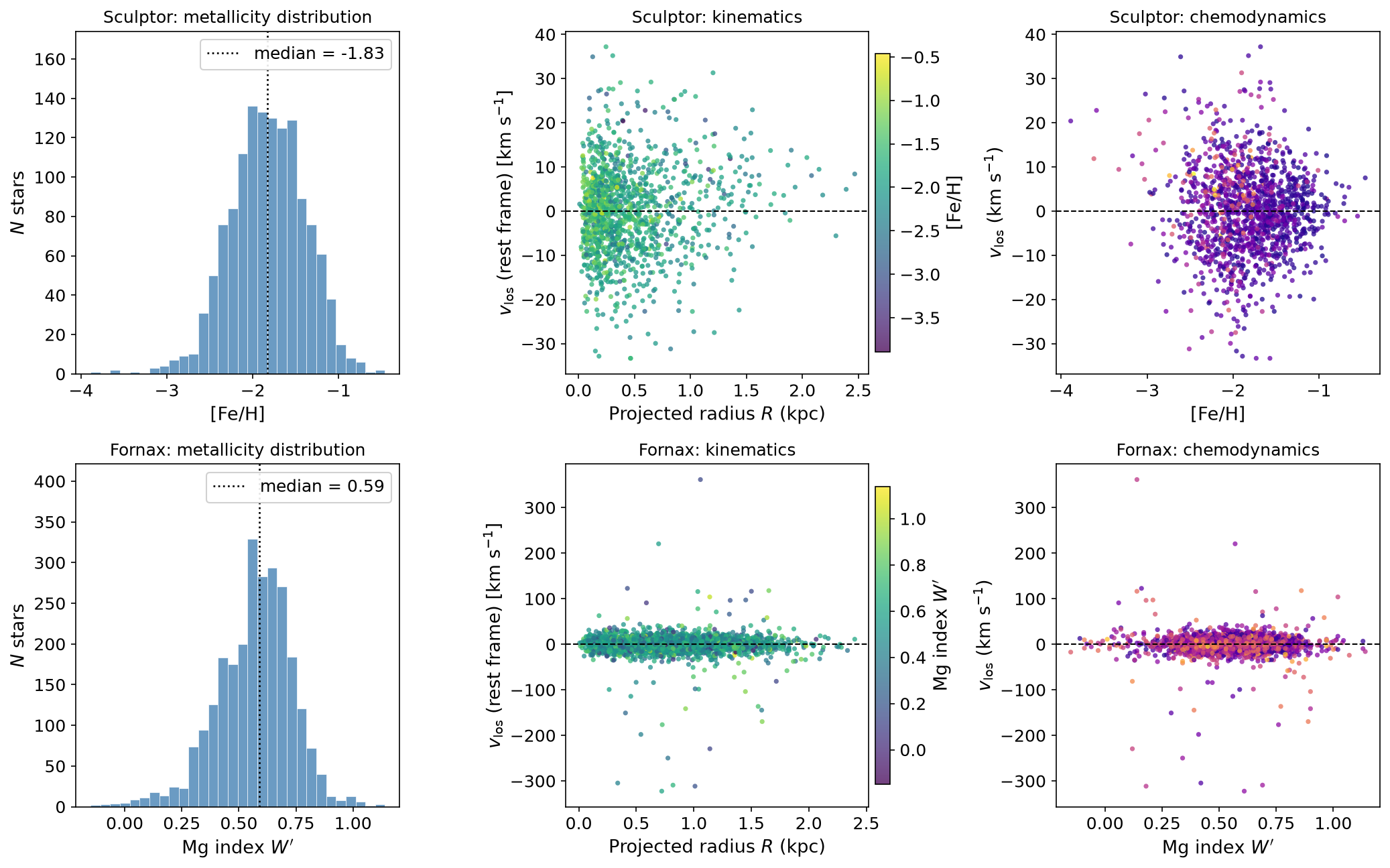}
    \caption{$2\times3$ grid, with Sculptor on the top row and Fornax on the bottom for comparison of datasets. Columns: metallicity distribution with median marked, velocity versus projected radius colored by metallicity, velocity versus metallicity. Note Fornax's metallicity axis is the Mg index $W'$, not $[\text{Fe}/\text{H}]$.}
    \label{fig:two-galaxy-overview}
\end{figure*}

The \citet{2009AJ....137.3100W} MMFS catalog is used for Fornax analyses. Structural parameters are adopted from \citet{1995MNRAS.277.1354I} (see Table~\ref{table:irwin-params}); the ellipticity and position angle enter the projected elliptical radius of Equation~\ref{eq:elliptical-radius}, and the distance sets the physical scale. The averaged $\langle\Sigma\text{Mg}\rangle$ column of the published per-star table carries an Mg index for only 459 of the 2633 Fornax rows, so the per-star sample is rebuilt by inverse-variance averaging the 3150 per-exposure measurements, recovering the 2603 stars \citet{2011ApJ...742...20W} modeled (their Table~1). Non-members are retained and the member fraction $f_\text{mem}$ fitted, as in \citet{2011ApJ...742...20W}; the Mg index $W'$ is used in lieu of $[\text{Fe}/\text{H}]$ (see Figure~\ref{fig:two-galaxy-overview}).

\begingroup
    \setlength{\tabcolsep}{10pt}
    \renewcommand{\arraystretch}{1.5}
    \setlength\extrarowheight{2pt}
    \begin{table}
        \centering
        \begin{tabular}{ c c c c }  
            $D$ (kpc) & $V_{\text{sys}}$ (km/s) & $e$ & $\text{PA}$ (deg) \\
            \hline \hline     
            $147$ & $55.3$ & $0.30$ & $41.9$ \\
            \hline \hline
        \end{tabular}  
        \caption{Structural parameters from \citet{1995MNRAS.277.1354I} APM observations.}
        \label{table:irwin-params}
    \end{table}
\endgroup

\subsubsection{Gaia proper motions}
\label{sec:gaia-pms}

\begin{figure}[!htbp]
    \centering
    \includegraphics[width=\columnwidth]{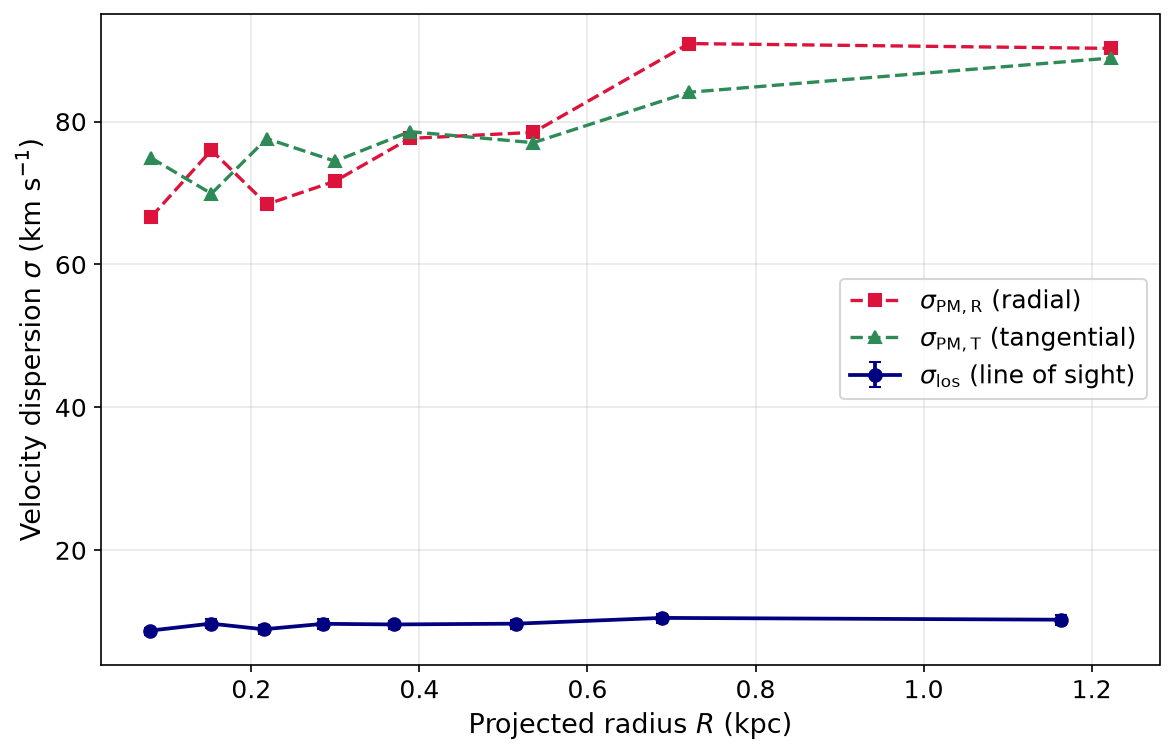}
    \caption{Sculptor; three curves versus projected radius: $\sigma_\text{los}\approx9$--$10.5\,\text{km/s}$ with error bars, and the two PM dispersions at 53--62 and 38--45$\,\text{km/s}$. The PM dispersions are five times larger than the LOS one, which is measurement error rather than kinematics, a visual demonstration of how Gaia PMs are too imprecise for individual-star dynamical modeling. Computed from the 1339-star $[\text{Fe}/\text{H}]$-selected sample.}
    \label{fig:disp-prof}
\end{figure}

\begin{figure*}[!htbp]
    \centering
    \includegraphics[width=\textwidth]{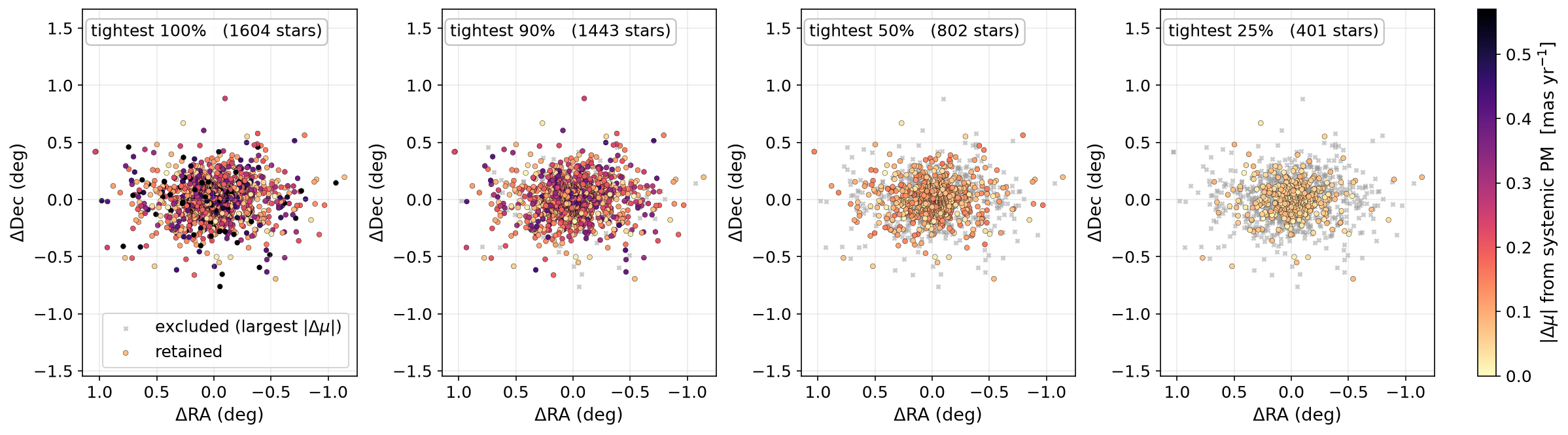}
    \caption{A set of projected skymaps of Gaia DR3 Sculptor data at 100\%, 90\%, 50\%, and 25\% cuts on $|\Delta\mu|$, colored by $|\Delta\mu|$ from systemic PM in $\text{mas}\,\text{yr}^{-1}$.}
    \label{fig:gaia-skymap}
\end{figure*}

Gaia DR3 is crossmatched with the \citet{Tolstoy_I} catalog for completeness across the field (see Figure~\ref{fig:gaia-skymap} for projected skymaps); the resulting PMs validate the spectroscopic membership selection and enable the action computation performed with the Action-based Galaxy Modeling Architecture \citep[AGAMA;][see Section~\ref{sec:action-space-grad}]{2019MNRAS.482.1525V}. Gaia PMs are too imprecise for individual-star dynamical modeling (see Figure~\ref{fig:disp-prof}) but adequate for membership and ensemble action statistics; see Section~\ref{sec:gmm-mem} to see why a fitted mixture is not used. Combining Gaia astrometry with an earlier HST epoch does reach the required precision: \citet{2018NatAs...2..156M} achieve per-star uncertainties below $0.07\,\text{mas}\,\text{yr}^{-1}$ over a 12.3-year baseline and measure Sculptor's internal transverse dispersions from 15 stars (see Section~\ref{sec:anisotropy-comparison}). No such multi-epoch imaging is available across the full \citet{Tolstoy_I} dataset, so this work uses Gaia PMs only for membership and for the action computation of Section~\ref{sec:action-space-grad}. A systemic proper motion of $(\mu_{\alpha,*}, \mu_\delta)=(0.100, -0.147)\,\mathrm{mas\,yr^{-1}}$ is adopted, obtained from the \citet{2022A&A...657A..54B} systemic measurement after applying their quasar-derived zero-point correction. The correction itself is small ($0.012\,\mathrm{mas\,yr^{-1}}$ in $\mu_\delta$), below the quoted systematic floor for this system, and cancels in differential quantities computed against individual stellar proper motions in the same astrometric frame.

\subsubsection{Membership and robustness}
\label{sec:mem+rob}

\begin{figure}[!htbp]
    \centering
    \includegraphics[width=\columnwidth]{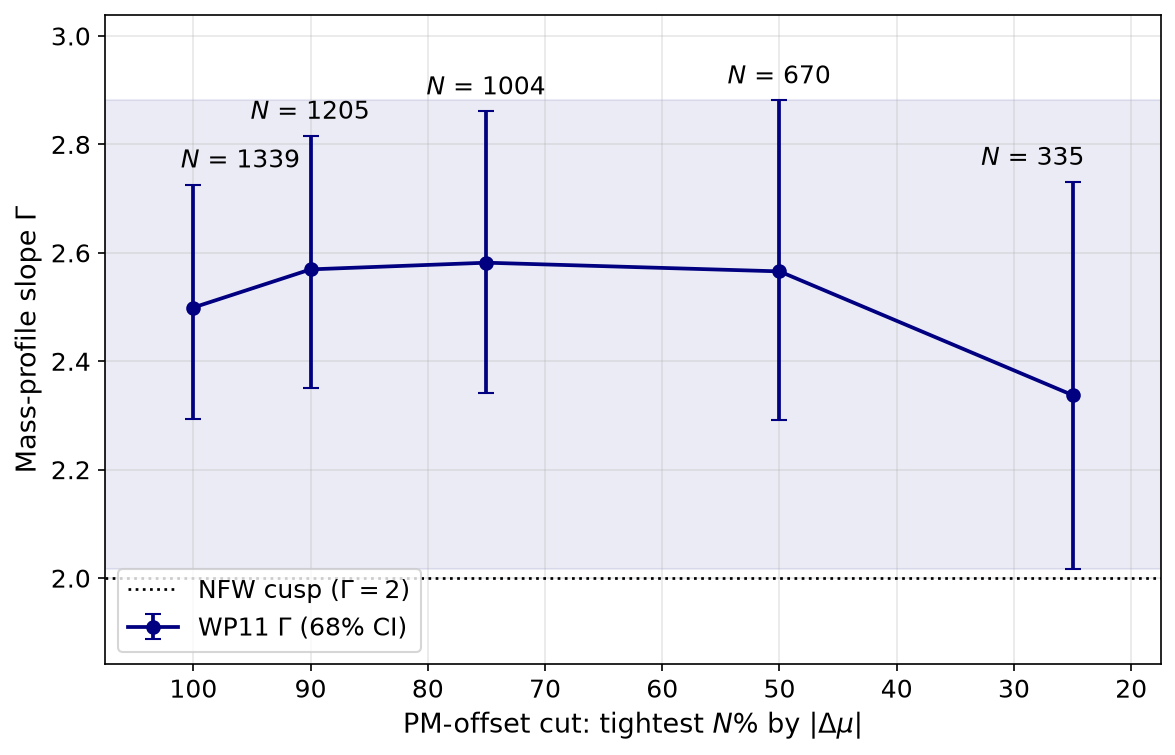}
    \caption{The effect of PM-offset percentile cut (tightest $N\%$ retained, tightening from left to right as the axis is inverted) on Sculptor $\Gamma$ from the two-population estimator, with $N$ annotated at each point and 68\% interval error bars. The dotted line at $\Gamma=2$ is the NFW cusp threshold; $\Gamma>2$ at every cut.}
    \label{fig:membership-robustness}
\end{figure}

Refitting the mass-profile slope on subsamples defined by PM offset from the systemic value gives $\Gamma = 2.49$, $2.57$, $2.58$, $2.57$, and $2.34$ when the tightest 100, 90, 75, 50, and 25\% of stars are retained ($N = 1339\to335$; see Figure~\ref{fig:membership-robustness}). All five exceed $\Gamma = 2$ at the 68\% confidence level, so the PM membership criterion does not drive the exclusion of an NFW cusp. A very-metal-poor sub-population ($\sim24$ stars, $[\text{Fe}/\text{H}]\approx-2.9$, offset $\sim15\,\text{km/s}$ from systemic) is present, with a candidate origin in a minor galactic merger \citep{2024A&A...692A.195A}. \citet{2011ApJ...742...20W} had previously noted a possible third ``ultrafaint'' subcomponent beyond their two-population model, though without characterizing it dynamically; \citet{2024A&A...692A.195A} provide the first statistical detection.

\subsection{Spherical Jeans (generalized Navarro--Frenk--White)}
\label{sec:spherical-jeans}

\begingroup
    \setlength{\tabcolsep}{5.5pt}
    \renewcommand{\arraystretch}{1.5}
    \setlength\extrarowheight{2pt}
    \begin{table}
        \centering
        \begin{tabular}{ c c c c c }  
            Method & $N$ & Walkers & Steps & Params. \\
            \hline \hline     
            Spherical Jeans & 1338 & 24 & 5000 & 3 \\
            GravSphere & 1338 & 24 & 40000 & 6 \\
            Two-pop. (Scl) & 1339 & 64 & 20000 & 9 \\
            Two-pop. (For) & 2603 & 64 & 100000 & 10 \\
            Cont. & 1339 & 28 & 486 & 13 \\
            Cont. (seeded) & 500 & 56 & 372 & 13 \\
            \hline \hline
        \end{tabular}
        \caption{\texttt{emcee} affine-invariant ensemble sampler configuration used throughout \citep{2013PASP..125..306F}. The two-population estimator follows \citet{2011ApJ...742...20W}. Scl and For represent Sculptor and Fornax, respectively. Neither continuous DF chain converged; the seeded run initialized 56 walkers at the higher-likelihood mode of the full-sample chain on a 500-star subsample (see Appendix~\ref{ap:mcmc-conv}).}
        \label{table:sampling-config}
    \end{table}
\endgroup

A LOS projection of the spherical Jeans equation (see Equation~\ref{eq:sph-jeans-unprojected}) is used as the base for our implementation: \begin{equation}\sigma_{\text{los}}^2(R) = \frac{2}{\Sigma(R)} \int_{R}^{\infty} \left( 1 - \beta(r) \frac{R^2}{r^2} \right) \frac{\nu(r) \sigma_r^2(r) \, r}{\sqrt{r^2 - R^2}} \, dr\end{equation} The velocity anisotropy parameter $\beta$ is defined as: \begin{equation}\beta(r) = 1 - \frac{\sigma_\theta^2(r)}{\sigma_r^2(r)}\end{equation} $\nu(r)$ is the 3D tracer number density and $\Sigma(R)$ its projected surface density. There are three free parameters used in the Markov Chain Monte Carlo (MCMC) simulation ($\gamma,\ \log_{10}\,r_s,\ \log_{10}\,M_{\text{DM}}$), with $\alpha=1$ and $\eta=3$ being fixed values. The \texttt{emcee} Python package \citep[see Appendix~\ref{ap:software} for full package list]{2013PASP..125..306F} is used, and all priors are uniform ($\gamma\in[0.02,1.88];\log_{10}r_s\in[-2.9,0.9];\log_{10}M_{\text{DM}}\in[7.2,10.8]$). The Osipkov-Merritt (OM) anisotropy profile ($\beta(r) = \frac{r^2}{r^2 + r_a^2}$) is used for simplicity, with $r_a=1.5$ fixed. The Jeans analysis is fit to the binned $\sigma_\text{los}$ of the two metallicity populations (see Table~\ref{table:sampling-config} for sampling configuration and Section~\ref{sec:sigma-los-degen} for degeneracy). A robust $4\sigma$ median-absolute-deviation clip on $v_\text{los}$ removes one star from the spherical Jeans and GravSphere samples, yielding $N=1338$ rather than the 1339 used by the two-population estimator. Those two fits carry no contamination component, so a single unresolved binary or foreground interloper at large $|v_\text{los}|$ would inflate the binned dispersion and bias $\gamma$ toward a spurious cusp. The two-population fit is not clipped, so the headline $\Gamma$ rests on the full catalog; applying the same clip to it returns $\Gamma=2.47^{+0.22}_{-0.20}$ against the unclipped $2.49^{+0.23}_{-0.20}$, a shift an order of magnitude below the statistical uncertainty, so the differing sample definitions do not affect the comparison between methods. Binned dispersions are error-deconvolved by subtracting the square of a constant $0.6\,\text{km}\,\text{s}^{-1}$, the mean velocity uncertainty of the \citet{Tolstoy_I} sample, from the raw sample variance in each bin, with a floor of $0.5\,\text{km}^2\,\text{s}^{-2}$ preventing unphysical negative variances. The correction is $\lesssim0.5\%$ of the variance at Sculptor's $\sigma_\text{los}\approx9\,\text{km}\,\text{s}^{-1}$. The dispersion profile of each population is computed in six equal-count bins in elliptical radius (seven for the GravSphere fit, which uses a single tracer population). Refitting at eight bins per population returns $\gamma=0.97^{+0.21}_{-0.31}$ against the headline $0.97^{+0.21}_{-0.33}$, a shift of $\Delta\gamma=0.003$, two orders of magnitude below the $\approx0.25$ anisotropy systematic described in Section~\ref{sec:model-assumpt}. This variant chain ran with a different walker count (60) and reached $\hat{R}=1.020$, so it is reported as an indicative check rather than a matched comparison. Coarser binning is not tested, since with three free parameters and two populations, it reduces the degrees of freedom of the fit rather than testing the sensitivity of $\gamma$ to the binning scale. Burn-in and thinning are not fixed; instead they are set adaptively from the integrated autocorrelation time ($n_{\text{burn}} = \min(3\tau_{\text{max}}, n_{\text{steps}}/2)$), with a fallback of $n_{\text{steps}}/3$ if $\tau$ does not converge. Non-rotation is assumed, supported by two independent checks: natural-log Bayes factors for rotation from \citet{2024A&A...692A.195A} ($B_{12}=0.6$ for all stars, $-0.9$ excluding the offset component; $-0.9$ for the metal-rich population and $-2$ for the metal-poor), which are inconclusive for the full sample (within the $[0,1]$ inconclusive range on Jeffreys' scale) but show no evidence of rotation in either population individually (negative $B_{12}$), and a test from \citet{2011ApJ...742...20W} attributing Sculptor's velocity gradient maximally to intrinsic rotation rather than solar-reflex or perspective motion, which shifts their $\Gamma$ value by only $0.01$ ($2.95\to2.94$).

\subsection{GravSphere}
\label{sec:gravsphere}

Virial Shape Parameters (VSPs), often represented by $\zeta_A$ and $\zeta_B$, are metrics to measure the shape of the LOS velocity distribution of tracers \citep{2014MNRAS.441.1584R}. Replacing kurtosis, they are a unique method to break $\beta$ degeneracy as they do not require the introduction of the orbital anisotropy parameter $\beta'$, making mass models more robust:\begin{equation}\label{eq:beta-prime}\beta'(r) = 1-\frac{3}{2}\frac{\langle v_r^2v_t^2\rangle}{\langle v_r^4\rangle}\end{equation} They are derived by taking higher-order moments of the CBE; while traditional Jeans equations are second-order velocity moments, VSPs are fourth-order velocity moments. $\zeta_A$ is a fourth-order virial equation ratio that is mostly invariant to changes in the density profile of the tracer, while $\zeta_B$ is a ratio that constrains the outer density slope. The structural projected velocity components (first fourth-order spatial moment integral $v_{s1}$ and second fourth-order spatial moment integral $v_{s2}$), which are unnormalized forms entering $\zeta_A$ and $\zeta_B$ respectively, are integrations across the radius $r$ \citep{2017MNRAS.471.4541R}:\begin{align}v_{s1} &= \frac{2}{5}\int_0^\infty GM(r) \nu(r)[5-2\beta(r)]\sigma_r^2(r) r \, dr\\v_{s2} &= \frac{4}{35}\int_0^\infty GM(r) \nu(r)[7-6\beta(r)]\sigma_r^2(r)r^3 \, dr\end{align} These are compared against their observational counterparts:\begin{align}\hat{v}_{s1}&=\int_0^\infty\Sigma\langle v_{\text{LOS}}^4\rangle R \, dR\\\hat{v}_{s2}&=\int_0^\infty\Sigma\langle v_{\text{LOS}}^4\rangle R^3\,dR\end{align} $\Sigma$ is normalized to unity, so no explicit star count appears.

The Baes--van Hese profile is a parametric velocity anisotropy profile that fixes the problem of unphysical sharp transitions from isotropic centers to radiality-dominated boundaries in the OM profile \citep{2007A&A...471..419B}:\begin{equation}\beta(r)=\frac{\beta_0+\beta_\infty(r/r_\beta)^{2\delta}}{1+(r/r_\beta)^{2\delta}}\end{equation} In this formulation, $\beta_0$ is the central velocity anisotropy value ($r\to0$), $\beta_\infty$ is the asymptotic velocity anisotropy value ($r\to\infty$), $r_\beta$ is the transition scale radius where orbital structure changes occur, and $\delta$ is a parameter controlling the sharpness of transitions between $\beta_0$ and $\beta_\infty$. This work samples the symmetrized anisotropy $\tilde{\beta}=\beta/(2-\beta)$, which maps $\beta\in(-\infty,1]$ onto $\tilde{\beta}\in[-1,1]$, so a uniform prior in $\tilde{\beta}$ does not diverge at radial anisotropy. This profile unifies older models, as it reproduces OM profile results by setting $\beta_0=0$, $\beta_\infty=1$, and $\delta=1$. This work fixes $\delta=1$ but leaves $\beta_0$ and $\beta_\infty$ free, so the model is OM with the endpoint anisotropies determined by the fit rather than imposed.

\begingroup
    \setlength{\tabcolsep}{10pt}
    \renewcommand{\arraystretch}{1.5}
    \setlength\extrarowheight{2pt}
    \begin{table}
        \centering
        \begin{tabular}{ l l }
            Parameter & Prior \\
            \hline \hline
            $\gamma$                & $\mathcal{U}(0.05,\,1.9)$ \\
            $\log_{10} r_s$         & $\mathcal{U}(-1.0,\,0.8)$ \\
            $\log_{10} \rho_s$      & $\mathcal{U}(6.0,\,9.0)$ \\
            $\tilde{\beta}_0$       & $\mathcal{U}(-0.95,\,0.95)$ \\
            $\tilde{\beta}_\infty$  & $\mathcal{U}(-0.95,\,0.95)$ \\
            $\log_{10} r_\beta$     & $\mathcal{U}(-1.2,\,0.8)$ \\
            \hline
            $\delta$                & fixed at 1 \\
            $\alpha$                & fixed at 1 \\
            $\eta$                  & fixed at 3 \\
            \hline \hline
        \end{tabular}
        \caption{Priors for the six free GravSphere parameters, all uniform, with the three fixed shape parameters below the rule.}
        \label{table:gravsphere-priors}
    \end{table}
\endgroup

An implementation of GravSphere \citep{2017MNRAS.471.4541R} is utilized, using a second-order Jeans model with two VSPs and a free Baes--van Hese $\beta(r)$, fit to binned kinematics (7 bins; distinct from the newer fourth-order, unbinned GravSphere2, \citealp{2026A&A...705A.212B}). It is cross-validated against the reference code ($\sigma_{\text{los}}$ to 0.32\%, VSPs to 0.10\%, see Appendix~\ref{ap:grav-cross}). The MCMC simulation uses six free parameters, all uniform, with $\delta$, $\alpha$, and $\eta$ fixed (see Table~\ref{table:gravsphere-priors}).

\subsection{Two-population estimator}
\label{sec:wp11-two-pop}

The \citet{2011ApJ...742...20W} two-population estimator is used, with two Plummer subcomponents, each with an individual Gaussian velocity and metallicity distribution:\begin{equation}M(r_h)=\frac{5r_h\sigma^2}{2G}\end{equation}$M(r_h)$ is the \citet{2009ApJ...704.1274W} mass estimator, applied at two half-light radii to give:\begin{align}\Gamma&\equiv\Delta\log{M}/\Delta\log{r}\\\Gamma&=3-\gamma\end{align}The estimator constant cancels in $\Gamma$, so absolute masses differ from Table 4 of \citet{2011ApJ...742...20W} by a constant offset while the slope reproduces it. $r_{h,1}$ and $r_{h,2}$ are fit as free parameters by the MCMC simulation, not fixed inputs; the free-fit half-light radii and velocity dispersions this yields for the real Sculptor data are what generate the $\Gamma=2.49$ result (see Section~\ref{sec:inner-slope-measures}), independent of any adopted structural-parameter table.

For Fornax, $N=2603$, with a Mg-index metallicity proxy in place of $[\text{Fe}/\text{H}]$. Two components of the \citet{2011ApJ...742...20W} model are not reproduced exactly: their spectroscopic completeness $w(r)$ is reconstructed against a Gaia DR3 photometric parent rather than the unpublished \citet{Walker_2007} target catalog; the reconstructed parent is broader than an RGB selection box, so $w(R)$ is flatter than theirs and the fitted half-light radii come out systematically larger, with a variant of the parent without an RGB color-cut bounding this systematic at $\approx0.2$ in $\Gamma$. Their perspective term is also omitted; the proper motion they fit, $(\mu_\alpha,\mu_\delta)=(63,-29)\,\text{mas}\,\text{century}^{-1}$, corresponds to a heliocentric transverse velocity of $483\,\text{km}\,\text{s}^{-1}$ at $D=147\,\text{kpc}$ and an apparent line-of-sight gradient of $8.4\,\text{km}\,\text{s}^{-1}\,\text{deg}^{-1}$ across a sample reaching $0.94^\circ$, contributing $1.1$ and $2.1\,\text{km}^2\,\text{s}^{-2}$ to the metal-rich and metal-poor variances when averaged over azimuth. Since comparable fractions inflate both subcomponents, and $\Gamma$ depends only on the ratio of their dispersions, removing this term shifts $\Gamma$ by 0.006, which is negligible relative to the quoted interval. The selection-function reconstruction is therefore the dominant modeling systematic of the two.

\subsection{Continuous $f(\boldsymbol{J},[\text{Fe}/\text{H}])$}
\label{sec:cont-method}

\begingroup
    \setlength{\tabcolsep}{5.75pt}
    \renewcommand{\arraystretch}{1.5}
    \setlength\extrarowheight{2pt}
    \begin{table}
        \centering
        \begin{tabular}{ l l r }
            Parameter & Prior & Component \\
            \hline \hline 
            $\log_{10} M_{\rm DM}$   & $\mathcal{U}(7.0,\,11.0)$  & DM potential \\
            $\log_{10} r_s$          & $\mathcal{U}(-3.0,\,1.0)$  & DM potential \\
            $\alpha$                 & $\mathcal{U}(0.0,\,7.0)$   & DM potential \\
            $\eta$                   & $\mathcal{U}(2.0,\,7.0)$   & DM potential \\
            $\gamma$                 & $\mathcal{U}(0.0,\,1.9)$   & DM potential \\
            $\Gamma_p$               & $\mathcal{U}(0.0,\,3.0)$   & stellar DF \\
            $B_p$                    & $\mathcal{U}(3.0,\,30.0)$  & stellar DF \\
            $g_{z,p}$                & $\mathcal{U}(0.0,\,1.5)$   & stellar DF \\
            $h_{z,p}$                & $\mathcal{U}(0.0,\,1.5)$   & stellar DF \\
            $\log_{10} J_{p,\rm ref}$ & $\mathcal{U}(-1.0,\,3.0)$ & metallicity--action \\
            $k_J$                    & $\mathcal{U}(-4.0,\,0.5)$  & metallicity--action \\
            $\mathcal{M}_z$          & $\mathcal{U}(-2.7,\,-1.3)$  & metallicity distribution \\
            $\sigma_z$               & $\mathcal{U}(0.05,\,0.8)$  & metallicity distribution \\
            \hline 
            $V_C$                    & fixed & contamination \\
            $\sigma_{V,C}$           & fixed & contamination \\
            $\mathcal{M}_C$          & fixed & contamination \\
            $\sigma_{M,C}$           & fixed & contamination \\
            $f_C$                    & fixed & contamination \\
            \hline \hline
        \end{tabular}
        \caption{Parameters of the continuous $f(\boldsymbol{J},[\mathrm{Fe/H}])$ model, including potential parameters. Thirteen are sampled; the five contamination parameters are held at fiducial values.}
        \label{table:cont-priors}
    \end{table}
\endgroup

\begingroup
    \setlength{\tabcolsep}{10pt}
    \renewcommand{\arraystretch}{1.5}
    \begin{table}
        \centering
        \begin{tabular}{ l l l }
            Parameter & Value & Description \\
            \hline \hline
            $V_C$            & $13.0\,\text{km/s}$ & mean velocity offset \\
            $\sigma_{V,C}$   & $7.0\,\text{km/s}$  & velocity dispersion \\
            $\mathcal{M}_C$  & $-2.8$              & mean $[\text{Fe}/\text{H}]$ \\
            $\sigma_{M,C}$   & $0.5$               & metallicity dispersion \\
            $f_C$            & $0.02$              & fraction of the sample \\
            \hline \hline
        \end{tabular}
        \caption{Fiducial values for the third-population component, held fixed rather than sampled. Values follow the population identified by \citet{2024A&A...692A.195A} (see Section~\ref{sec:mem+rob}).}
        \label{table:cont-fiducial}
    \end{table}
\endgroup

AGAMA \citep{2019MNRAS.482.1525V} is implemented to create a metallicity-dependent DF. The DM potential is a gNFW with a cutoff \citep[][their Equation~4]{2025A&A...699A.347A}:\begin{equation}
\begin{split}\rho(r)={}&\rho_{0}\,(r/r_{s})^{-\gamma}\left[1+(r/r_{s})^{\alpha}\right]^{(\gamma-\eta)/\alpha}\\&\times\exp\left[-(r/r_{\rm cut})^{\xi}\right]\end{split}\end{equation} Here $\rho_0$ is the density normalization factor, and $r_s$ is the scale radius. The inner logarithmic slope $\gamma$ governs the profile as $r\ll r_s$, while $\eta$ is the outer slope. As $r\gg r_s$, the bracketed term reduces the profile to $\rho\propto(r/r_s)^{-\eta}$. The transition sharpness between the two regimes is set by $\alpha$. The exponential factor truncates the halo beyond $r_{\text{cut}}$, with $\xi$ controlling abruptness. $\alpha$ and $\eta$ are sampled with the other potential parameters, recovering the NFW form at $\gamma=1$, with $\alpha=1$ and $\eta=3$. $r_{\text{cut}}=20\,\text{kpc}$ with $\xi=1$; the cutoff radius is $\sim10$ times the outermost member star, so it has negligible effect on the inner profile this work constrains. A double-power-law DF is implemented, with the reduced form being \citep[][their Equation~5]{2025A&A...699A.347A}:\begin{equation}\label{eq:double-power-law}\begin{split}f_{p}(\boldsymbol{J}) = \frac{f_{0,p} M_{p}}{\left(2\pi J_{p}\right)^{3}}&\left[1+\frac{J_{p}}{h_{p}(\boldsymbol{J})}\right]^{\Gamma_{p}}\\\times&\left[1+\frac{g_{p}(\boldsymbol{J})}{J_{p}}\right]^{-B_{p}},\end{split}\end{equation}\begin{align}g_{p}(\boldsymbol{J}) &\equiv g_{r,p} J_{r}+g_{z,p} J_{z}+g_{\phi,p}\left|J_{\phi}\right|,\nonumber\\h_{p}(\boldsymbol{J}) &\equiv h_{r,p} J_{r}+h_{z,p} J_{z}+h_{\phi,p}\left|J_{\phi}\right|\end{align} Here $\Gamma_p$ and $B_p$ are the inner and outer slopes of the DF in the action space, $J_p$ is the scale action that designates the transition between them, $M_p$ is the population's tracer mass (this work uses $M_p=1$), and $f_{0,p}$ is a dimensionless factor that normalizes the DF to $M_p$. Spherical symmetry is imposed by setting $h_{z,p}=h_{\phi,p}$ and $g_{z,p}=g_{\phi,p}$; thus, each triplet, summing to 3, reduces to a single free coefficient. Equation~\ref{eq:double-power-law} is the steepness-unity case of the general AGAMA double power law. The metallicity dependence of the scale action of the DF is as follows:\begin{equation}\log_{10}{J_p(z)}=\log_{10}{J_{p,\mathrm{ref}}}+k_J(z-\langle z \rangle)\end{equation} Here $z$ represents $[\text{Fe}/\text{H}]$ and $\langle z\rangle$ its sample mean, so $J_{p,\text{ref}}$ is the scale action at mean metallicity. The coupling $k_J$ is the slope of $\log_{10}{J_p}$ with metallicity: $k_J<0$ places metal-rich stars at smaller scale action, and thus on more tightly bound and less radially excursive orbits. It is a minimal, phenomenological metallicity-action coupling, with 18 total parameters, 13 being free (see Table~\ref{table:cont-priors} for parameters and priors). The five parameters describing the third population identified by \citet{2024A&A...692A.195A} (see Section~\ref{sec:mem+rob}) are held at the fiducial values in Table~\ref{table:cont-fiducial} rather than sampled. This component is modeled as uniform across the observed region, an approximation to its more extended observed distribution, with Gaussian velocity and metallicity distributions. There are $K=11$ metallicity nodes, with per-star likelihood marginalized over true metallicity. Median-split tracer labels are seeded from \citet{2024A&A...692A.195A} Table~C.2 ($f_{\text{MR}}=0.34$; $R_h=0.128^\circ$ and $0.272^\circ$; $0.1875$ and $0.3983\,\text{kpc}$ at $D=83.9\,\text{kpc}$). This method is included to measure $k_J$, not to improve $\gamma$. It also contrasts with the \citet{2024A&A...692A.195A} spatial-gradient continuum; theirs is a continuum in position with fixed kinematics, while this work is a continuum in action that carries the kinematics self-consistently.

\subsection{Mock validation framework}
\label{sec:mock-valid}

Four mock experiments were performed to validate this work: continuous-truth mocks (one Plummer tracer, $a_0=0.28\,\text{kpc}$, smooth $[\text{Fe}/\text{H}]$ gradient), discrete-truth mocks (two Plummer tracers matched to \citealt{2024A&A...692A.195A} Table C.2 -- metal-rich: $f=0.34,\,\langle[\text{Fe}/\text{H}]\rangle=-1.44,\,\sigma_M=0.26,\,a=0.1875\,\text{kpc}$; metal-poor: $f=0.66,\,\langle[\text{Fe}/\text{H}]\rangle=-2.02,\,\sigma_M=0.34,\,a=0.3983\,\text{kpc}$), null and label-scramble controls isolating tracer geometry from decomposition, and a chain-length sweep isolating prior shrinkage from non-convergence. The continuous-truth mock tests whether splitting a continuum biases $\gamma$, and the discrete-truth mock tests whether the continuous model recovers $\gamma$ when the truth is discrete. Each configuration uses 200 realizations.

\section{Results}
\label{sec:results}

\subsection{The $\sigma_\text{los}$ degeneracy}
\label{sec:sigma-los-degen}

\begin{figure}[!htbp]
    \centering
    \includegraphics[width=\columnwidth]{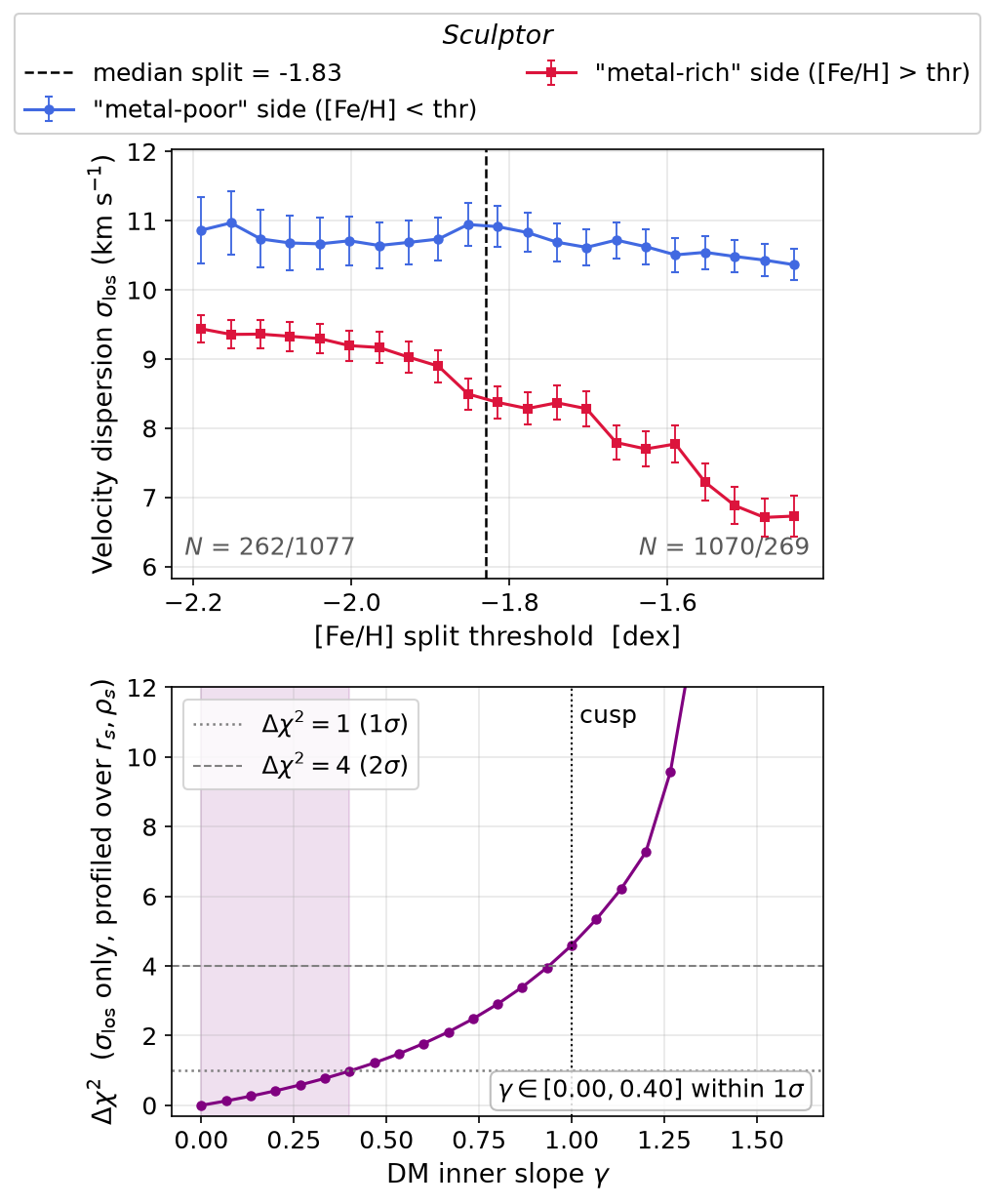}
    \caption{Sculptor; top: $\sigma_\text{los}$ for the metal-poor and metal-rich sides as the $[\text{Fe}/\text{H}]$ split threshold slides, with a dashed line at the median split. Bottom: $\chi^2$ versus $\gamma$, profiled over $r_s$ and $\rho_s$, with horizontal lines at $\Delta\chi^2=1$ and 4 ($1\sigma$ and $2\sigma$), a shaded band over the $1\sigma$ $\gamma$ range, and a marker at $\gamma=1$ labeled ``cusp''. The in-axes box gives the $1\sigma$ range. $\sigma_{\text{los}}$ varies smoothly with the threshold, so as to say there is no plateau marking a natural split. The $\Delta\chi^2$ curve is weakly constraining.}
    \label{fig:sliding-metallicity}
\end{figure}

The isotropic likelihood, the $\sigma_\text{los}$-only case with $\beta=0$, constrains $\gamma$ only weakly ($\Delta \chi^2<1$ for $\gamma\lesssim0.4$ and $\Delta\chi^2<4$ for $\gamma\lesssim0.9$, with the minimum at or below the lower scan limit, so the maximum likelihood estimator is not well determined, and all reported slopes are posterior medians; see Figure~\ref{fig:sliding-metallicity}). The isotropic spherical Jeans fit returns $\gamma=0.71^{+0.31}_{-0.40}$ (see Appendix~\ref{ap:mcmc-conv}), consistent with this range; the two differ in binning and in the two-population tracer treatment, not in the assumption of isotropy. Although \citet{2014MNRAS.441.1584R} avoid posteriors for this reason, comparing models by likelihood instead, this work retains posterior inference and tests the resulting prior sensitivity directly: refitting under $\gamma\in[0,1.5]$ (prior median 0.75, against 0.95 for the default $[0,1.9]$) returns $\gamma=0.95^{+0.21}_{-0.35}$, a shift of 0.02 against a 0.20 shift in the prior median. The posterior is therefore set by the data rather than by the prior, despite the weakly constraining likelihood.

\subsection{Non-identifiability of proper-motion membership models}
\label{sec:gmm-mem}

A two-component Gaussian mixture model applied to the PMs of this spectroscopically-selected sample does not yield an identifiable membership decomposition; the fit returns $P_{\text{mem}} = 1$ for every star (standard deviation of $8\times10^{-13}$ across the sample), and bounding the foreground dispersion does not resolve it. Across five initializations, the optimizer converges to four different corners of the bounded parameter space with a total log-likelihood range of $\Delta\ln{\mathcal{L}}=2.6$. Limiting $\sigma_f$ from above relocates the pathology, with the foreground collapsing to the lower bound at $(0.050,0.050)$ and an equally good likelihood. That non-identification is consistent with the sample already being largely free of foreground contamination before any PM-based cut. PM membership is therefore characterized by direct offset from the systemic value $|\Delta\mu|$ rather than by a fitted membership probability and should not be applied downstream of spectroscopic member selection without first checking identifiability.

\subsection{Marginal metallicity unimodality}
\label{sec:marginal-feh-uni}

Sculptor's $[\text{Fe}/\text{H}]$ distribution is unimodal: a bimodality coefficient of $0.30$ (bimodal if $>0.555$), $\Delta\text{BIC}=-8$ (favoring unimodality), and Hartigan's dip test $p=0.79$. Discrete mocks matched to \citet{2024A&A...692A.195A} return the same ($0.90 < p < 1.00$  across $5$ realizations, mean $\approx0.96$) because the components are separated in metallicity by only $1.35\sigma$ (from $0.58\,\text{dex}$ against a combined scatter of $0.43\,\text{dex}$). The marginal test, therefore, cannot distinguish a continuum from two overlapping populations. The real data sits slightly closer to bimodal than the two-population mock predictions ($p=0.79$ versus $p\approx0.96$), although both are far from rejecting unimodality.

\citet{Breddels_2014} find a bimodal distribution function in Sculptor from non-parametric orbit-based models, independent confirmation of the existence of a dual structure and discreteness.

\begin{figure}[!htbp]
    \centering
    \includegraphics[width=\columnwidth]{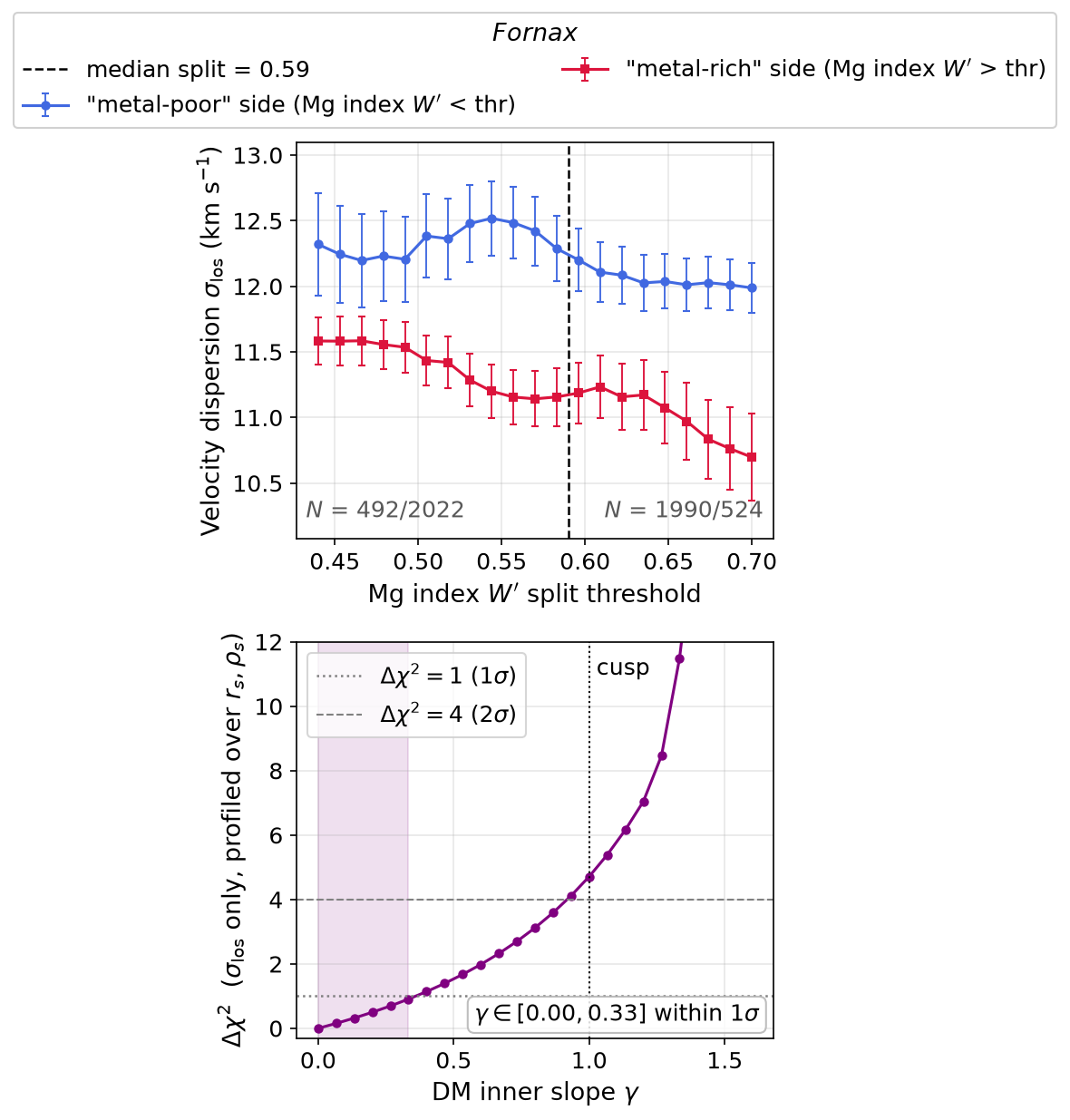}
    \caption{Fornax; top: $\sigma_\text{los}$ for the metal-poor and metal-rich sides as the $W'$ split threshold slides, with a dashed line at the median split; $N$ per side annotated at each end, since the two are strongly imbalanced across the sweep. Bottom: $\chi^2$ versus $\gamma$, profiled over $r_s$ and $\rho_s$, with horizontal lines at $\Delta\chi^2=1$ and 4 ($1\sigma$ and $2\sigma$), a shaded band over the $1\sigma$ $\gamma$ range, and a marker at $\gamma=1$ labeled ``cusp''. The in-axes box gives the $1\sigma$ range. Unlike Sculptor's weakly constraining curve (see Figure~\ref{fig:sliding-metallicity}), the Fornax curve rises monotonically from the lower scan boundary, so the $\sigma_\text{los}$-only constraint is an upper limit, $\gamma<0.33$ at $1\sigma$, with a cusp excluded at $\Delta\chi^2\approx4.7$. The minimum lies at or below $\gamma=0$, so no best-fit value is quoted.}
    \label{fig:fornax-sliding-metallicity}
\end{figure}

Computed on the 2514 stars with a membership probability $\geq0.5$, Fornax's Mg-index gives $\Delta\text{BIC}=+76.9$, favoring two Gaussians, and Hartigan's dip test $p=0.006$; unimodality is a statement about the member distribution, so non-members are excluded here even though the two-population fit retains them. Though not perfectly controlled due to the use of different discriminants and instruments, compared to Sculptor's values, there is a clean contrast: a dwarf whose chemical discriminant is strongly bimodal registers as such under these tests (see Figure~\ref{fig:fornax-sliding-metallicity}). Sculptor's unimodality is therefore not a limitation of the tests themselves. The bimodality coefficient disagrees with the other two statistics here, returning 0.31 against a bimodal threshold of 0.555. The bimodality coefficient is constructed from skewness and kurtosis and so responds to asymmetry rather than to separation between components; the Fornax $W'$ distribution is skewed, which suppresses it. The dip test and $\Delta\text{BIC}$, which test for a dip and for two-component fit quality respectively, both favor bimodality.

\subsection{Bias from population decomposition and tracer structure}
\label{sec:decomp-bias}

\begin{figure}[!htbp]
    \centering
    \includegraphics[width=\columnwidth]{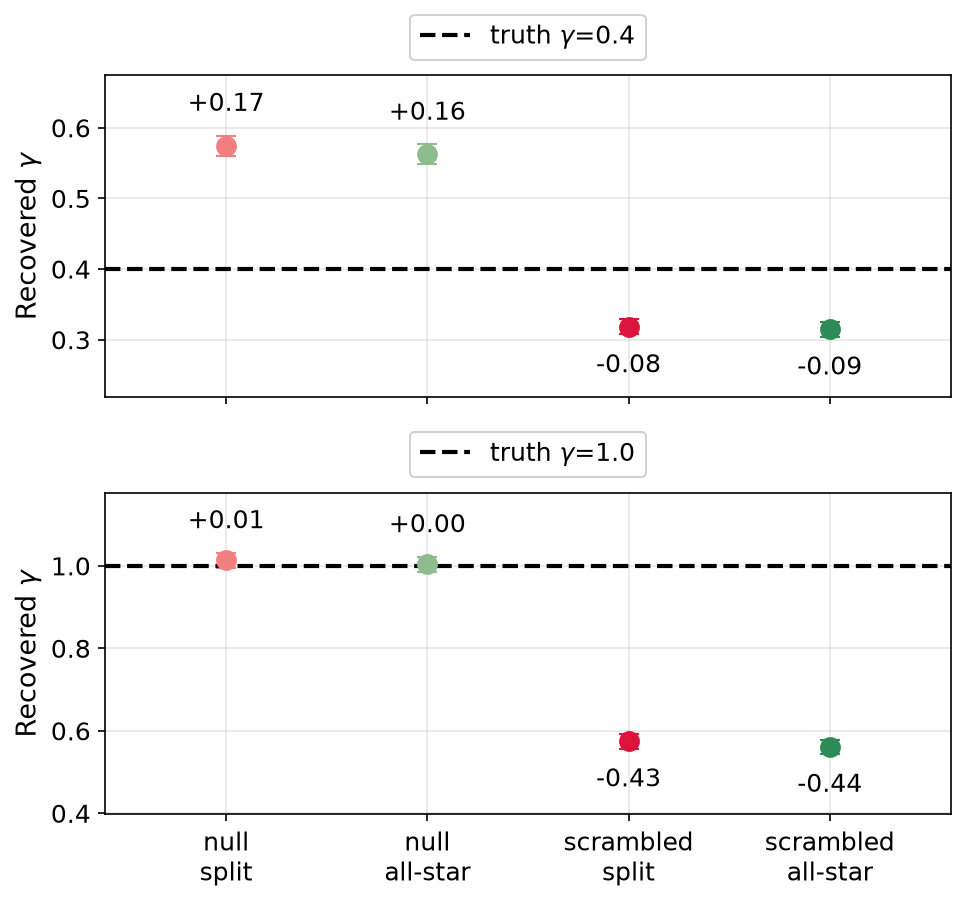}
    \caption{Four points per panel showing the recovered $\gamma$ bias under two control experiments: null (one Plummer tracer, no metallicity structure) and scramble (two Plummers, metallicity permuted); each fit both with a median split and as a single continuous population. Dashed line at true $\gamma$, one panel per $\gamma_{\text{true}}$ value. The bias values are annotated above each point.}
    \label{fig:gate-diagnostics}
\end{figure}

\begin{figure}[!htbp]
    \centering
    \includegraphics[width=\columnwidth]{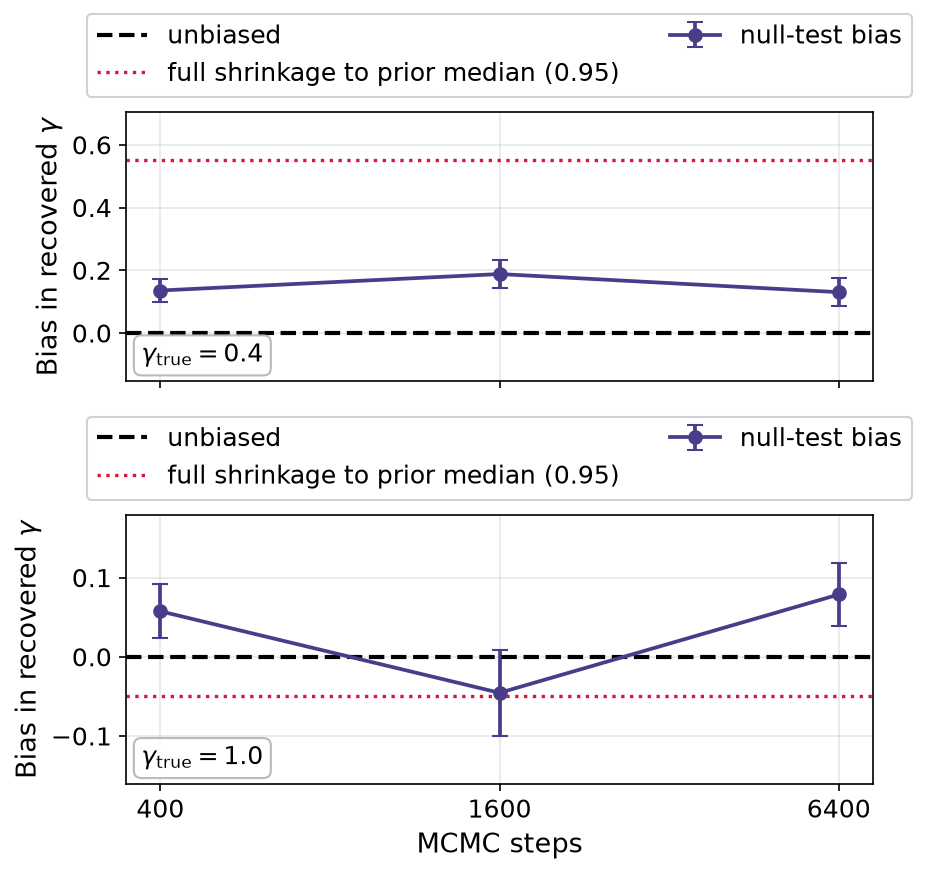}
    \caption{Effect of MCMC steps in log scale on bias in recovered $\gamma$. One panel per $\gamma_{\text{true}}$. The dashed line at zero is unbiased; the dotted line marks full shrinkage to the prior median (0.95). Points are null-mock results with standard errors. Bias is flat across chain length, indicating prior shrinkage rather than non-convergence. Its magnitude tracks the distance from the prior median: large at $\gamma_{\text{true}}=0.4$, consistent with zero at $\gamma_{\text{true}}=1.0$.}
    \label{fig:shrinkage-test}
\end{figure}

\begingroup
    \setlength{\tabcolsep}{10pt}
    \renewcommand{\arraystretch}{1.5}
    \setlength\extrarowheight{2pt}
    \begin{table}
        \centering
        \begin{tabular}{ c c c }  
            $\gamma_{\text{true}}$ & Split & Continuous \\
            \hline \hline     
            0.4 & $+0.459\pm0.013$ & $+0.165\pm0.014$ \\
            1.0 & $+0.296\pm0.011$ & $-0.009\pm0.018$ \\
            \hline \hline
        \end{tabular}
        \caption{Continuous truth (bias-gate mocks).}
        \label{table:cont-truth}
    \end{table}
\endgroup

\begingroup
    \setlength{\tabcolsep}{10pt}
    \renewcommand{\arraystretch}{1.5}
    \setlength\extrarowheight{2pt}
    \begin{table}
        \centering
        \begin{tabular}{ c c c }  
            $\gamma_{\text{true}}$ & Split & Continuous \\
            \hline \hline     
            0.4 & $-0.029\pm0.013$ & $-0.094\pm0.009$ \\
            1.0 & $-0.292\pm0.020$ & $-0.420\pm0.017$ \\
            \hline \hline
        \end{tabular}
        \caption{Discrete truth (mocks matched to \citealt{2024A&A...692A.195A}). The metallicity split is calculated at the median rather than by the adopted mixing fraction, as no analysis knows the true mixing fraction in advance.}
        \label{table:discrete-truth}
    \end{table}
\endgroup

Within the null (one Plummer, no metallicity structure) and label-scramble tests (two Plummers, permuted metallicity), the split and all-star fits agree closely (see Figure~\ref{fig:gate-diagnostics}): the null test yields $+0.174$ versus $+0.162$; the scramble test yields $-0.082$ versus $-0.085$. Both differences are consistent with zero at their standard errors. The decomposition therefore contributes negligibly to the recovered slope when tracer geometry is held fixed. This bias instead tracks the geometry; introducing the two-Plummer structure alone (null$\to$scramble) moves the recovery by $-0.26$ at $\gamma_\text{true}=0.4$ and $-0.44$ at $\gamma_\text{true}=1.0$, while restoring the true metallicity link moves it back by only $+0.05$ and $+0.13$. The bias sign flip between the continuous and discrete truths in Table~\ref{table:cont-truth} and Table~\ref{table:discrete-truth} is thus largely geometric rather than a property of the decomposition. A chain-length sweep confirms a prior-driven shrinkage of $\gamma$ toward the prior median, independent of chain length (see Figure~\ref{fig:shrinkage-test}). The shrinkage fraction is $\approx0.24$ ($f=\langle\hat\gamma-\gamma_\text{true}\rangle/(\gamma_\text{prior}-\gamma_\text{true})$, estimated from a single truth value), that is to say, the posterior is pulled 24\% of the way from the true value toward the prior median; correction is negligible here because 0.97 sits essentially at the prior median of 0.95. The model whose assumptions match the underlying truth recovers the slope best in both directions; for Sculptor, which independent photometric and chemodynamical evidence indicates is discrete \citep{1999ApJ...520L..33M,Tolstoy_2004,2024A&A...692A.195A,Breddels_2014}, the two-population split is the better-matched model for $\gamma$.

Matching the recovery split to the mock's own mixing fraction ($f_\text{MR}=0.34$, a cut at the 66th percentile rather than the median) reduces the split-arm bias under discrete truth from $-0.29$ to $-0.26$ at $\gamma_\text{true}=1.0$, about $1.6\sigma$. A median split over-populates the metal-rich arm with stars drawn from the more extended component and flattens the recovered slope, but this accounts for only part of the bias; the remainder is geometric. The median split is retained as the headline because no analysis of real data knows the true mixing fraction in advance.

\subsection{Inner-slope measurements across methods}
\label{sec:inner-slope-measures}

\begin{figure}[!htbp]
    \centering
    \includegraphics[width=\columnwidth]{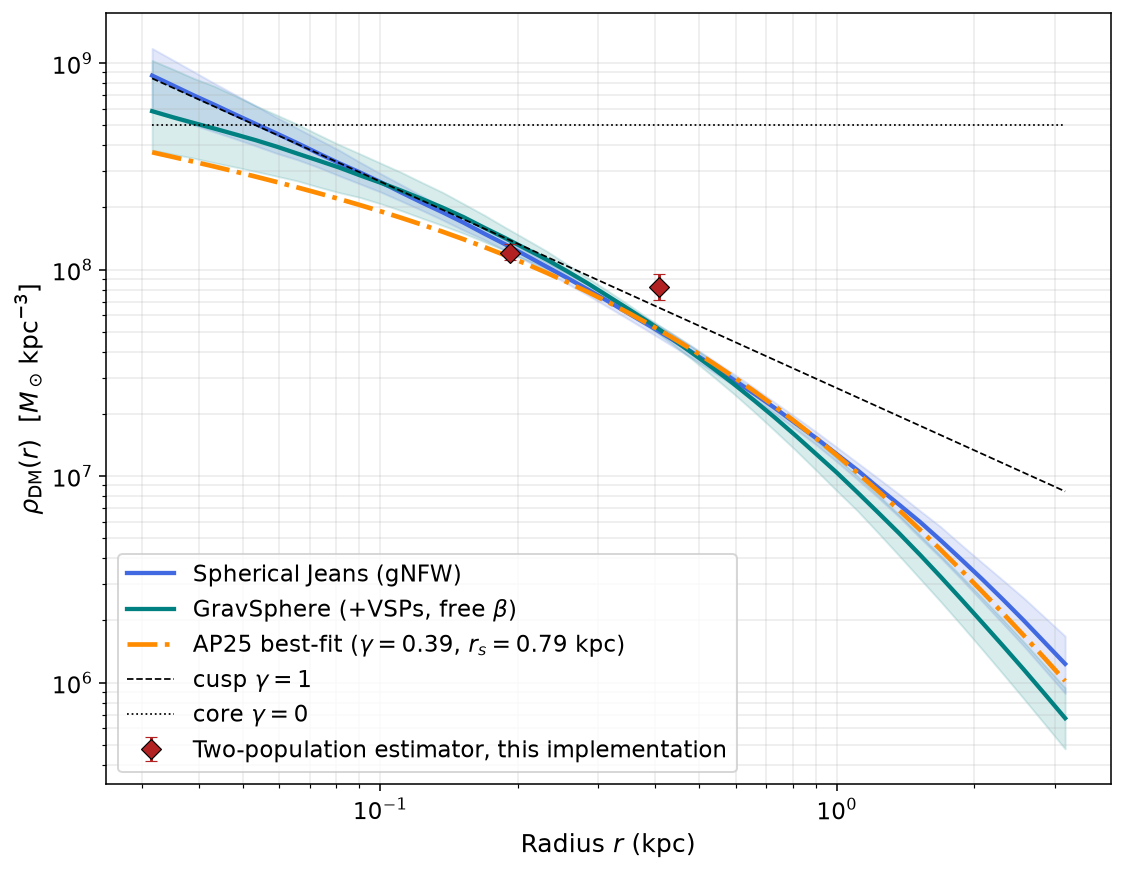}
    \caption{Sculptor DM density profiles across frameworks with the \citealt{2025A&A...699A.347A} curve overlaid, with \citealt{2011ApJ...742...20W} density points for reference.}
    \label{fig:fig4-all-chains}
\end{figure}

\begin{figure}[!htbp]
    \centering
    \includegraphics[width=\columnwidth]{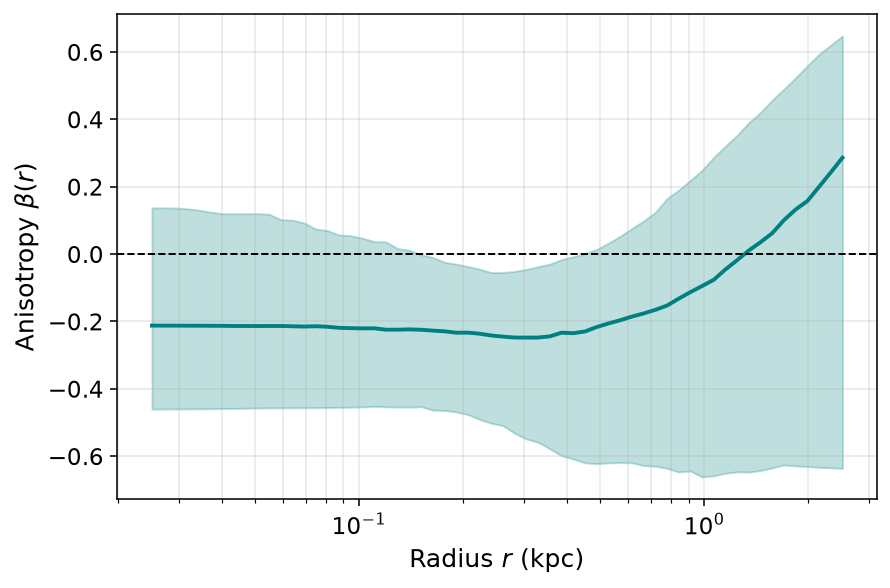}
    \caption{$\beta(r)$ from the GravSphere posterior: the median line with a 68\% band, dashed line at $\beta=0$ for isotropy, log radius axis. $\beta\approx-0.03$ flat inside $\sim0.3\,\text{kpc}$, rising to $\sim0.36$ by $2\,\text{kpc}$, with the band spanning roughly $-0.3$ to $+0.7$ at both ends. These are $\tilde{\beta}_0$ and $\tilde{\beta}_\infty$ in physical units; the inner region is tightly constrained near isotropy while the outer band is wide, since $\tilde{\beta}_\infty$ is bimodal.}
    \label{fig:grav-beta}
\end{figure}

\begingroup
    \setlength{\tabcolsep}{8pt}
    \renewcommand{\arraystretch}{1.5}
    \setlength\extrarowheight{2pt}
    \begin{table}
        \centering
        \begin{tabular}{ l l }  
            Method & $\gamma$ or $\Gamma$ \\
            \hline \hline     
            Spherical Jeans & $0.97^{+0.21}_{-0.33}$ \\
            GravSphere & $0.46^{+0.40}_{-0.28}$ \\
            Two-population & $\Gamma=2.49^{+0.23}_{-0.20}\to\gamma=0.50^{+0.20}_{-0.22}$ \\
            Continuous DF & Non-converging; see Appendix~\ref{ap:mcmc-conv} \\
            AP25 (published) & $\gamma=0.39^{+0.23}_{-0.26}$ \\
            \hline \hline
        \end{tabular}
        \caption{Inner-slope measurements across methods. Values from \citet{2025A&A...699A.347A} shown for comparison. 99.6\% of the posterior lies above $\Gamma=2$ (84.4\% after $\lambda$ calibration) for two-population model from \citet{2011ApJ...742...20W}. Note that $\Gamma=3-\gamma$ holds only if the profile is a local power law between the two radii.}
        \label{table:inner-slope-measurements}
    \end{table}
\endgroup

Three measurements (GravSphere, two-population via \citealp{2011ApJ...742...20W}, and \citealp{2025A&A...699A.347A}) cluster at $\gamma\approx0.39$--$0.50$, and spherical Jeans sits higher at 0.97, though its 68\% interval extends down to 0.64 and so remains close to the others (see Table~\ref{table:inner-slope-measurements}). The GravSphere $\gamma$ result has a one-sided posterior, so the constraint is an upper limit on cuspiness rather than a detection at 0.46. The two fitted profiles and the \citet{2025A&A...699A.347A} published curve converge near $r\approx0.2\,\text{kpc}$, roughly the metal-rich half-light radius, then diverge outward (see Figure~\ref{fig:fig4-all-chains}); the two-population estimator contributes discrete points at its two fitted radii rather than a profile, since it constrains the mass at those radii alone. It follows that the enclosed mass is well constrained at the half-light radius and poorly constrained elsewhere; the three curves constrain the same quantity well and differ due to different extrapolations from it (the two-population estimator constrains the mass only at its two fitted radii). The agreement is quantitative: the spherical Jeans and GravSphere posteriors give $M(<0.2\,\text{kpc})=6.41^{+0.58}_{-0.56}\times10^6$ and $6.48^{+1.45}_{-1.15}\times10^6\,M_\odot$, consistent to 1\% in the median despite inner slopes differing by 0.5. The two-population estimator gives a dynamical mass of $4.32^{+0.63}_{-0.59}\times10^6\,M_\odot$ within its fitted $r_{h,1}=0.192^{+0.013}_{-0.012}\,\text{kpc}$; this is not directly comparable, being a total rather than a dark-matter mass and carrying the \citet{2009ApJ...704.1274W} estimator's own normalization. GravSphere fits the same $\sigma_\text{los}$ data but adds VSPs, which break the mass-anisotropy degeneracy and tighten the constraint; the Jeans fit, lacking them, is the least constrained.

\subsection{Two-population estimator reproduction}
\label{sec:wp11-reprod}

\begin{figure*}[!htbp]
    \centering
    \includegraphics[width=\textwidth]{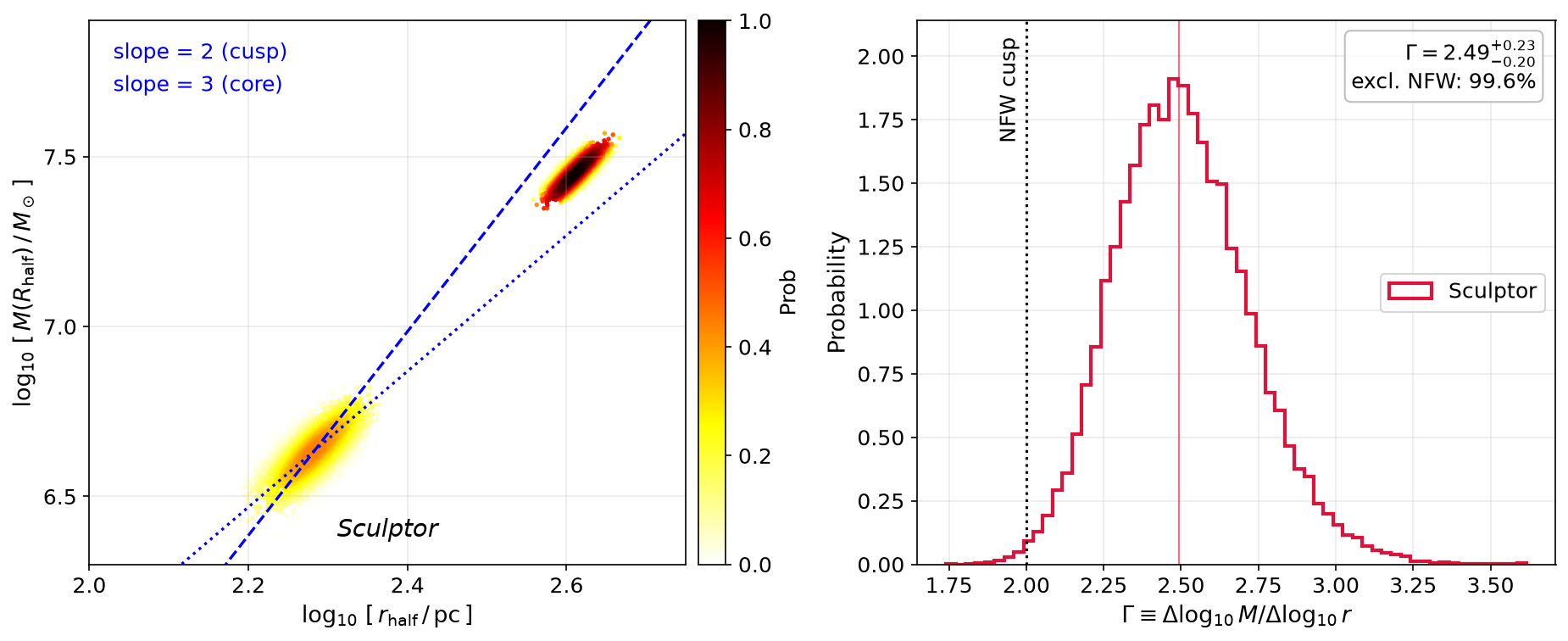}
    \caption{Sculptor; left panel: $\log_{10}{M(r_{\text{half}})}$ versus $\log_{10}{r_\text{half}}$, showing the two subcomponents as posterior density clouds, with dashed and dotted reference lines for slope = 2 (cusp) and slope = 3 (core). Right panel: the $\Gamma$ posterior, with the NFW cusp marked at $\Gamma=2$ and the median at 2.49. Lower-left cloud in left panel is the more concentrated metal-rich population.}
    \label{fig:wp11}
\end{figure*}

This work compares to two values from \citet{2011ApJ...742...20W} (see Figure~\ref{fig:wp11}): their elliptical-radius value ($\Gamma=2.40$, NFW exclusion $\geq93.9\%$) and their circular headline ($\Gamma=2.95$, exclusion $\geq99.8\%$). The elliptical value uses the \citet{1995MNRAS.277.1354I} ellipticity rather than the \citet{2018ApJ...860...66M} Plummer fit adopted throughout this work. The present study measures $\Gamma=2.49$ with 99.6\% of the posterior above $\Gamma=2$. \citet{2018MNRAS.474.1398G} independently recalculate the circular measurement's exclusion at 93.6\% once simulation-calibrated bias is accounted for, a different analysis from the \citet{2011ApJ...742...20W} posterior-integral figure rather than a correction of arithmetic. Applying the same calibration to the present measurement, convolving the $\Gamma$ posterior with the $\lambda$ distribution gives $\Gamma=2.61^{+0.74}_{-0.60}$ with 84.4\% of the corrected posterior above $\Gamma=2$. The median shifts slightly towards a core, as expected from $\langle\lambda\rangle<1$, but the interval widens by roughly a factor of three; the calibration uncertainty dominates the statistical uncertainty at this sample size. Four realizations at $\Gamma_\text{true}=2.94$ recover $2.35$--$3.62$, mean $3.15$, scatter $0.56$, consistent with no systematic bias but indicating realization-to-realization scatter exceeding the per-fit 68\% intervals. Their synthetic-data tests find their method systematically underestimates $\Gamma$, biasing toward apparent cuspiness due to a mechanism (mass overestimation scaling with stellar-subcomponent embeddedness) independent of the misalignment mechanism that \citet{2018MNRAS.474.1398G} later identify. Therefore, two independent analyses find the same conservative bias direction.

\subsection{Robustness to very-metal-poor contamination}
\label{sec:very-metal-poor-contam}

\begin{figure}[!htbp]
    \centering
    \includegraphics[width=\columnwidth]{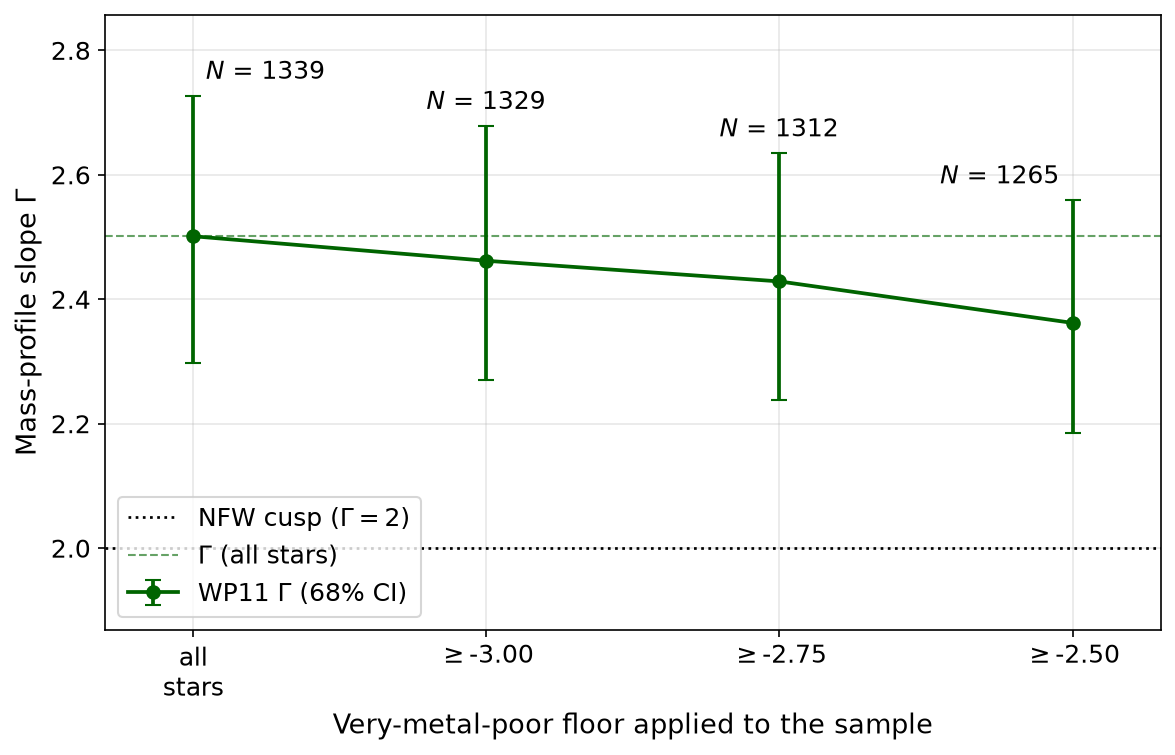}
    \caption{$\Gamma$ from the Sculptor two-population estimator against progressively stricter $[\text{Fe}/\text{H}]$ floors (cf. Figure~\ref{fig:membership-robustness}, which cuts on proper-motion offset), with $N$ annotated at each point, 68\% error bars, a dotted line at $\Gamma=2$ representing the NFW cusp, and a dashed reference line at the full-sample value. The strictest cuts remove the third population characterized by \citet{2024A&A...692A.195A} ($\sim24$ stars at $[\text{Fe}/\text{H}]\approx-2.9$). $\Gamma$ declines monotonically as very-metal-poor stars are removed but stays above 2 at every cut, so the contaminant's presence biases toward a core, meaning the full-sample value is the least conservative of the four and $\Gamma=2.36$ is the conservative choice. $\Gamma>2$ at every cut regardless.}
    \label{fig:pop3-robustness}
\end{figure}

\begingroup
    \setlength{\tabcolsep}{10pt}
    \renewcommand{\arraystretch}{1.5}
    \setlength\extrarowheight{2pt}
    \begin{table}
        \centering
        \begin{tabular}{ c c c }  
            $[\text{Fe}/\text{H}]$ floor & $N$ & $\Gamma$ \\
            \hline \hline     
            none & 1339 & $2.49^{+0.23}_{-0.20}$ \\
            $\geq-3.00$ & 1329 & $2.46^{+0.22}_{-0.19}$ \\
            $\geq-2.75$ & 1312 & $2.43^{+0.21}_{-0.19}$ \\
            $\geq-2.50$ & 1265 & $2.36^{+0.20}_{-0.18}$ \\
            \hline \hline
        \end{tabular}
        \caption{Robustness of $\Gamma$ to contamination by the very-metal-poor subcomponent characterized by \citet{2024A&A...692A.195A}.}
        \label{table:feh-floor-v-gamma}
    \end{table}
\endgroup

Table~\ref{table:feh-floor-v-gamma} shows $\Gamma$ measurements are robust to contamination by the very-metal-poor subcomponent measured by \citet{2024A&A...692A.195A} ($\sim24$ stars, $[\text{Fe}/\text{H}]\approx-2.9$, $\sim15\,\text{km/s}$ offset). Maximum $|\Delta\Gamma|=0.14$, and $\Gamma>2$ at every cut. Removing these stars lowers $\Gamma$, meaning their presence biases towards a core (see Figure~\ref{fig:pop3-robustness}).

\subsection{$k_J$ and the action-space gradient}
\label{sec:action-space-grad}

\begin{figure*}[!htbp]
    \centering
    \includegraphics[width=\textwidth]{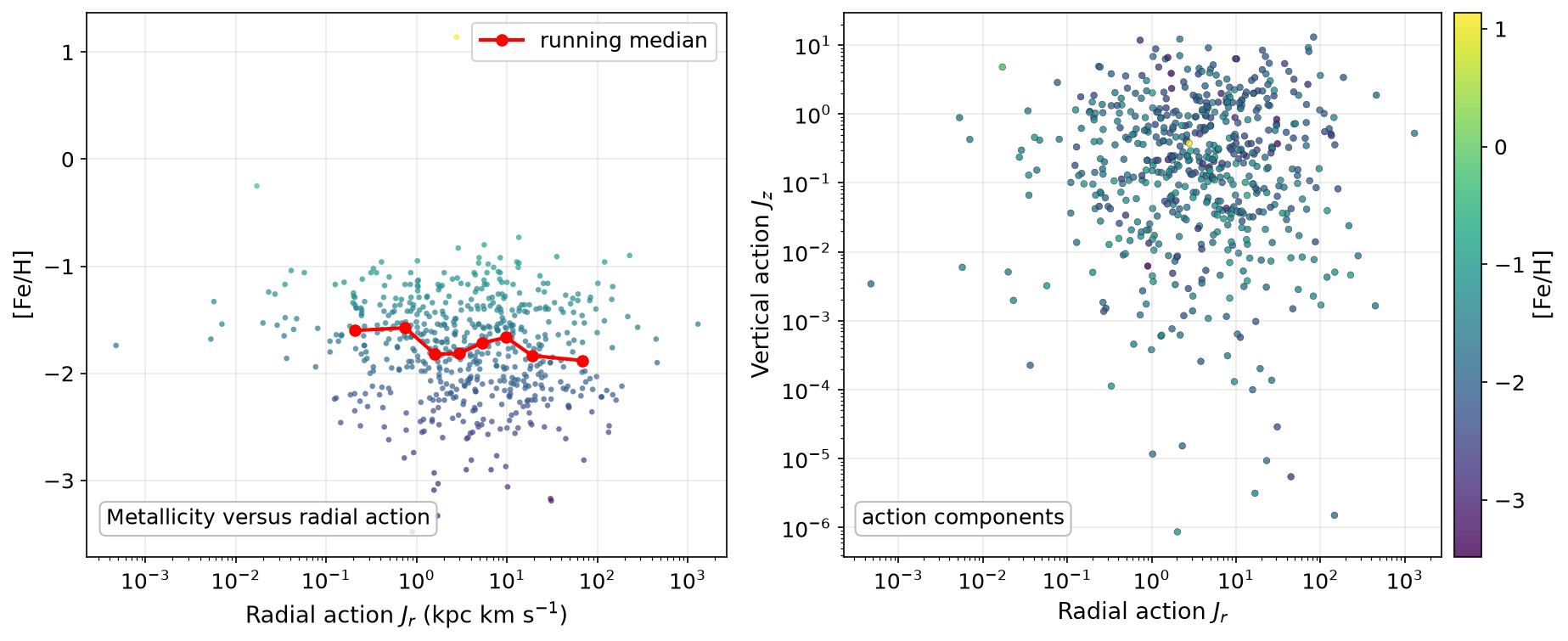}
    \caption{Sculptor; left: $[\text{Fe}/\text{H}]$ versus radial action $J_r$ in log scale, points colored by $[\text{Fe}/\text{H}]$, with a red running-median line. Right: $J_z$ versus $J_r$, both log, colored by $[\text{Fe}/\text{H}]$, with a colorbar. The right shows the sample spans a wide range of $J_r$ and $J_z$ and no comparable trend with $J_z$. The running median declines from $\approx-1.6$ to $\approx-1.9$ across three decades in $J_r$, the gradient that $k_J<0$ encodes. There are 575 stars with valid actions.}
    \label{fig:action-space}
\end{figure*}

The metallicity--action gradient that $k_J$ parameterizes is measured directly rather than through the fit, by computing actions for the 575 stars with valid Gaia astrometry (see Figure~\ref{fig:action-space}); running-median $[\text{Fe}/\text{H}]$ declines from $\approx-1.6$ at $J_r\approx0.2$ to $\approx-1.9$ at $J_r\approx70\,\text{kpc}\,\text{km}\,\text{s}^{-1}$ ($\sim0.3\,\text{dex}$ over $\sim3$ decades in $J_r$). The continuous DF chain is non-converging (see Appendix~\ref{ap:mcmc-conv}), so no posterior value of $k_J$ is quoted; the sign and magnitude of the gradient rest on the direct computation, which is independent of the fit. The gradient is the dSph analog of the \citet{2015MNRAS.449.3479S} action-space metallicity analysis for the Galactic disc.

\subsection{Fornax}
\label{sec:fornax}

\begin{figure*}[!htbp]
    \centering
    \includegraphics[width=\textwidth]{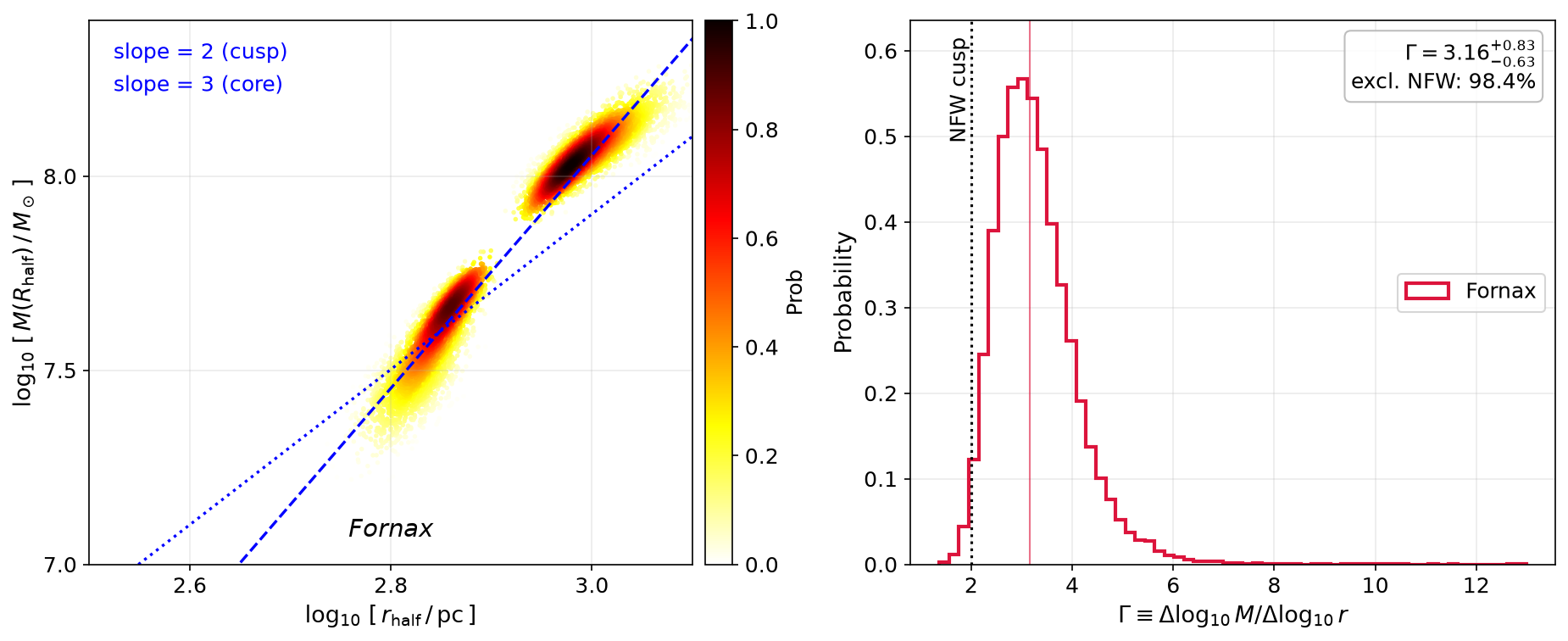}
    \caption{Fornax; left panel: $\log_{10}{M(r_{\text{half}})}$ versus $\log_{10}{r_\text{half}}$, showing the two subcomponents as posterior density clouds, with dashed and dotted reference lines for slope = 2 (cusp) and slope = 3 (core). Right panel: the $\Gamma$ posterior, with the NFW cusp marked at $\Gamma=2$ and the median at $3.16$; 98.4\% of the posterior lies above $\Gamma=2$. The subcomponents separate, with the metal-rich population the more concentrated of the two. Non-members are included in the sample.}
    \label{fig:fornax-wp11}
\end{figure*}

\citet{2011ApJ...742...20W} report $\Gamma=2.61^{+0.43}_{-0.37}$ for Fornax, excluding an NFW cusp at $\geq96\%$. Applying the implementation of Section~\ref{sec:wp11-two-pop} to the same catalog, Mg-index discriminant, and 2603-star sample gives $\Gamma=3.16^{+0.83}_{-0.63}$, with 98.4\% of the posterior above $\Gamma=2$, agreement at $0.8\sigma$ on both the slope and the exclusion. The fitted nuisance parameters also track theirs. This work obtains $f_\text{mem}=0.953\pm0.006$ versus their $f_\text{mem}$ of 0.96; $f_\text{sub}=0.560^{+0.154}_{-0.190}$ versus their $f_\text{sub}$ of 0.60; $\Delta W'=0.150\pm0.038$ against their 0.16; and velocity dispersions of $9.9$ and $14.0\,\text{km}\,\text{s}^{-1}$ against their 10.0 and 14.5.

This work finds systematically larger half-light radii of $705^{+36}_{-48}\,\text{pc}$ and $976^{+89}_{-62}\,\text{pc}$ (radius ratio of $0.722^{+0.049}_{-0.047}$), compared with the \citet{2011ApJ...742...20W} values of $552\,\text{pc}$ and $891\,\text{pc}$ (radius ratio of 0.62). This is the expected direction for the reconstructed selection function, whose Gaia DR3 photometric parent is broader than the \citet{Walker_2007} target list \citet{2011ApJ...742...20W} used and which is therefore flatter in $R$ than theirs. Since the parent is a reconstruction rather than the original selection function, two variations are possible. Restricting the parent in $BP$--$RP$ to $(0.7,2.2)$, limiting it in the direction of the RGB \citet{Walker_2007} targeted, gives the quoted $\Gamma=3.16$. A magnitude-only parent, which additionally admits main-sequence foreground, gives $\Gamma=2.97^{+0.71}_{-0.56}$ with radii $716^{+38}_{-52}$ and $1019^{+98}_{-70}\,\text{pc}$. The reconstruction thus carries a systematic of $\approx0.2$ in $\Gamma$, larger than any other modeling choice examined here and comparable to a third of the statistical interval.

The $0.55$ offset in $\Gamma$ from the \citet{2011ApJ...742...20W} value thus decomposes as $\approx0.2$ from the selection-function reconstruction, with the remaining $\approx0.35$ inside the realization-to-realization scatter of $0.56$ measured on mock data in Appendix~\ref{ap:mock-recovery}. The agreement is therefore partially accounted for and partially absorbed by scatter, rather than resting on the width of the interval alone.

The recovered $\Gamma=3.16$ formally implies $\gamma=3-\Gamma=-0.16$, which is unphysical, and 59.1\% of the Fornax posterior lies above $\Gamma=3$. The relation $\Gamma=3-\gamma$ only holds where the density profile is a single power law between the two half-light radii, and the estimator carries no prior enforcing it. $\Gamma$ is a finite-difference slope between two independent mass measurements and unbounded above. A posterior with most of its mass beyond $\Gamma=3$ thus indicates that the local-power-law assumption fails across the radial separation of the Fornax subcomponents ($r_{h,1}=705\,\text{pc}$, $r_{h,2}=976\,\text{pc}$, both far outside Sculptor's fitted radii), rather than a physical inner slope, and the narrow fitted radius ratio ($r_{h,1}/r_{h,2}=0.722$ against the 0.62 of \citealt{2011ApJ...742...20W}) inflates $\Gamma$ for the same dispersion ratio, since $\Gamma$ scales inversely with $\log_{10}{r_{h,2}/r_{h,1}}$. The Fornax fit is accordingly reported as a constraint on $\Gamma$ and on the exclusion of $\Gamma=2$, an NFW cusp, and no $\gamma$ is inferred from it. This does not affect the Sculptor measurement, where only 2.0\% of the posterior lies above $\Gamma=3$, so the conversion to $\gamma$ is well defined there.

\section{Discussion}
\label{sec:discussion}

\subsection{Robustness of the core-like result}
\label{sec:core-like-robustness}

Of the four methodologically distinct estimators applied to the same $\sigma_\text{los}$ data, two converge on a shallow slope: GravSphere at $\gamma=0.46$ and the two-population estimator at $\gamma=0.50$; the continuous DF is non-converging, and the spherical Jeans fit sits higher at $\gamma=0.97$, though a prior-sensitivity test shows this is a data-driven result rather than a prior artifact, and roughly half of the discrepancy is attributable to the fixed OM profile. Refitting at an effectively isotropic $r_a=100\,\text{kpc}$ gives $\gamma=0.71$. Both shallow values agree with the published \citet{2025A&A...699A.347A} measurement of 0.39 on the same catalog and the \citet{2016MNRAS.463.1117Z} $0.5\pm0.3$ from independent data. They differ in tracer treatment (single- versus two-population tracers), in whether fourth moments are used (LOS-only versus VSP-included), and in whether a profile is fitted at all, leading to the argument that no single method's systematics drive the conclusion, though their data systematics are not independent: all four share the same membership selection and velocity measurements from \citet{Tolstoy_I}, so a systematic in the catalog would propagate to every result. This is supported by $\Gamma=2.49$ surviving PM-offset cuts down to 25\% of the sample ($\Gamma>2$) and very-metal-poor exclusion ($|\Delta\Gamma|\leq0.14$, $\Gamma>2$ throughout). The fitted half-light radii ($192^{+13}_{-12}$/$408^{+15}_{-14}$ pc) independently reproduce the \citet{2024A&A...692A.195A} decomposition (187.5/398.3 pc).

\subsection{Response to inference-artifact critiques}
\label{sec:inference-artifacts}

\citet{2018MNRAS.474.1398G} find $\Gamma\geq2$ in 17.8\% of cases for simulated dwarfs that are cuspy by construction and recalculate the Sculptor exclusion from \citet{2011ApJ...742...20W} at 93.6\% rather than 99\%. However, the mean bias runs the other way. $\lambda=\Gamma_{\text{est}}/\Gamma_{\text{true}}=0.95^{+0.28}_{-0.19}$ for the \citet{2009ApJ...704.1274W} estimator, which is used in this work. Since $\Gamma=3-\gamma$, underestimating $\Gamma$ means biasing cuspier. A core measured with a cusp-biased estimator is conservative; the 17.8\% is the tail of the $\lambda$ distribution, not its expectation.

Additionally, the failure described requires misalignment rather than elongation. Their Galaxies 1 and 2 are elongated but the subpopulation major axes are aligned; thus, the errors cancel in $\Gamma$, and the cusp is recovered. Only Galaxy 4, which is misaligned, produces a spurious core: its inferred $\Gamma$ ranges from $1.12$, viewed perpendicular to the metal-poor major axis, to $2.45$, viewed along it, and is measured as cored for viewing angles between roughly $0$ and $50\,\mathrm{deg}$. Sculptor also passes their radial-separation cut: they discard $\log_{10}{r_2/r_1}<0.06$, and Sculptor's fitted radii give $0.328^{+0.030}_{-0.029}$, with no posterior mass below the threshold. Fornax's fitted radii give $\log_{10}{r_2/r_1}=0.149^{+0.028}_{-0.029}$, with 0.09\% below. Neither galaxy falls in the regime they exclude by construction, though Fornax sits much closer to the threshold than Sculptor does, and that narrow separation is also what inflates its recovered $\Gamma$ (see Section~\ref{sec:fornax}).

\citet{2019MNRAS.484.1401R} flag the \citet{2011ApJ...742...20W} Sculptor result as an outlier, finding that its enclosed mass near $300\,\mathrm{pc}$ exceeds that of every other analysis they compare against by a factor of $\sim2$: the steep slope required to reach that outer mass from a more widely agreed inner value drives the incompatibility with an NFW cusp, and they lack a satisfactory explanation for the resulting $\sim2\sigma$ discrepancy. This work reproduces that measurement independently on a different catalog, obtaining $\Gamma=2.49^{+0.23}_{-0.20}$, and can locate where it does and does not sit apart from the literature. At $150\,\mathrm{pc}$, the profiles of this work agree with theirs at $0.74\sigma$ (see Section~\ref{sec:central-density}). The discrepancy is therefore not in the inner mass, which every method here recovers in common with the published values, but in the outer point of the two-population pair. Three checks argue against a geometric origin for it: the subcomponent major axes are aligned to $-4.1\pm4.3\,\mathrm{deg}$, excluding the misalignment mechanism of \citet{2018MNRAS.474.1398G}; the fitted radius ratio $\log_{10}r_2/r_1 = 0.328^{+0.030}_{-0.029}$ sits well clear of their discard threshold with no posterior mass below it; and $\lambda$ calibration shifts the median only towards a core. The origin of the outer-mass excess remains open, but it is localized and is not attributable to projection.

The critique targets a geometry utilized by two of the four methods: the two-population estimator and the spherical Jeans fit both rely on the metallicity split, though only the former derives $\Gamma$ from the ratio of half-light radii. GravSphere, which fits a single tracer profile with VSPs and free $\beta$, and the continuous DF, which imposes no split, do not use it and agree with the split-based results. Their failure mode requires misalignment rather than elongation, a testable condition. Splitting the sample at the median metallicity and measuring the orientation of each subcomponent from the second moments of its projected positions gives $\text{PA}=84.6^{+3.4}_{-3.5}\,\text{deg}$ for the metal-rich population and $88.7^{+2.5}_{-2.7}\,\text{deg}$ for the metal-poor population, a misalignment of $-4.1\pm4.3\,\text{deg}$. Both are genuinely elongated, so the orientations are well-defined. Bootstrapping the split gives an 8.5\% probability that the misalignment exceeds $10\,\text{deg}$ and none that it exceeds $20\,\text{deg}$, placing Sculptor with their Galaxies 1 and 2, which are elongated but aligned and for which the cusp is recovered, rather than with the misaligned Galaxy 4 that produces a spurious core. The measurement uses the spectroscopic sample, which is not spatially complete; both subcomponents are drawn from the same targeting, so a selection-imposed geometry largely cancels in their difference, and the full sample returns $\text{PA}=87.5\pm2.1\,\text{deg}$ against the $92.0\,\text{deg}$ of the spatially complete \citet{2018ApJ...860...66M} photometry, a $2\sigma$ offset that bounds how much orientation the selection can be imposing. Sculptor's intrinsic three-dimensional shape remains unobservable from a single viewing angle, so this does not exclude the 17.8\% tail outright, but it does test the specific mechanism.

Synthetic analyses from \citet{2011ApJ...742...20W} find that their method systematically underestimates $\Gamma$ through mass overestimation scaling with subcomponent embeddedness, a mechanism unrelated to misalignment. The \citet{2018MNRAS.474.1398G} simulations and \citet{2011ApJ...742...20W} self-test agree the bias is conservative. \citet{2018MNRAS.474.1398G} Appendix A2 finds geometry, not metallicity mixing, drives their spurious cores; Section~\ref{sec:decomp-bias} independently attributes $77$--$83\%$ of the mock bias to tracer geometry through null and scramble controls (at $\gamma_\text{true}=0.4$, the chemical contribution is $+0.053\pm0.017$, and at 1.0 it is $+0.134\pm0.026$), finding the same conclusion.

\subsection{Relation to Arroyo-Polonio et al. (2024)}
\label{sec:other-one-pops}

\citet{2024A&A...692A.195A} describe their single-population model as having one surface number density profile, a metallicity gradient, and a constant $\sigma_{\text{los}}$ by construction, making their $\Delta\text{BIC}$ a test of kinematics-metallicity correlation rather than continuity. They note the same limitation themselves (their Section~5.2.5), observing that the measured dispersion profiles decline by $\geq2\,\text{km}\,\text{s}^{-1}$ across the radial range in both populations and propose distribution-function modeling as the solution, which is the approach taken in \citet{2025A&A...699A.347A} and, in action space, here. Additionally, their reliability test never checks continuous truth against false two-population preference; the continuous DF here is a different kind of continuum (action, not position) that they did not test.

\subsection{The VSP method applied to Sculptor}
\label{sec:vsp-sculptor}

\citet{2014MNRAS.441.1584R}, who introduce $\zeta_A$ and $\zeta_B$, applied them to Sculptor and concluded the halo ``may be cuspy after all, in contrast to the conclusions of other approaches'', plotting their result directly against the \citet{2011ApJ...742...20W} headline $\Gamma$. This work uses the same VSP machinery via GravSphere and recovers a core-like $\gamma=0.46$. Both analyses fix $\nu(r)$ to a Plummer profile (by construction in this work), and thus the tracer-profile assumption cannot account for the opposite conclusions. However, \citet{2014MNRAS.441.1584R} use MMFS data \citep{2009AJ....137.3100W}, while this work uses VLT/FLAMES \citep{Tolstoy_I}, though \citet{2016MNRAS.463.1117Z} also use MMFS data and produce a core, making the data an unlikely explanation. \citet{2014MNRAS.441.1584R} themselves anticipate the VLT/FLAMES sample in their Appendix B, noting that its dispersion profile ``rises out to a greater radius $R$'' than the data they consider, and expecting it to behave like their 12-bin case; however, they never run the VSPs against the data. Moreover, they compare discrete model families (NFW versus Burkert versus fixed-($\alpha,\beta,\gamma$) Zhao) by $p$-value, while this work fits gNFW with $\gamma$ free and reports a posterior; the experimental questions were different. Additionally, \citet{2014MNRAS.441.1584R} deliberately avoid posteriors; they state that Jeans analyses have weak likelihood surface peaks, making prior choice important, and due to prior uncertainty, they use likelihood in place of posterior distribution. This is the concern quantified in Section~\ref{sec:decomp-bias} at a mock shrinkage fraction of 0.24, and tested directly in Section~\ref{sec:sigma-los-degen}: on the real data, a 0.20 shift in the prior median moves the posterior by 0.02. Retaining posterior inference is therefore defensible here provided the prior sensitivity is measured rather than assumed, which is what \citet{2014MNRAS.441.1584R} could not do without such a test. Furthermore, they concede that a modestly steeper outer tracer slope ($\beta_*=5.5$ rather than the traditional Plummer 5) makes cored models consistent with their $\zeta$ measurements again (their Section 4.1.2, Figure 10) and that with full freedom in $\nu(r)$, $\beta(r)$, and $\beta'(r)$, the fourth moments are not sufficient to solve the degeneracy fully.

\subsection{Comparison with directly measured anisotropy}
\label{sec:anisotropy-comparison}

\citet{2018NatAs...2..156M} measure Sculptor's internal transverse motions directly, combining a 2002 HST epoch with Gaia astrometry over a 12.3-year baseline. From 15 stars with the best proper motions, they obtain projected dispersions $\sigma_R=11.5\pm4.3$ and $\sigma_T=8.5\pm3.2\,\text{km}\,\text{s}^{-1}$ at $R_\text{HST}\approx185\,\text{pc}$. Propagating these through the spherical Jeans equations gives a posterior for $\hat{\beta}$ with maximum a posteriori (MAP) $0.86^{+0.12}_{-0.83}$ and median $\approx0.46$.

\citet{2018ApJ...860...56S} directly reanalyze the \citet{2018NatAs...2..156M} measurements and find that cored and cuspy potentials predict ``almost identical'' transverse dispersion, with $\sigma_R$ differing by $\sim1\sigma$; these dispersions cannot discriminate between a core and cusp at the projected radius utilized. 

The GravSphere posterior of this work gives $\beta\approx-0.03$ inside $0.3\,\text{kpc}$ (see Figure~\ref{fig:grav-beta}), below the \citet{2018NatAs...2..156M} median and far below their MAP value, but their lower bound extends nearly to isotropy, so the two are not formally inconsistent. The comparison is limited by their sample size, as 15 stars at a single projected radius yield 36\% fractional errors on each dispersion, and the resulting $\hat{\beta}$ posterior is broad enough to accommodate most of the published range. \citet{2018NatAs...2..156M} note that their metal-rich value $\hat{\beta}_\text{MR}=0.95^{+0.04}_{-0.27}$ agrees with predictions for a cuspy halo, while published cored models predict lower anisotropy. They also observe that $\beta\approx0.9$ cannot hold at constant radius alongside a shallow central light profile, since $\gamma\geq2\beta$ in the spherical limit \citep{2006ApJ...642..752A}, and take this as evidence that Sculptor's halo may be axisymmetric or triaxial rather than spherical (see Section~\ref{sec:model-assumpt}).

Two orbit-based analyses independently support the near-isotropic inner anisotropy found here. \citet{2013MNRAS.433.3173B}, using a Schwarzschild orbit superposition model rather than a Jeans-family model, find a tangentially biased profile, nearly constant with radius. \citet{2013A&A...558A..35B} find $\beta\approx-0.5\pm0.3$ for Sculptor, overlapping the inner band of Figure~\ref{fig:grav-beta}. Both sit far from the \citet{2018NatAs...2..156M} MAP value and close to the value of this work. 

The most direct constraint now available comes from \citet{Vitral_2026}, whose 119-star HST PM catalog supersedes the 15-star sample of \citet{2018NatAs...2..156M} by an order of magnitude in size and resolves the transverse dispersions radially rather than at a single projected radius. They measure a ratio of tangential to plane-of-sky radial dispersions of $1.19\pm0.19$ and a ratio of LOS to plane-of-sky dispersion of $0.93\pm0.08$, and they recover a globally averaged anisotropy that is mildly radial but statistically consistent with isotropy at their fiducial inclination. That is closer to the near-isotropic inner value found here than to the strongly radial \citet{2018NatAs...2..156M} MAP value, though their anisotropy, like their density slope, correlates with the assumed inclination and becomes strongly tangential in edge-on cases.

\subsection{Central density and the core-cusp diagnostic}
\label{sec:central-density}
\citet{2019MNRAS.484.1401R} classify dwarfs by the amplitude of the DM density at a fixed small radius, $\rho_\mathrm{DM}(150\,\mathrm{pc})$, rather than by inner slope, because, with $\sim500$ LOS velocities, GravSphere constrains the DM density robustly while its inference of inner slope is prior-dependent. That effect is quantified in Section~\ref{sec:decomp-bias} at a mock shrinkage fraction of $0.24$ and tested directly on the data in Section~\ref{sec:sigma-los-degen}; it is independent confirmation of the prior sensitivity that this work measures rather than assumes.

Evaluating the posteriors of this work at their radius gives $\rho_\mathrm{DM}(150\,\mathrm{pc}) = 1.77^{+0.33}_{-0.26}\times10^{8}\,M_\odot\,\mathrm{kpc^{-3}}$ from GravSphere and $1.70\pm0.11\times10^{8}$ from spherical Jeans, against their Sculptor value of $1.49^{+0.28}_{-0.23}\times10^{8}$ in their Table~1: agreement at $0.74\sigma$, from an independent implementation, a different photometric catalog and a different tracer-profile treatment. \citet{10.1093/mnras/stae1716} compare five independent Sculptor measurements of this quantity, spanning $1.47$ to $2.14\times10^8$ and taken from Jeans, axisymmetric Jeans, and phase-space distribution-function analyses. Their DF-based value of $1.47^{+0.18}_{-0.20}\times10^8$ agrees with the GravSphere posterior of this work at $0.95\sigma$ and with the Jeans posterior at $1.1\sigma$. Both values of this work sit inside that range of values rather than at either edge of it. Every posterior sample of both chains lies above their $10^8\,M_\odot\,\mathrm{kpc^{-3}}$ threshold, placing Sculptor as cusp-like, as its truncated star formation predicts.

This does not contradict the shallow $\gamma$ reported here, because the two are different quantities. The $\gamma$ of Equation~\ref{eq:inner-slope} is the asymptotic slope as $r\to0$; the slope attained at $150\,\mathrm{pc}$ is $-1.00^{+0.19}_{-0.23}$ for the GravSphere posterior and $-1.11^{+0.16}_{-0.13}$ for the Jeans posterior, against the $-0.83^{+0.30}_{-0.25}$ of \citet{2019MNRAS.484.1401R}. All three agree at the radius where the data constrains the profile, and all three are consistent with an NFW cusp under spherical modeling. The two fits of this work agree on $\rho_\mathrm{DM}(150\,\mathrm{pc})$ to $4\%$ while their asymptotic slopes differ by $0.51$, and the Jeans density interval is $\pm6\%$ while its $\gamma$ spans $0.64$--$1.18$: the $\gamma$--$r_s$--$M_\mathrm{DM}$ degeneracy of Figure~\ref{fig:dm5-corner} runs almost exactly along the direction that preserves the density near the constrained radius. The core-like $\gamma$ is thus an extrapolation inward of the innermost kinematic constraint, and the data bears on it only weakly.

The local slope itself, however, is where this work disagrees most sharply with a recent analysis using resolved 3D velocities; \citet{Vitral_2026} combine 119 Hubble Space Telescope PMs over a 20-year baseline with 1760 LOS velocities to obtain the first radially resolved 3D velocity dispersion profiles for Sculptor and fit them with axisymmetric Jeans models. At their fiducial inclination of $57.1\,\mathrm{deg}$, they recover a DM density slope averaged over the radial extent of the 3D data ($\sim120$--$240\,\rm pc$) of $0.29^{+0.41}_{-0.31}$, excluding a cusp at $99.8\%$ confidence. The local slopes reported above, $-1.00$ and $-1.11$, sit at or beyond the boundary they exclude, over a radial range that overlaps the one they constrain, and with the mass-anisotropy degeneracy broken by transverse motions rather than by fourth moments or a metallicity split. Their result thus contradicts this work's conclusion that a shallow asymptotic slope coexists with a near-NFW slope at $150\,\mathrm{pc}$.

However, their constraint is inclination-dependent: their Table~2 shows the recovered slope essentially follows the full cusp-to-core range as the assumed inclination changes from face-on to edge-on. The inclination-averaged value is correspondingly unconstrained. The $99.8\%$ exclusion follows from adopting the flattened configuration favored by comparing the observed LOS kurtosis against \citet{10.1093/mnras/282.3.909} models, a comparison they describe as offering educated guesses rather than a precise inclination fit. Against that, the degeneracy they identify is invisible to every method applied in this work, all four of which assume spherical symmetry (see Section~\ref{sec:model-assumpt}), so it is not a systematic that the internal consistency of the four methods could have detected; thus, the local slope at $150\,\rm pc$ is the axis on which the present LOS-only analysis is least secure, and resolving it requires either the inclination constraint their method lacks or the transverse velocities this work lacks.

\subsection{Literature comparison and tension}
\label{sec:lit-comp}

Four analyses using metallicity information cluster at $\gamma=0.39$--$0.50$ across two independent datasets. \citet{2016MNRAS.463.1117Z} find $\gamma=0.5\pm0.3$ using MMFS data, and \citet{2025A&A...699A.347A} find $\gamma=0.39$ using VLT/FLAMES data, bracketing this work's GravSphere ($\gamma=0.46$) and two-population ($\gamma=0.50$) results. \citet{2012MNRAS.419..184A} reach a similar conclusion earlier from Michie--King DFs fit to the \citet{Battaglia_2008} two-population data, finding that while an NFW potential gives an acceptable $\chi^2$, a cored profile is preferred by a likelihood ratio test. The spread in results is framed by \citet{2016MNRAS.463.1117Z}; they find $\sim6000$ discrete stars are needed to distinguish $\gamma=0$ from $\gamma=1$ when Sculptor has $N=1339$, an independent confirmation of the implications of the weakly constraining likelihood in Section~\ref{sec:sigma-los-degen}. \citet{Battaglia_2008} show that a pseudo-isothermal fit is better ($\chi^2=6.9$ versus $10.8$) and note NFW ``tends to overpredict the central values'', with both fitting acceptably. \citet{10.1093/mnras/stw713} coreNFW simulations predict that ``pristine dark matter cusps will be found either in systems that have truncated star formation and/or at radii $r > r_{1/2}$''. Sculptor is ancient and quenched: \citet{2019MNRAS.487.5862B} find Sculptor stopped forming stars $\sim11.3\,\mathrm{Gyr}$ ago with a duration of $\sim2.2\,\mathrm{Gyr}$; \citet{2012A&A...539A.103D} conversely find an extended, continuous star formation of $6$--$7\,\mathrm{Gyr}$. \citet{2019MNRAS.487.5862B} also find a radial trend, with the innermost region showing $\sim1.5\,\mathrm{Gyr}$ of star formation and the outermost confined to $\sim0.5\,\mathrm{Gyr}$. The \citet{10.1093/mnras/stw713} criterion turns on duration, so under \citet{2019MNRAS.487.5862B}, the study with deeper photometry, a cusp is firmly predicted; under \citet{2012A&A...539A.103D}, the tension is much weaker. That prediction is met on the central-density axis and not the slope axis: Section~\ref{sec:central-density} finds $\rho_\mathrm{DM}(150\,\mathrm{pc})$ above the threshold of \citet{2019MNRAS.484.1401R} in every posterior sample, while the two-population and GravSphere asymptotic slopes are shallow. The distinction is the one \citet{10.1093/mnras/stv1504} draw in reframing the core-cusp problem as an inner mass deficit rather than a slope mismatch; Sculptor as measured here has no inner mass deficit and a shallow asymptotic slope, so any tension with feedback-driven core formation rests on the slope alone. The spherical Jeans result of $\gamma=0.97$ does not support this reading, though a prior-sensitivity test shows this is a data-driven result rather than just a prior artifact (see Section~\ref{sec:sigma-los-degen}). Refitting with effective isotropy across the fitted range (anisotropy radius set to $100\,\text{kpc}$) finds $\gamma=0.71^{+0.31}_{-0.40}$, so $\sim50\%$ of the offset between the spherical Jeans value and the 0.39--0.50 cluster is attributable to the fixed OM profile, with the remainder unexplained. Two other differences between the Jeans fit and the $\sigma_\text{los}$-only profile scan from Section~\ref{sec:sigma-los-degen} are not isolated by this test, as the Jeans fit uses two binned dispersion profiles rather than the full unbinned sample. The error interval of the refit widened from a 68\% interval of width 0.54 to 0.71; thus, fixing $r_a=1.5$ tightened the posterior as well as raising it.

The alternative DM explanations of Section~\ref{sec:intro} are not favored by this measurement either. SIDM \citep{1992ApJ...398...43C,Spergel_2000} produces a core by transferring energy outward, lowering the central density; the $\rho_\text{DM}(150\,\mathrm{pc})$ of Section~\ref{sec:central-density} is instead consistent with an unmodified cusp. Sculptor's shallow asymptotic slope is not accompanied by the central-density deficit that either feedback-driven or self-interaction core formation would produce.

\subsection{Relation to multi-component distribution function work}
\label{sec:multi-df}

\subsubsection{Inner slope diversity in dwarf spheroidals}

\citet{Pascale_2026} find that the Draco and Ursa Minor (UMi) dSphs span cusp-to-core with discrete two-population action-based DFs ($\gamma_{\text{Draco}}=0.98,\gamma_{\text{UMi}}=0.37$); that two systems modeled with identical machinery span nearly the full cusp-to-core range indicates real diversity in dSph inner slopes, so Sculptor's shallow value should not be read as a universal dwarf-galaxy property. \citet{2013A&A...558A..35B} concur, showing that it is unlikely that all galaxies are embedded in the same type of cored profiles of the form $\rho_\text{DM}\propto1/(1+r^2)^{\beta/2}$, arguing against a universal core.

\citet{10.1093/mnras/stae1716} reach the same conclusion on the density axis rather than the slope axis, applying a single phase-space DF pipeline with uniform priors to nine bright Milky Way dSphs and recovering central densities that vary by a factor of $\sim5$, bracketed by Draco and Leo~I at the dense end and Fornax and Carina at the shallow end. Such diversity surviving a uniform analysis of a homogeneous sample suggests it is physical rather than an artifact of heterogeneous methods.

\subsubsection{Response to lack of core preference in separable distribution functions}

\citet{Strigari_2017} use a separable $f(E,J)$ DF on Sculptor's two populations and find the data is consistent with populations in equilibrium within an NFW potential with structural parameters in the range expected in $\Lambda$CDM, with no statistical preference for a potential with a core. That result has since been tested against a broader DF family. \citet{10.1093/mnras/stae1716} apply a generalized separable DF to the same class of systems and find, contrary to previous findings for some of these objects, that the DF analysis yields results consistent with the standard Jeans analysis; their Appendix~A reproduces the \citet{Strigari_2017} Sculptor inference under a matching configuration and recovers close agreement. The disagreement between DF and Jeans inferences for Sculptor is therefore specific to the model family rather than intrinsic to DF methods, which is the perspective this work adopts below.

The \citet{Strigari_2017} model family fixes the anisotropy structure through the separability ansatz, which Section~\ref{sec:model-assumpt} and Table~\ref{table:anisotropy-sensitivity} show moves $\gamma$ by 0.25. Additionally, they re-derive the half-light radii from the data themselves and get $0.18$ and $0.22\,\mathrm{kpc}$ for the metal-rich and metal-poor populations, respectively, with the metal-rich value in good agreement with \citet{2011ApJ...742...20W} but the metal-poor value significantly smaller than the \citet{2011ApJ...742...20W} value of $0.30\,\mathrm{kpc}$. As $\Gamma$ scales inversely with $\log_{10}{r_2/r_1}$, the discrepancy in the radius values is the source of the disagreement. This work's fitted radii ($192$/$408\,\mathrm{pc}$) reproduce the \citet{2024A&A...692A.195A} decomposition values of $187.5$ and $398.3\,\mathrm{pc}$ independently, from VLT/FLAMES with $[\text{Fe}/\text{H}]$ as a discriminant. Contrariwise, the \citet{Strigari_2017} $0.22\,\mathrm{kpc}$ value comes from an Mg-index split of MMFS data by a Kolmogorov--Smirnov-optimized cut, a weaker discriminant than metallicity.

They instead fault the \citet{2011ApJ...742...20W} model directly: the metal-rich and metal-poor populations are each represented by a circularly symmetric Plummer profile, with Gaussian velocity and metallicity distributions independent of radius, assumptions they note are not consistent with the \citet{Battaglia_2008} data. This work's implementation makes the same assumptions, with two Plummer subcomponents, each with a Gaussian velocity and metallicity distribution. They find that, in the absence of any prior on the shape of the inner potential, the Sculptor data prefers a shallower profile than NFW but find this preference too weak to exclude the cosmological prediction. They weaken the constraint by allowing more freedom through the choice of their family of DFs; this work narrows it by imposing a parametric form. This work's $\rho(150\,\mathrm{pc})$ value is in agreement with \citet{2019MNRAS.484.1401R} regardless.

Methodologically, \citet{Strigari_2017} take the geometric mean of the major and minor axes as the radial coordinate, with ellipticity $0.30$ from \citet{1995MNRAS.277.1354I}; this work uses the semi-major axis with $e=0.33$ from \citet{2018ApJ...860...66M}.

\citet{Strigari_2017} diagnose why \citet{2012MNRAS.419..184A} prefer a core; the preference is driven by the lower prediction for the innermost points of the count profiles and, to a lesser extent, a somewhat larger predicted difference in velocity dispersion between the two populations. With their more flexible DF, the count discrepancy at small radius disappears for the NFW potential.

\subsection{The continuous method in context}
\label{sec:cont-context}

The $f(\boldsymbol{J},[\text{Fe}/\text{H}])$ continuous DF does not measure $\gamma$ more accurately than other methods. Section~\ref{sec:decomp-bias} shows the split is the better-matched model for Sculptor, a system taken to be discrete on independent grounds. However, it targets $k_J$, a continuous metallicity-action coupling inaccessible to any split-based method, which the non-converged chain does not deliver here, though the underlying gradient is corroborated in action space.

\section{Limitations}
\label{sec:limitations}

\subsection{Modeling assumptions}
\label{sec:model-assumpt}

All four methods implemented in this work assume spherical symmetry while using the projected elliptical radius (semi-major axis) of Equation~\ref{eq:elliptical-radius}. Sculptor's projected ellipticity is 0.33; its intrinsic three-dimensional shape and any subpopulation misalignment are unobservable from a single viewing angle, so the geometric systematic identified by \citet{2018MNRAS.474.1398G} cannot be fully tested, though the misalignment mechanism specifically is (see Section~\ref{sec:inference-artifacts}). The direction of this systematic is unsettled: \citet{2018MNRAS.474.1398G} report that $\Gamma$ may be over- or underestimated depending on viewing angle, with major-axis views tending to overestimate it. \citet{2016MNRAS.463.1117Z} relax spherical symmetry and still obtain $\gamma=0.5\pm0.3$, consistent with this work's spherical result.

In contrast, \citet{Vitral_2026} model Sculptor as an oblate axisymmetric system and find that the unknown inclination introduces a degeneracy which admits acceptable fits across the full range of profiles from cuspy to cored, with the recovered slope and the velocity anisotropy changing together as the assumed inclination changes. They note that this degeneracy is not detectable under spherical symmetry, as a spherical model has no inclination to marginalize over and thus reports whichever solution the assumed geometry selects. The spherical fits here return well-defined posteriors, but the width of those posteriors does not include the geometric freedom an axisymmetric model would. The sphericity systematic should accordingly be treated as unquantified rather than small, and the internal agreement among the four methods applied here cannot bound it, since all four share the assumption.

The OM anisotropy radius is fixed at $r_a=1.5\,\text{kpc}$ in the spherical Jeans fit. Refitting at $100\,\text{kpc}$, which is effectively isotropic across the fitted range ($\beta<10^{-3}$ at the outermost star), returns $\gamma=0.71^{+0.31}_{-0.40}$ against $0.97^{+0.21}_{-0.33}$. The choice therefore carries a systematic of $\approx0.25$ in $\gamma$, larger than any other modeling choice examined for this method; it also tightens the posterior. Part of the Jeans fit's apparent precision comes from the fixed anisotropy rather than the data.

GravSphere's Plummer scale is fixed from the median observed radius of $0.333\,\text{kpc}$, 22\% above the \citet{2018ApJ...860...66M} photometric half-light radius of $0.273\,\text{kpc}$, because spectroscopic targeting is not spatially random. Refitting at the photometric value gives $\gamma=0.439$ versus $\gamma=0.458$; both chains converged, so the systematic is real but $\approx6\%$ of the statistical uncertainty.

$k_J$ is phenomenological, not derived from physical processes the way \citet{2015MNRAS.449.3479S} derived theirs from disc dynamics. The method is a dSph adaptation of their framework, not a new construction.

$\alpha$ and $\eta$ are fixed ($\alpha=1,\eta=3$) for the Jeans and GravSphere fits but sampled in the continuous DF; the methods are not strictly comparable on identical parameterizations.

\citet{Strigari_2017} note that the assumption of a circularly symmetric Plummer profile with Gaussian velocity and metallicity distributions independent of radius is not consistent with the \citet{Battaglia_2008} data (see Section~\ref{sec:multi-df}). Additionally, \citet{2024A&A...692A.195A} find dispersion profiles declining by $\geq2\,\mathrm{km/s}$ across the radial range in both populations, so the constant-dispersion assumption is inconsistent with this work's data source.

\subsection{Statistical limitations}
\label{sec:stat-limit}

All primary MCMC chains converged within the adopted threshold of $\hat{R}\leq1.01$ (two-population: $\hat{R}\leq1.003$, GravSphere: $\hat{R}\leq1.010$, spherical Jeans: $\hat{R}\leq1.007$; the continuous DF chain is non-converging and is excluded from this list. See Appendix~\ref{ap:mcmc-conv} for convergence diagnostics for all chains). The GravSphere maximum is reached by $\tilde\beta_0$, the parameter with the longest autocorrelation time in Table~\ref{table:mcmc-convergence}; $\gamma$ itself reaches $\hat R=1.004$. The seed-1 replicate of the spherical Jeans method reaches 1.012, and the seed-7 replicate reaches 1.009; see Table~\ref{table:seed-stability}. The $\sigma_\text{los}$-only likelihood is weakly constraining in $\gamma$ (see Section~\ref{sec:sigma-los-degen}), which is why the prior matters rather than the sampler.

Null mocks give a prior shrinkage fraction of $f\approx0.24$, independent of chain length (see Section~\ref{sec:decomp-bias}). Applied naively to the reported slope this would imply a correction of $\approx+0.005$, negligible because 0.97 sits essentially at the prior median of 0.95. In spite of that, the mocks overstate the shrinkage affecting the real fit: refitting the data under a median of 0.75 moves the posterior by only 0.02 (see Section~\ref{sec:sigma-los-degen}), against the 0.05 that $f=0.24$ would predict. Two differences plausibly account for this. The shrinkage fraction is measured on single-population null mocks, whereas the spherical Jeans fit uses two binned dispersion profiles and is correspondingly better constrained; and it is estimated from a single truth value ($\gamma_\text{true}=0.4$), because the $\gamma_\text{true}=1.0$ panel sits almost at the prior median and cannot constrain $f$ where the prior exerts no pull. $f\approx0.24$ should therefore be treated as an upper bound.

\citet{2016MNRAS.463.1117Z} estimate $\sim6000$ discrete stars are needed to separate $\gamma=0$ from $\gamma=1$. This work's Sculptor sample has 1339, bounding what any method here can conclude, including this work's. This is supported by the finding that, with current data, it is not possible to distinguish dark matter density profiles (as per the \citealt{2013A&A...558A..35B} analysis of four dSphs using orbit-based models).

Four realizations of the \citet{2011ApJ...742...20W} mock at $\Gamma_\text{true}=2.94$ recover $2.35$--$3.62$ with a scatter of 0.56, while the per-fit 68\% intervals span only $\pm0.2$ to $\pm0.5$, and only one of four covers truth. The mean deviation ($+0.22$) is $0.8\sigma$, so no systematic bias exists, but the realization-to-realization scatter exceeds what any single posterior reports. The quoted $\Gamma=2.49^{+0.23}_{-0.20}$ somewhat understates the true uncertainty at this sample size, which independently supports the \citet{2016MNRAS.463.1117Z} $\sim6000$-star requirement.

\subsection{Data and sample}
\label{sec:data-sample}

The result in Section~\ref{sec:marginal-feh-uni} concerns the marginal $[\text{Fe}/\text{H}]$ distribution. Mocks matched to the \citet{2024A&A...692A.195A} two-population parameters return the same (dip $p\approx0.96$ versus $0.79$ for the real data), so the test cannot discriminate a continuum from two overlapping populations.

$\sim24$ stars with a $\sim15\,\text{km/s}$ offset from a very-metal-poor subpopulation remain in the sample; $|\Delta\Gamma|\leq0.14$ (see Section~\ref{sec:very-metal-poor-contam}), but the component's nature (minor merger versus Gaussian-fitting artifact) is unresolved. \citet{2024A&A...692A.195A} concede it may be an artifact of forcing Gaussians onto a metal-poor tail.

The selection function is reconstructed from a Gaia DR3 photometric parent rather than the unpublished target catalog of \citet{Walker_2007}, and the \citet{2011ApJ...742...20W} perspective term is omitted; the resulting half-light radii are systematically larger than theirs, and the variant without the RGB color-cut bounds this systematic at $\approx0.2$ in $\Gamma$.

\subsection{Method scope}
\label{sec:method-scope}

The continuous DF does not improve $\gamma$. Section~\ref{sec:decomp-bias} shows the split is the better-matched model for Sculptor, taken to be discrete on independent grounds. Its contribution is $k_J$.

Mock-based decomposition claims must control for geometry. The null and scramble controls in Section~\ref{sec:decomp-bias} attribute $77$--$83\%$ of the mock bias to tracer spatial structure. This applies to this work's own bias-gate results as much as to anyone else's.

\section{Conclusions}
\label{sec:conclusions}
\begin{enumerate}
  \item Two of the four methodologically distinct estimators applied here agree that Sculptor's asymptotic inner slope is shallow: $\Gamma=2.49^{+0.23}_{-0.20}$, 99.6\% of the posterior lies above $\Gamma=2$ (84.4\% after $\lambda$ calibration); $\gamma_{\text{GravSphere}}=0.46^{+0.40}_{-0.28}$. Both are consistent with the \citet{2025A&A...699A.347A} value of 0.39 and the \citet{2016MNRAS.463.1117Z} $0.5\pm0.3$ value from independent data.
  \item The spherical Jeans $\gamma$ value sits higher at $\gamma_{\text{Jeans}}=0.97^{+0.21}_{-0.33}$. Although the $\sigma_\text{los}$-only likelihood is weakly constraining, refitting under a median of 0.75 rather than 0.95 moves the posterior by only 0.02, so this is not a result of the prior choice. It is partly an anisotropy artifact, however, as refitting with the OM radius raised to $100\,\text{kpc}$, effectively isotropic, gives $\gamma=0.71^{+0.31}_{-0.40}$, roughly half the way to the 0.39--0.50 cluster. The residual disagreement is unexplained.
  \item At $150\,\mathrm{pc}$, where the data constrains the profile, the posteriors give $\rho_\mathrm{DM}(150\,\mathrm{pc}) = 1.77^{+0.33}_{-0.26}\times10^8\,M_\odot\,\mathrm{kpc^{-3}}$ (GravSphere) and $1.70\pm0.11\times10^8$ (spherical Jeans), agreeing with the independent GravSphere measurement of \citet{2019MNRAS.484.1401R} at $0.74\sigma$ and placing Sculptor in their cusp-like class. The local logarithmic slope there is $-1.00^{+0.19}_{-0.23}$ under the spherical models used here; \citet{Vitral_2026} find a shallower slope over an overlapping range from resolved 3D velocities, and Section~\ref{sec:central-density} discusses the disagreement. The core-like asymptotic $\gamma$ is an extrapolation inward of the innermost constraint; the two methods agree on the density to $4\%$ while their asymptotic slopes differ by $0.51$.
  \item The slope from the two-population estimator is robust to membership definition ($\Gamma>2$ across PM-offset cuts retaining 100\% down to 25\% of the sample) and to very-metal-poor contamination ($|\Delta\Gamma|\leq0.14$, $\Gamma>2$ throughout). The subcomponents are also aligned to $-4.1\pm4.3\,\text{deg}$, so the misalignment mechanism that produces spurious cores in \citet{2018MNRAS.474.1398G} is not present here.
  \item The continuous $f(\boldsymbol{J},[\text{Fe}/\text{H}])$ model targets $k_J$, a metallicity--action coupling no two-population decomposition can produce. Its chain does not converge on this sample, but the underlying gradient is measured directly in action space ($\sim0.3\,\text{dex}$ over three decades in $J_r$).
  \item On mocks, the model whose assumptions match the underlying truth recovers the slope best in both directions; for Sculptor, a system taken to be discrete on independent grounds, the two-population split is better matched for $\gamma$.
  \item Tracer spatial structure contributes $77$--$83\%$ of the mock decomposition bias. Mock-based claims about population decomposition must control for geometry.
  \item Sculptor's two metallicity subcomponents overlap at $1.35\sigma$: the marginal $[\text{Fe}/\text{H}]$ distribution is unimodal, but so are two-population mocks matched with \citet{2024A&A...692A.195A}, so the marginal test cannot discriminate between a continuum and two overlapping populations.
  \item Applying the same implementation to the \citet{2009AJ....137.3100W} Fornax sample ($N=2603$, Mg index in place of $[\text{Fe}/\text{H}]$) recovers $\Gamma=3.16^{+0.83}_{-0.63}$, with $f_\text{mem}$, $f_\text{sub}$, the chemical separation, and both velocity dispersions agreeing within their uncertainties. This validates the estimator on a second galaxy, though the Fornax posterior sits largely above $\Gamma=3$ and does not yield an interpretable inner slope (see Section~\ref{sec:fornax}).
\end{enumerate}

\section*{Acknowledgments}

I would like to express my sincere gratitude to Dr. Ryan Farber for his invaluable guidance throughout this research. His expertise and insightful advice were crucial in driving this research. Large language model assistance (Anthropic Claude Opus 5) was used for code review, debugging, and editorial review. No text or references in this manuscript were generated by such tools. Steps taken to test and validate the analysis code are described in Appendix~\ref{ap:software}. The author retains full responsibility for the accuracy, integrity, and content of this work.\\

\section*{Data Availability}

The \citet{Tolstoy_I} and \citet{2009AJ....137.3100W} catalogs are publicly available via VizieR (tables \texttt{J/A+A/675/A49} and \texttt{J/AJ/137/3100}); note that mirrors are not always synchronized, and the pipeline accepts a \texttt{VIZIER\_SERVER} environment variable to select one. Gaia DR3 data is available at the European Space Agency COSMOS portal. The data pipeline is available at the GitHub URL (\url{https://github.com/IndyRishi/gammascope}) under CC-BY and archived on Zenodo \citep{sanjeev_2026_zenodo}, along with the raw chain files.

\bibliographystyle{apsrev4-1}

\bibliography{gammascope}

\begin{appendix}

\section{Data overview figures}
\label{ap:data-overview-figs}

\begin{figure*}[!htbp]
    \centering
    \includegraphics[width=\textwidth]{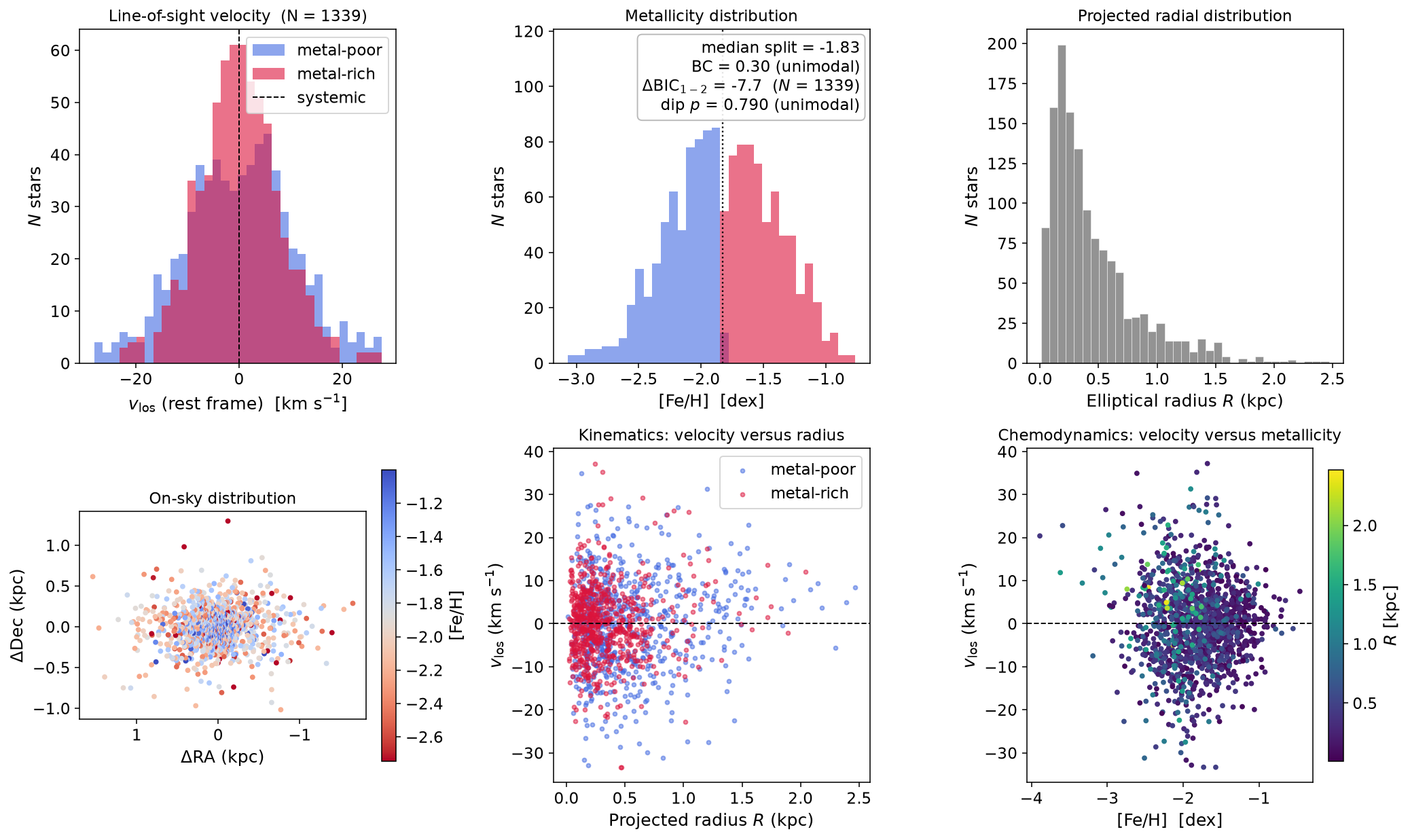}
    \caption{An overview of VLT/FLAMES LR8 Sculptor data, showing LOS velocities; metallicity, projected radial, and on-sky distributions; velocity versus radius kinematics; and velocity versus metallicity kinematics.}
    \label{fig:data-overview}
\end{figure*}

\begin{figure*}[!htbp]
    \centering
    \includegraphics[width=\textwidth]{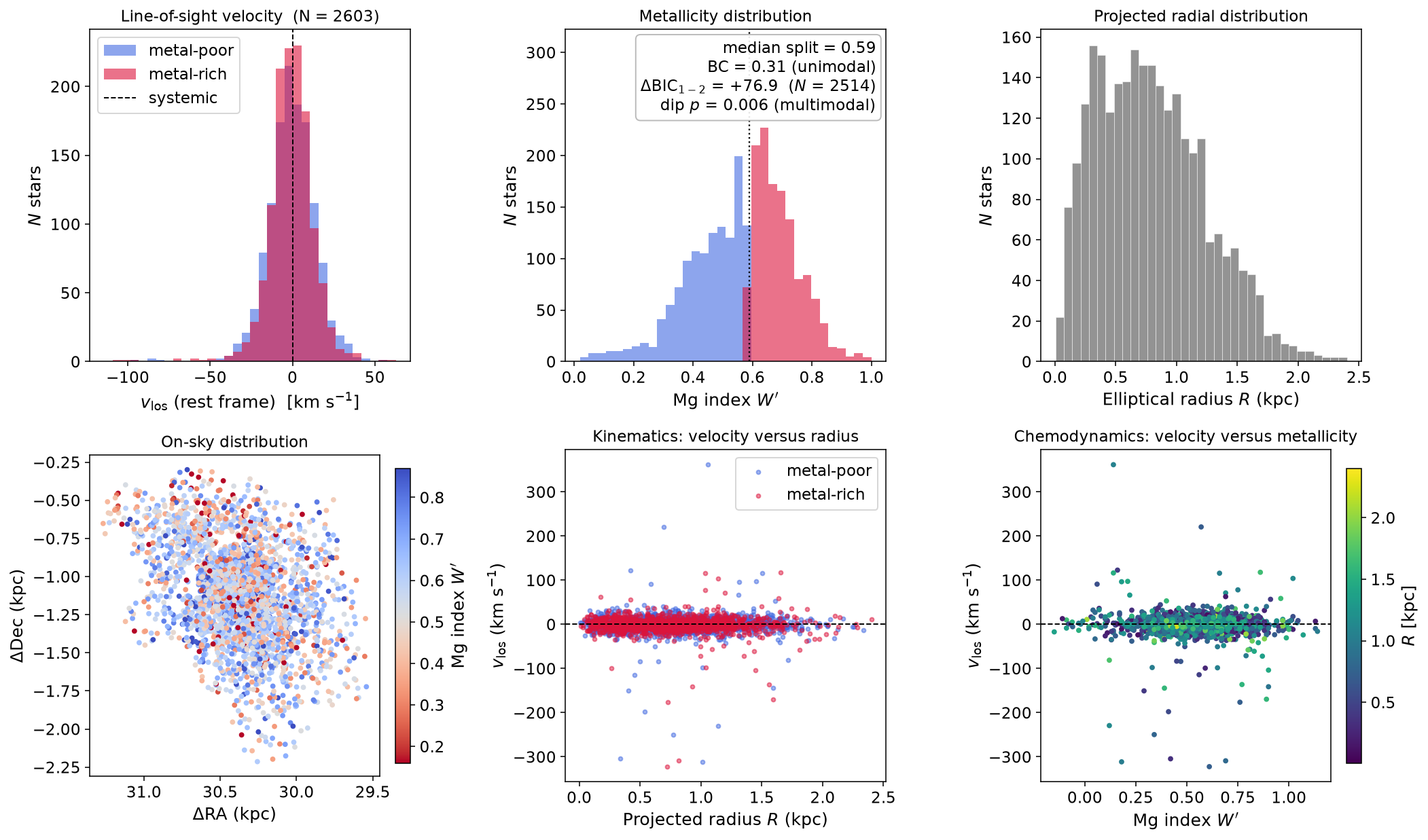}
    \caption{Six panels with a velocity histogram, Mg-index distribution with median, radial distribution, on-sky map, velocity versus radius, and velocity versus Mg index for Fornax; $N=2603$, non-members included.}
    \label{fig:fornax-data-overview}
\end{figure*}

Figures~\ref{fig:data-overview} and~\ref{fig:fornax-data-overview} show the two spectroscopic samples as they enter the analysis. The Sculptor panels use the 1339 \citet{Tolstoy_I} stars with both a velocity and a reliable $[\text{Fe}/\text{H}]$; the metallicity distribution is the one tested for bimodality in Section~\ref{sec:marginal-feh-uni}, the radial distribution sets the Plummer scales discussed in Section~\ref{sec:model-assumpt}, and the velocity-versus-metallicity panel shows the chemodynamical structure the two-population estimator decomposes. The Fornax panels use all 2603 stars recovered from the per-exposure MMFS table, non-members included, since the fit models the foreground rather than cutting it (see Section~\ref{sec:fornax-data}); the velocity histogram's tails are Galactic contamination, and the fitted member fraction $f_\text{mem}=0.953$ quantifies it. The chemical axis of Fornax is the Mg index $W'$ rather than $[\text{Fe}/\text{H}]$, so the two galaxies' metallicity panels are not on a common scale.

\section{Software}
\label{ap:software}

Python 3.11.15 is used with the \texttt{emcee} version 3.1.6 \citep{2013PASP..125..306F} and \texttt{h5py} version 3.16.0 packages for MCMC simulations. \texttt{agama} \citep{2019MNRAS.482.1525V} version 1.0 is used for the continuous DF implementation, QuasiSpherical DFs in the Jeans likelihood, and the potentials throughout, alongside general astronomical and scientific packages \texttt{numpy} version 2.4.6, \texttt{scipy} version 1.17.1, \texttt{astropy} version 8.0.0, \texttt{corner} version 2.2.3, \texttt{astroquery} version 0.4.11, \texttt{matplotlib} version 3.11.0, and \texttt{pandas} version 3.0.3. \texttt{diptest} version 0.11.0 is used for Hartigan's dip test in Section~\ref{sec:marginal-feh-uni}.

Catalog queries use \texttt{astroquery}'s VizieR interface. Mirrors occasionally fall out of sync and return an empty result rather than an error; the pipeline reports this explicitly, and the mirror can be set through the \texttt{VIZIER\_SERVER} environment variable. The row counts printed by each loader (1339 of 1701 rows for Sculptor, 2603 stars rebuilt from 3150 Fornax exposures) confirm that a given mirror has returned the expected sample.

Large-language-model assistance was utilized in the development and debugging of the data analysis pipeline. Every quantitative result this work reports derives from independently validated code. The GravSphere implementation is cross-checked against the reference code of \citet{2017MNRAS.471.4541R} (see Appendix~\ref{ap:grav-cross}), the two-population estimator is validated on synthetic data with a known slope (see Appendix~\ref{ap:mock-recovery}) and reproduces the published Fornax measurement of \citet{2011ApJ...742...20W} (see Section~\ref{sec:fornax}), and the spherical Jeans posterior is checked for seed and prior stability (see Appendix~\ref{ap:mcmc-conv}).

\section{GravSphere cross-validation}
\label{ap:grav-cross}

\begingroup
    \setlength{\tabcolsep}{8pt}
    \renewcommand{\arraystretch}{1.4}
    \begin{table}
        \centering
        \begin{tabular}{ l c c }
            Parameter & Case A & Case B \\
            \hline \hline
            $\gamma$              & 0.6    & 1.0 \\
            $r_s$ (kpc)           & 0.7    & 1.0 \\
            $\rho_s$ ($M_\odot\,\text{kpc}^{-3}$) & $1.5\times10^{8}$ & $6.0\times10^{7}$ \\
            $\tilde{\beta}_0$     & $-0.10$ & 0.00 \\
            $\tilde{\beta}_\infty$ & 0.50   & 0.00 \\
            $\log_{10} r_\beta$   & 0.0    & 0.0 \\
            $a$ (kpc)             & 0.33   & 0.28 \\
            \hline \hline
        \end{tabular}
        \caption{Cross-validation test cases. Case A is Sculptor-like and anisotropic; Case B is an NFW cusp with isotropic orbits.}
        \label{table:gsx-cases}
    \end{table}
\endgroup

\begingroup
    \setlength{\tabcolsep}{5pt}
    \renewcommand{\arraystretch}{1.4}
    \begin{table*}
        \centering
        \begin{tabular}{ l c c c c c c c c }
            & \multicolumn{7}{c}{$R$ (kpc)} & \\
            & 0.05 & 0.10 & 0.20 & 0.33 & 0.50 & 0.80 & 1.20 & max $|$dev$|$ \\
            \hline \hline
            A, this work & 10.725 & 10.992 & 11.159 & 11.303 & 11.450 & 11.388 & 10.919 & \\
            A, reference & 10.748 & 11.015 & 11.178 & 11.315 & 11.456 & 11.385 & 10.913 & 0.21\% \\
            \hline
            B, this work & 12.501 & 11.938 & 11.315 & 11.180 & 11.358 & 11.691 & 11.876 & \\
            B, reference & 12.541 & 11.972 & 11.339 & 11.195 & 11.363 & 11.687 & 11.867 & 0.32\% \\
            \hline \hline
        \end{tabular}
        \caption{$\sigma_\text{los}(R)$ in km\,s$^{-1}$ against the \citet{2017MNRAS.471.4541R} reference implementation, tolerance 1\%.}
        \label{table:gsx-sigma}
    \end{table*}
\endgroup

\begingroup
    \setlength{\tabcolsep}{8pt}
    \renewcommand{\arraystretch}{1.4}
    \begin{table}
        \centering
        \begin{tabular}{ l c c c }
            Case & $v_{s1}$ ratio & $v_{s2}$ ratio & max $|$dev$|$ \\
            \hline \hline
            A & 1.0008 & 1.0010 & 0.10\% \\
            B & 1.0008 & 1.0014 & 0.14\% \\
            \hline \hline
        \end{tabular}
        \caption{Theory VSP ratios (this work/reference), tolerance 3\%.}
        \label{table:gsx-vsp}
    \end{table}
\endgroup

GravSphere is cross-validated against the reference code: two test cases (see Table~\ref{table:gsx-cases}) are evaluated at seven radii ($0.05$ to $1.20\,\text{kpc}$). Case A is Sculptor-like: $\gamma=0.6,\,r_s=0.7\,\text{kpc},\,\rho_s=1.5\times10^8$, anisotropic ($\tilde{\beta_0}=-0.10,\,\tilde{\beta_\infty}=0.50$), $a=0.33\,\text{kpc}$. Case B is an NFW cusp, isotropic: $\gamma=1.0,\,r_s=1.0,\,\rho_s=6\times10^7,\,\tilde{\beta}=0,\,a=0.28$. The first check is a forward model: $\sigma_\text{los}(R)$ compared pointwise; the 0.32\% is the maximum fractional deviation across both cases and all seven radii, against a 1\% tolerance (see Table~\ref{table:gsx-sigma}). The second check is of the theory VSPs; $v_{s1}$ and $v_{s2}$ are compared as ratios, with 0.14\% being the max deviation against a 3\% tolerance (see Table~\ref{table:gsx-vsp}). The third check puts the observed-VSP estimators against the reference on 5,000 synthetic stars; they agree to 1 part in $10^{12}$, and VSP1 is identical. However, VSP2 differs by a known convention: the reference code normalizes by $\langle R^2\rangle$, while this work's test is the direct unbiased estimator of the same integral. Likelihood surfaces were also compared end-to-end; over 60 posterior draws, there was a mean offset of $+0.101$, rms residual of 0.122, max of 0.681, all well below $\Delta\chi^2=1$.

\section{Mock recovery tests}
\label{ap:mock-recovery}

\begin{figure*}[!htbp]
    \centering
    \includegraphics[width=\textwidth]{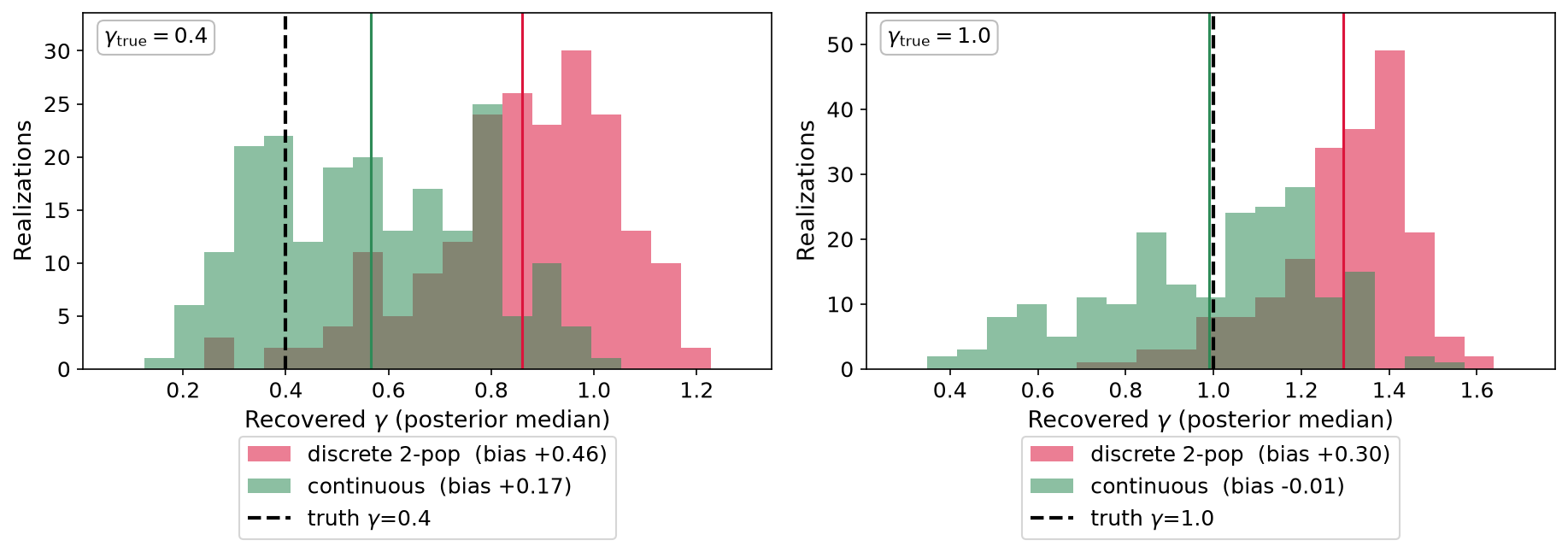}
    \caption{Two panels, one per $\gamma_\text{true}$, each with the recovered-$\gamma$ distributions from 200 realizations under continuous truth; one arm fit with a median metallicity split, one fit as a continuous population, with a dashed line at the true value and the mean bias annotated. The figure shows whether the two arms' distributions actually separate or just have different centers. Split biases $+0.459\pm0.013$ at $\gamma_\text{true}=0.4$ and $+0.296\pm0.011$ at $\gamma_\text{true}=1.0$; continuous $+0.165\pm0.014$ and $-0.009\pm0.018$. Splitting a continuum introduces a large positive bias where the continuous fit is nearly unbiased.}
    \label{fig:bias-gate-fig}
\end{figure*}

\begin{figure*}[!htbp]
    \centering
    \includegraphics[width=\textwidth]{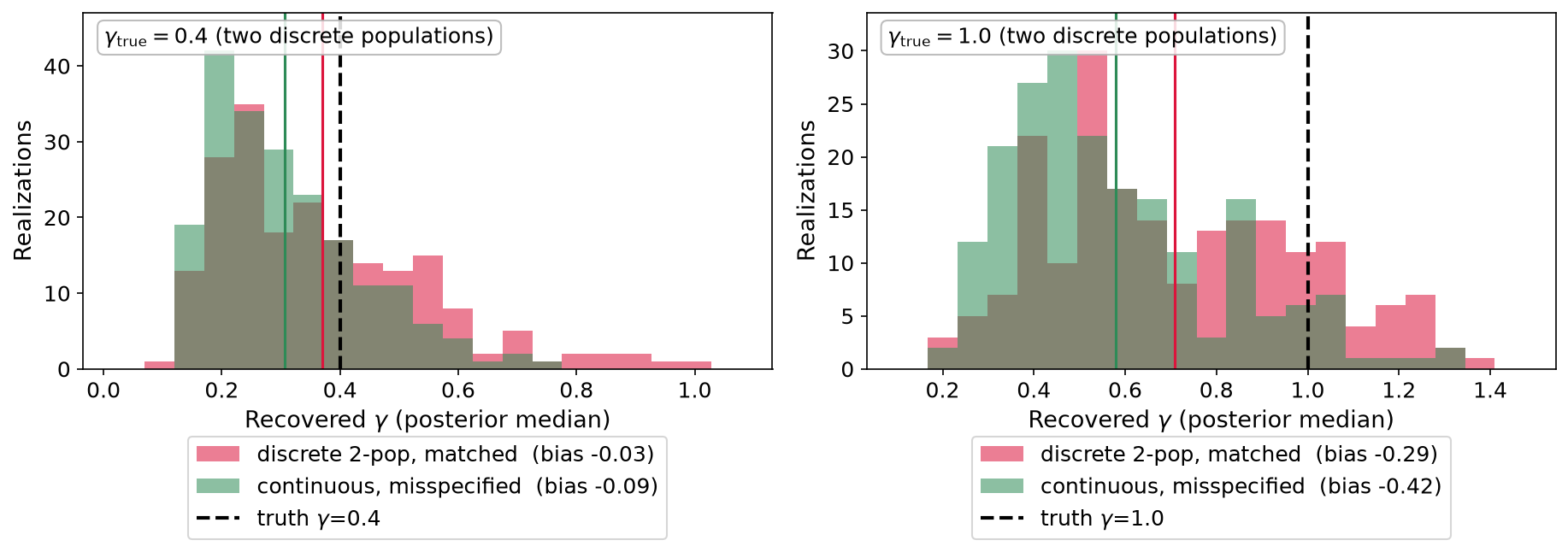}
    \caption{Two panels, one per $\gamma_\text{true}$, each with the recovered-$\gamma$ distributions from 200 realizations under discrete truth; one arm fit with a median metallicity split, one fit as a continuous population, with a dashed line at the true value and the mean bias annotated. The figure shows whether the two arms' distributions actually separate or just have different centers. Split biases $-0.029\pm0.013$ at $\gamma_\text{true}=0.4$ and $-0.292\pm0.020$ at $\gamma_\text{true}=1.0$; continuous $-0.094\pm0.009$ and $-0.420\pm0.017$. The continuous model carries an extra 0.06--0.13 of bias when the truth is discrete, and both arms are biased low.}
    \label{fig:rev-bias-gate-fig}
\end{figure*}

All mocks used in mock recovery tests use $N=1338$ stars in a common gNFW potential with $r_s=1.0\,\text{kpc},\,\rho_s=6\times10^7\,M_\odot\,\text{kpc}^{-3},\,\alpha=1,\,\eta=3$, at $\gamma_\text{true}=0.4$ and $1.0$. There are two truth values with 200 realizations each, including 20 for the shrinkage sweep. Recovery is by the posterior median from the same short isotropic-gNFW $\sigma_{\text{los}}$ MCMC as the real analysis; the mocks share the real fit's degeneracy rather than testing an idealized one. A forward gate experiment was run (continuous truth, one Plummer at $a_0=0.28\,\text{kpc}$ with a smooth gradient; see Figure~\ref{fig:bias-gate-fig}) as well as a reverse gate (discrete truth, two Plummers at $a=0.1875$ and $0.3983\,\text{kpc}$ matching \citet{2024A&A...692A.195A} Table C.2; see Figure~\ref{fig:rev-bias-gate-fig}) and null and scramble controls. The chain-length sweep was at 400/1600/6400 steps.

The scatter across realizations is consistently larger at $\gamma_\text{true}=1.0$ than at 0.4, by roughly 50\% in the standard error. Two effects contribute: a cuspier profile concentrates mass at small radii, so $\sigma_\text{los}(R)$ over the fitted range is less sensitive to $\gamma$ near 1 than near 0.4, and a flatter likelihood lets the posterior median wander further between realizations; this is the same degeneracy documented on the real data in Section~\ref{sec:sigma-los-degen}. In addition, $\gamma_\text{true}=1.0$ sits essentially at the prior median of 0.95, so prior shrinkage supplies no consistent pull to stabilize the recovery, whereas at $\gamma_\text{true}=0.4$ it pushes every realization in the same direction.

\begingroup
    \setlength{\tabcolsep}{10pt}
    \renewcommand{\arraystretch}{1.5}
    \setlength\extrarowheight{2pt}
    \begin{table}
        \centering
        \begin{tabular}{ l c c }
            Experiment & $\gamma_{\text{true}}=0.4$ & $\gamma_{\text{true}}=1.0$ \\
            \hline \hline
            Null, split        & $+0.174\pm0.014$ & $+0.013\pm0.018$ \\
            Null, all-star     & $+0.162\pm0.014$ & $+0.004\pm0.018$ \\
            Scramble, split    & $-0.082\pm0.011$ & $-0.426\pm0.017$ \\
            Scramble, all-star & $-0.085\pm0.010$ & $-0.440\pm0.016$ \\
            \hline \hline
        \end{tabular}
        \caption{Recovered $\gamma$ bias in the null and label-scramble controls. The change from
        null to scramble isolates the effect of tracer geometry; the residual difference between
        split and all-star fits within each control isolates the decomposition. These are the
        values behind the split-vs-all-star comparison and null$\to$scramble shift in Section~\ref{sec:decomp-bias}.}
        \label{table:gate-diagnostics-table}
    \end{table}
\endgroup

\subsection{Two-population estimator recovery}

The two-population estimator is validated on a synthetic dSph with a known slope: 1000 stars in two Plummer subcomponents sharing one potential, with $f_{\text{sub}}=0.5$, $r_{h,1}/r_{h,2}=0.55$, $r_{h,2}=300\,\text{pc}$, $\sigma_{V,1}=6.5\,\text{km/s}$, $\sigma_{V,2}=11.6\,\text{km/s}$, $\langle[\text{Fe}/\text{H}]\rangle_1=-1.5$, $\Delta[\text{Fe}/\text{H}]=0.5$, and fixed uncertainties of $2.0\,\text{km/s}$ and $0.1\,\text{dex}$. Velocities and metallicities are Gaussians; projected radii are drawn from the Plummer inverse cumulative distribution. The slope truth value comes from the initial parameters rather than being arbitrary:\begin{equation}\Gamma_{\text{true}} = 1 + \frac{\log_{10}(\sigma_{V,2}^2/\sigma_{V,1}^2)}{\log_{10}(r_{h,2}/r_{h,1})} = 2.94\end{equation} which places the test near the real Sculptor data value (see Section~\ref{sec:wp11-reprod}) rather than at an arbitrary point in parameter space. The fit recovers a mean $\Gamma$ of 3.15, with a range of 2.35--3.62 and a scatter of 0.56.

\section{MCMC convergence}
\label{ap:mcmc-conv}

\begin{figure*}[!htbp]
    \centering
    \includegraphics[width=\textwidth]{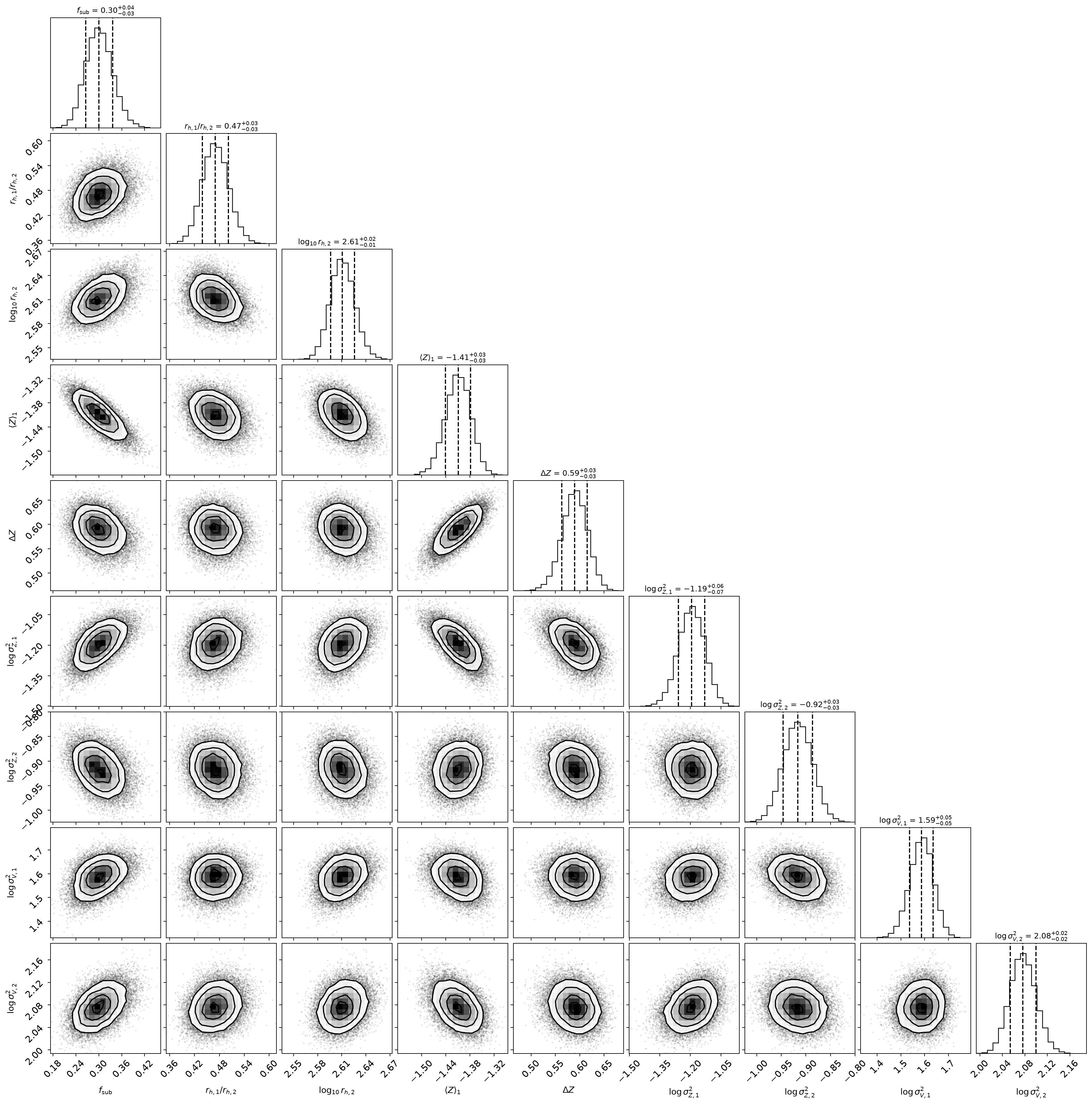}
    \caption{Corner plot for the \citet{2011ApJ...742...20W} two-population estimator on Sculptor: nine parameters ($f_\text{sub},\,r_{h,1}/r_{h,2},\,\log_{10}{r_{h,2}},\,\langle[\text{Fe}/\text{H}]\rangle_1,\,\Delta[\text{Fe}/\text{H}]$, and the four log variances) with 1D marginals on the diagonal showing median and 16th/84th percentiles as dashed lines, and 2D contours below. The contours are clean and elliptical with no multimodality. The two parameters that matter most for $\Gamma$ are $r_{h,1}/r_{h,2}$ and the velocity variances; they are well constrained. There is a visible anticorrelation between $f_\text{sub}$ and $\langle[\text{Fe}/\text{H}]\rangle_1$, as a larger metal-rich fraction pulls its mean metallicity down. $\Gamma$ is not a sampled parameter; instead, it is derived from $r_{h,1},\,r_{h,2},\,\sigma^2_{V,1}$ and $\sigma^2_{V,2}$ per sample, and thus it does not appear in the corner.}
    \label{fig:wp11-corner}
\end{figure*}

\begin{figure*}[!htbp]
    \centering
    \includegraphics[width=\textwidth]{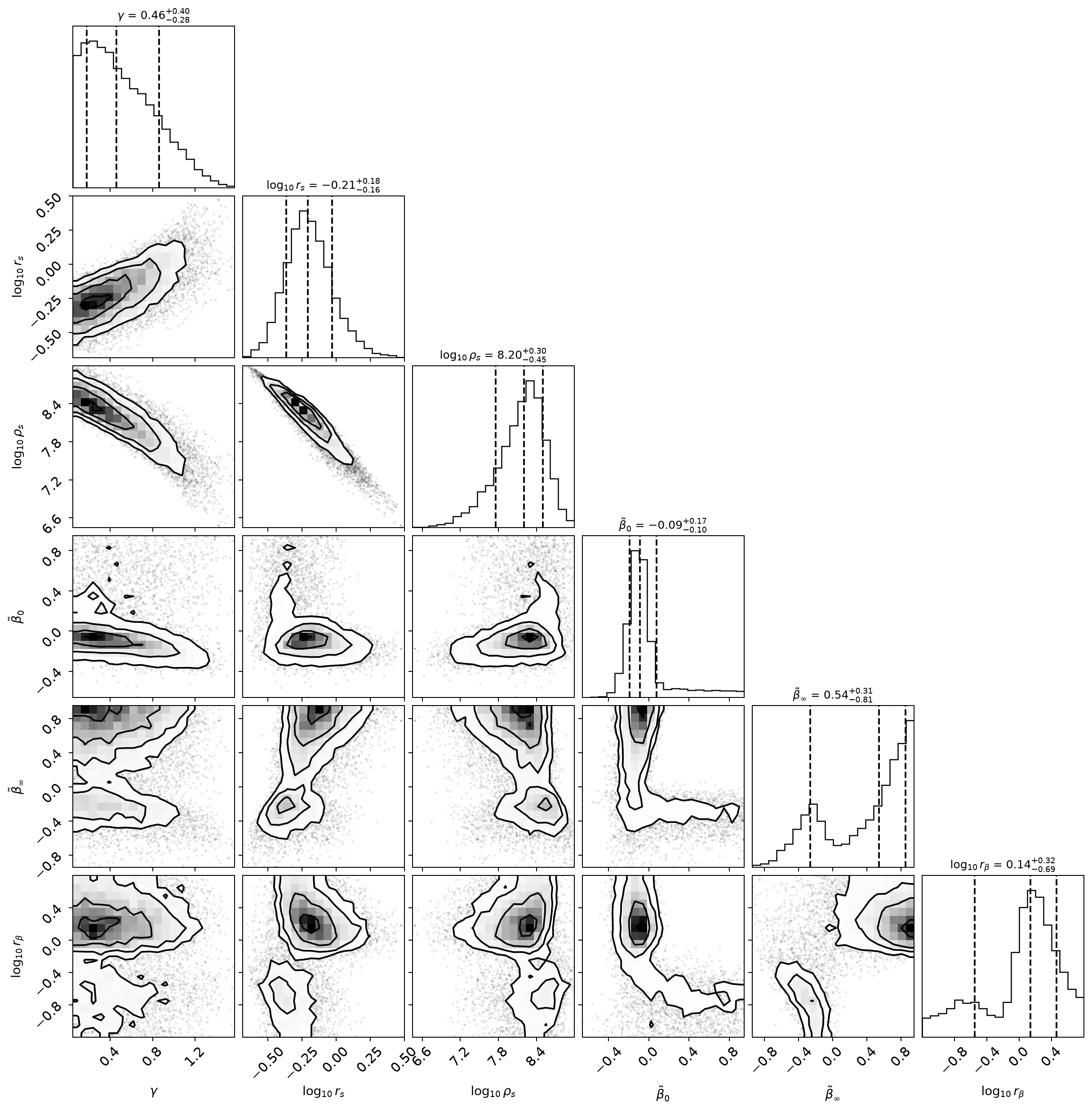}
    \caption{Corner plot for GravSphere: six parameters ($\gamma,\,\log_{10}{r_s},\,\log_{10}{\rho_s},\,\tilde{\beta_0},\,\tilde{\beta}_\infty,\,\log_{10}{r_\beta}$) with medians and 16th/84th percentiles as dashed lines. There is a strong $\rho_s$-$r_s$ anticorrelation in the third row, a sign of halo degeneracy; a denser, smaller halo and a shallower, larger halo both fit, and $\gamma$ correlates with both, so its posterior is wide. $\tilde{\beta}_\infty$ is bimodal, with two clear peaks near $-0.2$ and $+0.8$, which is why its interval is asymmetric ($^{+0.31}_{-0.81}$) and why it has $\tau\approx399$ in Table~\ref{table:mcmc-convergence}. The outer anisotropy is not well determined as both tangential and radial outer orbits are allowed. $\gamma$'s posterior is one-sided, peaking near 0.2 and falling toward higher values, with a tail to $\sim1.2$, so the constraint is an upper limit on cuspiness rather than a detection at 0.46. $\tilde{\beta}_0$ is tightly peaked near zero; the inner orbits are well constrained as near-isotropic, which contrasts sharply with $\tilde{\beta}_\infty$ and explains the different $\tau$ values between them.}
    \label{fig:gravsphere-corner}
\end{figure*}

\begin{figure*}[!htbp]
    \centering
    \includegraphics[width=\textwidth]{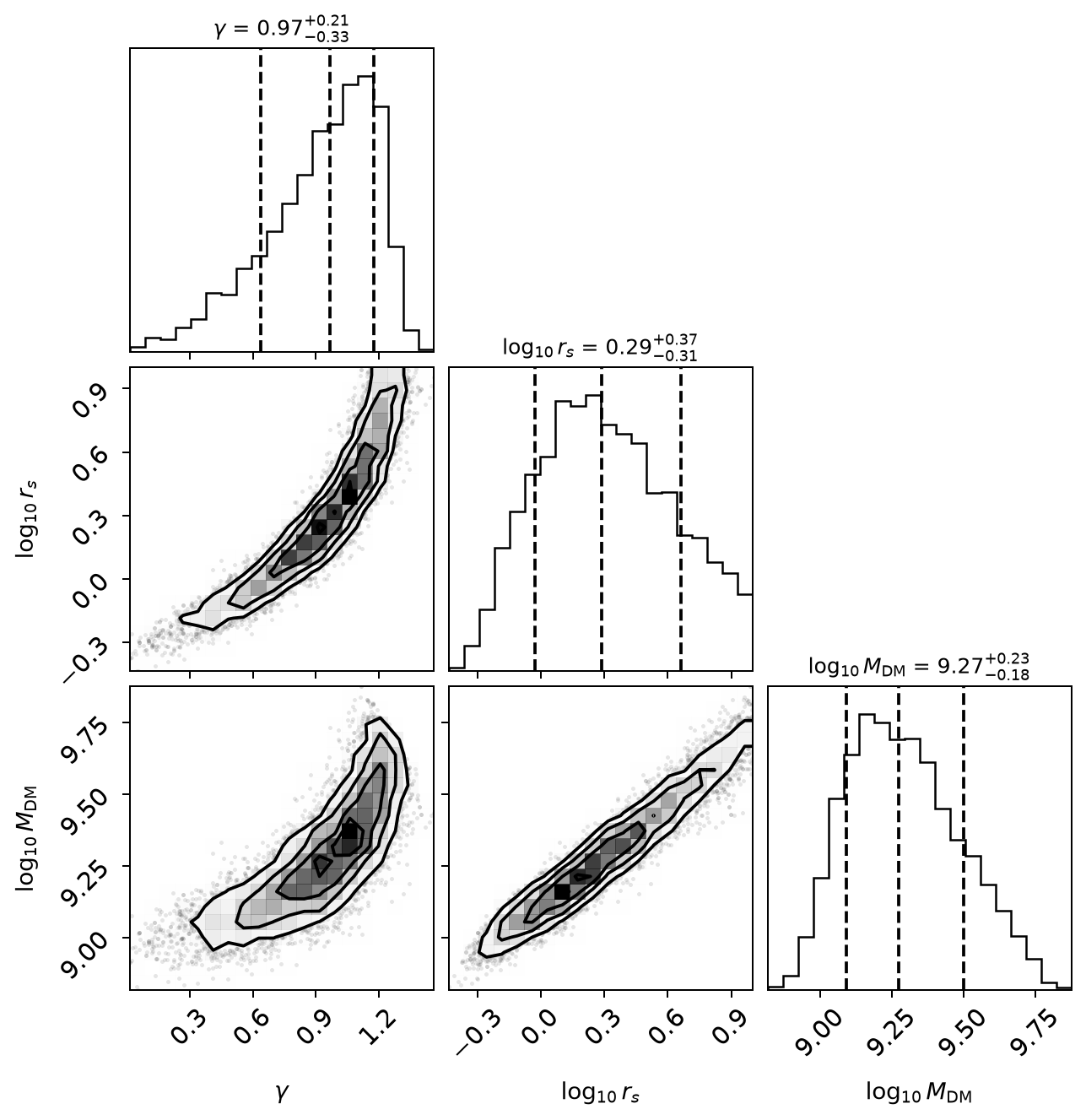}
    \caption{Corner plot for the spherical Jeans method, with three parameters: $\gamma=0.966^{+0.209}_{-0.329},\,\log_{10}{r_s}=0.286^{+0.374}_{-0.313},\,\log_{10}{M_{\rm DM}}=9.274^{+0.227}_{-0.183}$ with medians and 16th/84th percentiles as dashed lines. There is a correlation between $\gamma$ and $r_s$, and the $\gamma$ posterior spans nearly the full prior (see Section~\ref{sec:sigma-los-degen}).}
    \label{fig:dm5-corner}
\end{figure*}

\begin{figure*}[!htbp]
    \centering
    \includegraphics[width=\textwidth]{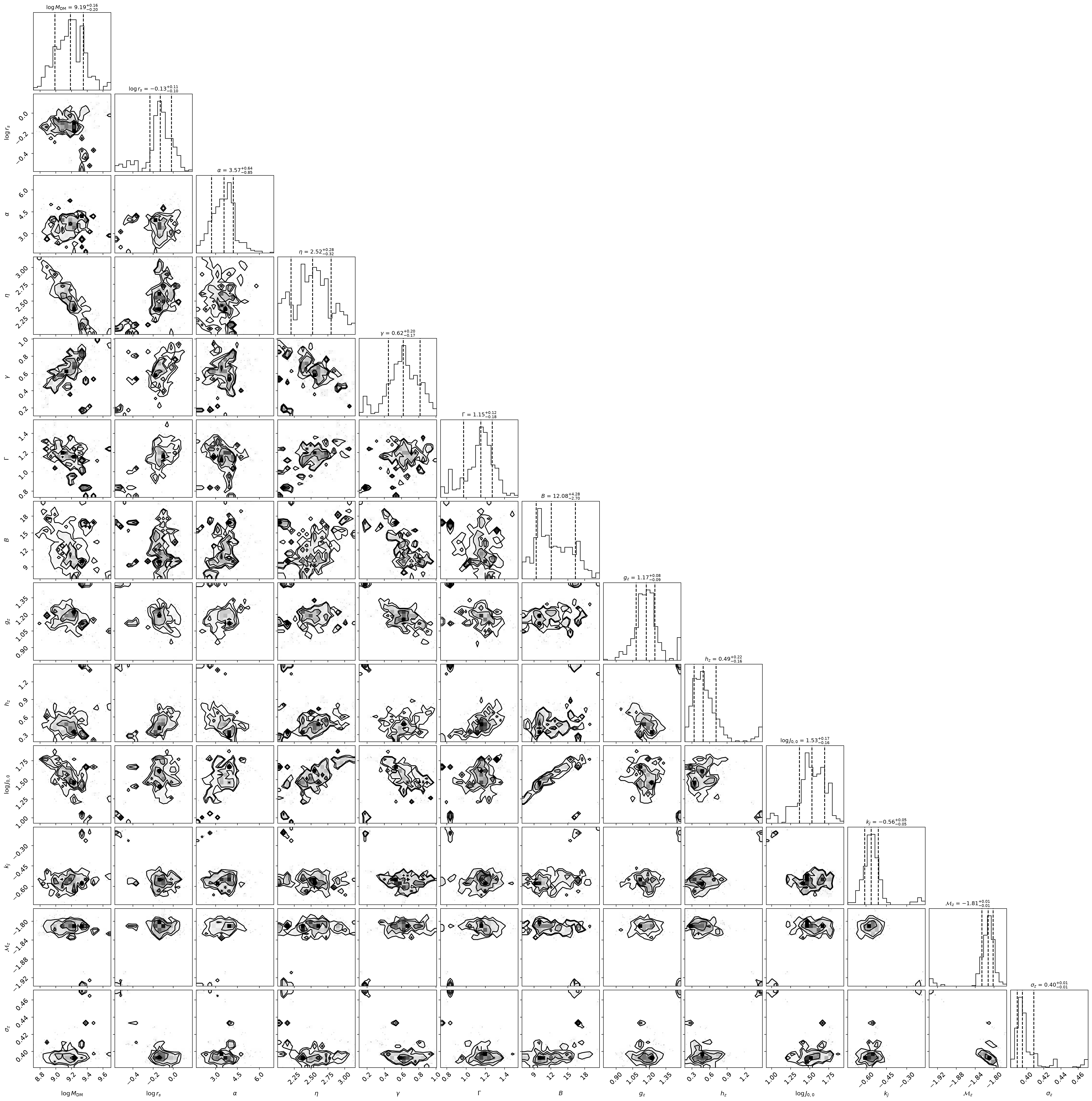}
    \caption{Corner plot for the continuous DF method, with 13 sampled parameters plus medians and 16th/84th percentiles as dashed lines. The chain is non-converging; the contours should not be read as a posterior measurement.}
    \label{fig:cont-corner}
\end{figure*}

\begin{figure*}[!htbp]
    \centering
    \includegraphics[width=\textwidth]{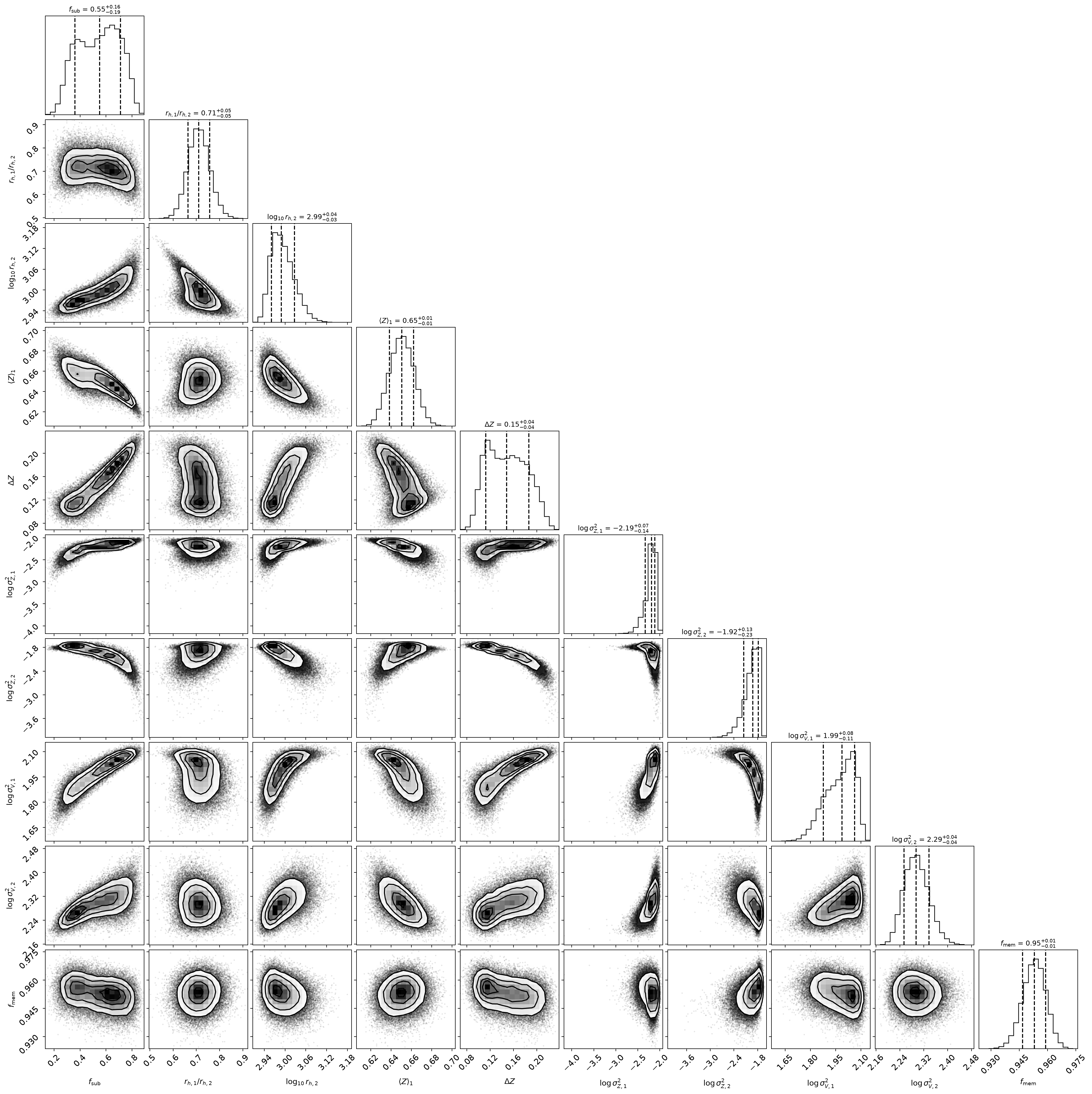}
    \caption{Corner plot for the \citet{2011ApJ...742...20W} two-population estimator on Fornax: ten parameters ($f_\text{sub},\,r_{h,1}/r_{h,2},\,\log_{10}{r_{h,2}},\,\langle W'\rangle_1,\,\Delta W',\,f_\text{mem}$, and the four log variances) with 1D marginals on the diagonal showing median and 16th/84th percentiles as dashed lines, and 2D contours below. $f_\text{mem}=0.953\pm0.006$ is the best-constrained parameter of the fit, and the chemical separation $\Delta W'=0.150^{+0.038}_{-0.035}$ is resolved rather than collapsing to zero. The radius ratio $r_{h,1}/r_{h,2}=0.722^{+0.049}_{-0.047}$ sits above the \citet{2011ApJ...742...20W} value of 0.62; since $\Gamma$ depends on the logarithm of this ratio, it is the parameter to which the recovered slope is most sensitive (see Section~\ref{sec:fornax}).}
    \label{fig:fornax-corner}
\end{figure*}

Convergence is assessed by the Gelman--Rubin statistic $\hat{R}$, the integrated autocorrelation time $\tau$, the effective sample size (ESS), and the mean acceptance fraction. The adopted threshold is $\hat{R}\leq1.01$ and a chain length of at least $50\tau$.

\begingroup
    \setlength{\tabcolsep}{8pt}
    \renewcommand{\arraystretch}{1.4}
    \begin{table}
        \centering
        \begin{tabular}{ l c c c c }
            Parameter & $\tau$ & $N/\tau$ & $\hat{R}$ & ESS \\
            \hline \hline
            \multicolumn{5}{l}{\textit{Two-population, Sculptor} (20{,}000 steps, 64 walkers, acceptance 0.44)} \\
            $f_{\rm sub}$              & 100.1 & 199.8 & 1.005 & 12041 \\
            $r_{h,1}/r_{h,2}$          & 101.9 & 196.3 & 1.005 & 12157 \\
            $\log_{10}r_{h,2}$         & 95.9 & 208.5 & 1.005 & 12715 \\
            $\langle$[\text{Fe}/\text{H}]$\rangle_1$ & 103.0 & 194.2 & 1.005 & 11793 \\
            $\Delta$[\text{Fe}/\text{H}]             & 101.4 & 197.2 & 1.007 & 11920 \\
            $\log\sigma^2_{Z,1}$       & 94.6 & 211.5 & 1.005 & 12852 \\
            $\log\sigma^2_{Z,2}$       & 102.8 & 194.6 & 1.005 & 11960 \\
            $\log\sigma^2_{V,1}$       & 101.3 & 197.5 & 1.006 & 11998 \\
            $\log\sigma^2_{V,2}$       & 103.8 & 192.7 & 1.006 & 10463 \\
            \hline
            \multicolumn{5}{l}{\textit{Two-population, Fornax} (100{,}000 steps, 64 walkers, acceptance 0.33)} \\
            $f_{\rm sub}$              & 587.0 & 170.4 & 1.007 & 10015 \\
            $r_{h,1}/r_{h,2}$          & 230.5 & 433.9 & 1.002 & 26655 \\
            $\log_{10}r_{h,2}$         & 478.0 & 209.2 & 1.005 & 12704 \\
            $\langle W'\rangle_1$      & 409.2 & 244.4 & 1.004 & 14665 \\
            $\Delta W'$                & 572.2 & 174.8 & 1.007 & 10263 \\
            $\log\sigma^2_{Z,1}$       & 462.7 & 216.1 & 1.004 & 13185 \\
            $\log\sigma^2_{Z,2}$       & 601.1 & 166.4 & 1.006 & 9779 \\
            $\log\sigma^2_{V,1}$       & 532.4 & 187.8 & 1.006 & 11363 \\
            $\log\sigma^2_{V,2}$       & 336.6 & 297.1 & 1.003 & 18525 \\
            $f_{\rm mem}$              & 232.4 & 430.4 & 1.002 & 26429 \\
            \hline
            \multicolumn{5}{l}{\textit{GravSphere} (40000 steps, 24 walkers, acceptance 0.12)} \\
            $\gamma$                   & 206.0 & 194.2 & 1.004 & 4060 \\
            $\log_{10}r_s$             & 214.8 & 186.2 & 1.004 & 4313 \\
            $\log_{10}\rho_s$          & 201.1 & 198.9 & 1.003 & 4625 \\
            $\tilde\beta_0$            & 519.1 & 77.1 & 1.010 & 1665 \\
            $\tilde\beta_\infty$       & 398.8 & 100.3 & 1.008 & 2278 \\
            $\log_{10}r_\beta$         & 433.6 & 92.3 & 1.007 & 2124 \\
            \hline
            \multicolumn{5}{l}{\textit{Spherical Jeans} (5000 steps, 24 walkers, acceptance 0.22)} \\
            $\gamma$                   & 42.5 & 117.7 & 1.007 & 2736 \\
            $\log_{10}r_s$             & 41.6 & 120.1 & 1.007 & 2776 \\
            $\log_{10}M_{\rm DM}$      & 37.3 & 134.2 & 1.007 & 3070 \\
            \hline
            \multicolumn{5}{l}{\textit{Continuous} $f(\boldsymbol{J},[\text{Fe}/\text{H}])$} \\
            \multicolumn{5}{l}{Did not reach $\hat{R}\leq1.01$; see text.} \\
            \hline \hline
        \end{tabular}
        \caption{Convergence diagnostics for all chains.}
        \label{table:mcmc-convergence}
    \end{table}
\endgroup

All reported chains were run entirely under the final likelihood, with no resumption across parameter changes.

The continuous DF chain is the one fit that does not meet the convergence criteria; a bimodal posterior rather than a slow sampler. On the full 1339-star sample, 486 steps at 28 walkers left $\hat{R}$ between 1.26 and 1.60 with no downward trend from step 90 onward, despite an acceptance fraction of 0.20 and a move set that already mixes the affine-invariant stretch proposal with differential-evolution and snooker proposals \citep[see][]{terBraak2008}. Inspecting the per-walker likelihoods reveals a clean separation: 26 walkers sit near $\ln{P}=-6753$ and two near $-6695$, a gap of 57 log units. The two groups occupy physically distinct solutions, with $\gamma=0.23$ against $0.62$ and $k_J=-0.23$ against $-0.56$, and the ensemble does not migrate between them.

A restart seeded at the higher-likelihood mode, using 56 walkers on a 500-star subsample, diagnoses the failure further. The seeded ensemble remained near that mode over 372 steps, with $\hat{R}$ falling from 1.82 and then stalling between $1.37$ and $1.76$, and two parameters explain why. The stellar-DF coefficients $g_{z,p}$ and $h_{z,p}$ sit at  1.486 and 1.499 against a prior ceiling of 1.5, and remained there at every checkpoint. Spherical symmetry fixes the radial coefficient at $g_{r,p}=3-2g_{z,p}$ so $g_{z,p}=1.5$ is where the radial-action dependence of the double power law fails, and \citet{2018arXiv180208255V} requires the three coefficients to be non-negative. The data therefore prefers a weaker dependence on $J_r$ than the spherically-constrained double-power-law family admits, and walkers accumulate against that boundary rather than exploring freely.

The non-convergence describes model flexibility rather than sampler tuning, as the failure would persist at any chain length, and relaxing the constraint would reintroduce the scale degeneracy between the action coefficients and $J_{p,\rm ref}$ that the sum-to-three convention exists to remove. Every parameter is nonetheless constrained by the data, with 68\% posterior widths between 2\% and 25\% of their prior ranges, so the failure is not one of identifiability. No posterior is quoted for $k_J$ (see Section~\ref{sec:action-space-grad}).

Chains were run on two machines. The spherical Jeans, GravSphere, and both two-population fits ran on a six-core CPU; the continuous DF chain ran on a 22-vCPU Google Cloud Platform instance (c3-standard-22), where a single step of the 1339-star likelihood takes roughly thirteen minutes across 22 workers, and the 500-star subsample seeded restart roughly six. That cost is what bounds the length of the DF chain, but it is not what limits their convergence. The boundary behavior described above would persist at any chain length.

Chains are checkpointed to per-configuration backends, and resumption is guarded against configuration changes, so the sensitivity runs of Table~\ref{table:mcmc-convergence}, Table~\ref{table:seed-stability}, and Table~\ref{table:prior-sensitivity} cannot contaminate the headline posteriors.

\subsection{Seed stability of the spherical Jeans posterior}

Because the $\sigma_{\text{los}}$-only likelihood is weakly constraining in $\gamma$ (see Section~\ref{sec:sigma-los-degen}), the position at which an ensemble sampler settles depends on walker origin. The Jeans fit was therefore repeated from independent walker initializations with all other settings held fixed.

\begingroup
    \setlength{\tabcolsep}{10pt}
    \renewcommand{\arraystretch}{1.4}
    \begin{table}
        \centering
        \begin{tabular}{ c c c c }
            Seed & $\gamma$ & $\hat{R}$ & $\tau$ \\
            \hline \hline
            42 & $0.97^{+0.21}_{-0.33}$ & 1.007 & 42.5 \\
            1  & $0.96^{+0.20}_{-0.33}$ & 1.012 & 43.3 \\
            7  & $0.96^{+0.21}_{-0.31}$ & 1.009 & 40.0 \\
            \hline \hline
        \end{tabular}
        \caption{Spherical Jeans posterior from independent walker initializations, 5000 steps each on separate backends. The three seeds agree to within 0.01 in $\gamma$, well inside the credible interval, so the posterior is determined by the data and prior rather than by the starting ensemble. This isolates the prior sensitivity of Section~\ref{sec:sigma-los-degen} from sampler variance.}
        \label{table:seed-stability}
    \end{table}
\endgroup

\subsection{Prior sensitivity of the spherical Jeans posterior}

The weakly constraining $\sigma_\text{los}$-only likelihood (see Section~\ref{sec:sigma-los-degen}) raises the concern that the reported slope reflects the prior rather than the data, which is the reason \citet{2014MNRAS.441.1584R} avoid posterior inference entirely. The fit was therefore repeated under a narrower uniform prior, moving the prior median from 0.95 to 0.75 while holding the seed, walker count, and chain length fixed, so that any shift in the posterior is attributable to the prior alone. The posterior falls off well inside both bounds, so the narrower prior's truncation of the upper tail is not what produces the small shift.

\begingroup
    \setlength{\tabcolsep}{10pt}
    \renewcommand{\arraystretch}{1.4}
    \begin{table}
        \centering
        \begin{tabular}{ c c c c c }
            Prior on $\gamma$ & Median & $\gamma$ & $\hat{R}$ & $\tau$ \\
            \hline \hline
            $\mathcal{U}(0,1.9)$ & 0.95 & $0.97^{+0.21}_{-0.33}$ & 1.007 & 42.5 \\
            $\mathcal{U}(0,1.5)$ & 0.75 & $0.95^{+0.21}_{-0.35}$ & 1.012 & 46.9 \\
            \hline \hline
        \end{tabular}
        \caption{Spherical Jeans posterior under two uniform priors on $\gamma$, 5000 steps and 24 walkers each on separate backends, seed 42 throughout. Moving the prior median by 0.20 moves the posterior median by 0.02, so the reported slope is set by the data rather than by the prior despite the weakly constraining $\sigma_\text{los}$-only likelihood (see Section~\ref{sec:sigma-los-degen}). The posterior falls off well inside both bounds, so the narrower prior's truncation of the upper tail is not what produces the small shift.}
        \label{table:prior-sensitivity}
    \end{table}
\endgroup

\subsection{Anisotropy sensitivity of the spherical Jeans posterior}

The OM anisotropy radius is fixed rather than sampled (see Section~\ref{sec:spherical-jeans}), so its effect on the recovered slope is measured by refitting at a second value. Setting $r_a=100\,\text{kpc}$ effectively makes the profile isotropic throughout the fit range, as $\beta=r^2/(r^2+r_a^2)<10^{-3}$ for the outermost star.

\clearpage

\begingroup
    \setlength{\tabcolsep}{10pt}
    \renewcommand{\arraystretch}{1.4}
    \begin{table}
        \centering
        \begin{tabular}{ c c c c }
            $r_a$ (kpc) & $\gamma$ & $\hat{R}$ & $\tau$ \\
            \hline \hline
            1.5 & $0.97^{+0.21}_{-0.33}$ & 1.007 & 42.5 \\
            100 & $0.71^{+0.31}_{-0.40}$ & 1.007 & 36.2 \\
            \hline \hline
        \end{tabular}
        \caption{Spherical Jeans posterior under two Osipkov--Merritt anisotropy radii,
        5000 steps and 24 walkers each on separate backends, seed 42 throughout. Removing
        the anisotropy moves the slope down by 0.26, toward the GravSphere and
        two-population values, and widens the interval by roughly 40\%: the fixed $r_a$
        contributes to the apparent precision of the headline Jeans fit as well as to its
        value. Both chains converged.}
        \label{table:anisotropy-sensitivity}
    \end{table}
\endgroup

\end{appendix}

\end{document}